\documentclass{aa}  

\usepackage{graphicx}
\usepackage{subcaption}
\usepackage{geometry}
\usepackage{pdflscape}
\usepackage{natbib}
\usepackage{multirow}
\usepackage{cellspace}
\usepackage{footnote}
\usepackage{txfonts}
\usepackage{float}
\usepackage{hyperref} 
\bibpunct{(}{)}{;}{a}{}{,}
\def\reff{$r_e$}

\begin{document} 
   \title{Structural analysis of massive galaxies using HST
deep imaging at $z < 0.5$}
      \author{Sandra N. dos Reis
          \inst{1,2}
          \and
          Fernando Buitrago\inst{1,2}
          \and
          Polychronis Papaderos\inst{1,2,3}
          \and
          Israel Matute\inst{1}
          \and
          José Afonso\inst{1,2}
          \and
          Stergios Amarantidis\inst{1,2}
          \and
          Iris Breda\inst{3}
          \and
          Jean M. Gomes\inst{3}
          \and
          Andrew Humphrey\inst{3}
          \and
          Catarina Lobo\inst{3,4}
          \and
          Silvio Lorenzoni\inst{1}
          \and
          Cirino Pappalardo\inst{1}
          \and
          Ana Paulino-Afonso\inst{1}
          \and
          Tom Scott\inst{3}
          }
   \institute{Instituto de Astrofísica e Ciências do Espaço, Universidade de Lisboa, OAL, Tapada da Ajuda, PT1349-018 Lisboa, Portugal
              \and
              Departamento de Física, Faculdade de Ciências da Universidade de Lisboa, Edifício C8, Campo Grande, PT1749-016 Lisboa, Portugal\\
              \email{sreis@oal.ul.pt}
              \and
              Instituto de Astrofísica e Ciências do Espaço, Universidade do Porto, CAUP, Rua das Estrelas, PT4150-762 Porto, Portugal
              \and
              Departamento de F\'{\i}sica e Astronomia, Faculdade de Ci\^encias, Universidade do Porto, Rua do Campo Alegre 687, PT4169-007 Porto, Portugal
             }
   \date{Received ; accepted 26-Nov-2019}

  \abstract 
  {The most massive ($M_\textrm{stellar} \geq 10^{11} M_\odot$) galaxies in the local Universe are characterized by having a bulge-dominated morphology and old stellar populations, in addition to be confined to a tight mass-size relation. Identifying their main components can provide insights into their formation mechanisms and subsequent mass assembly.}
{Taking advantage of Hubble Space Telescope (HST) CANDELS data, we analyze the lowest redshift ($z < 0.5$) massive galaxies in the $H$ and $I$-band in order to disentangle their structural constituents and study possible faint non-axis-symmetric features.}
{Our final sample consists of 17 massive galaxies. Due to the excellent HST spatial resolution for intermediate redshift objects, they are hard to model by purely automatic parametric fitting algorithms. We performed careful single and double (bulge-disk decompositions) Sérsic fits to their galaxy surface brightness profiles.
We also compare the model color profiles with the observed ones and also derive multi-component global effective radii attempting to obtain a better interpretation of the mass-size relation.
Additionally, we test the robustness of our measured structural parameters via simulations.}
{We find that the Sérsic index does not offer a good proxy for the visual morphological type for our sample of massive galaxies.
Our derived multi-component effective radii give a better description of the size of our sample galaxies than those inferred from single Sérsic models with \textsc{GALFIT}.
Our galaxy population lays on the scatter of the local mass-size relation, indicating that these massive galaxies do not experience a significant growth in size since $z \sim$ 0.5. 
Interestingly the few outliers are late-type galaxies, indicating that spheroids must reach the local mass-size relation earlier.
For most of our sample galaxies, both single and multi-component Sérsic models with \textsc{GALFIT} show substantial systematic deviations from the observed surface brightness profiles in the outskirts. These residuals may be partly due to several factors, namely a non-optimal data reduction for low surface brightness features, the existence of prominent stellar haloes for massive galaxies and could also arise from conceptual shortcomings of parametric 2D image decomposition tools. They consequently propagate into galaxy color profiles.
This is a significant obstacle to the exploration of the structural evolution of galaxies that calls for a critical assessment and refinement of existing surface photometry techniques.
}
{}
\titlerunning{Structural analysis of massive galaxies at $z<0.5$}
   \maketitle

\section{Introduction}

The most massive ($M_\textrm{stellar} \geq 10^{11} M_\odot$) galaxies in the Universe appear to have undergone a dramatic transformation in their structural properties across cosmic time, from compact star-forming disks to huge red and dead spheroidal galaxies \citep[e.g.][]{trujillo07,buitrago08,vandokkum10,buitrago13,huertas15}. However, how galaxies acquire their mass and how they evolve morphologically are still open questions. Therefore, the study of their structural properties is indispensable for a thorough understanding of the formation and evolution of these galaxies.

The currently most favored galaxy formation model for massive galaxies involves a two-phase build-up scenario. This scenario predicts a rapid formation phase at $2 < z < 6$ dominated by \textit{in-situ} star formation \citep{oser}, and a subsequent phase of stellar mass growth through multiple minor mergers \citep[e.g.][]{sanjuan10,sanjuan11,bluck12,sanjuan12,marmol,ferreras,ferreras17} that may transform them into present-day spheroids.
Following this scenario, \cite{hopkins} studied the radial surface density profiles of high ($z > 2$) and low-redshift ($z \sim 0$) massive galaxies, comparing directly the observed profiles at the same physical radii. They demonstrate that the central components of local massive spheroids are not different to the high-redshift systems. Inside the same physical radii, the stellar surface mass densities of many of the local ellipticals are comparable to those of high-redshift objects, differing in effective radius (\reff) mainly due to an extended low surface-brightness envelope in the low-redshift spheroids, as opposed to the steep fall of profiles of high-redshift objects.
A similar study from \cite{bezanson} strengthens this conclusion: the central surface densities of the high-redshift systems are comparable with the average densities within 1 kpc of the low-redshift ellipticals, implying an inside-out growth scenario whereby the compact high-redshift ($z \geq 2$) galaxy remnants are contained within the cores of nearby early-type galaxies. So far, however, there is no direct detection of compact old cores within elliptical galaxies, and the main drivers of the dramatic morphological and structural evolution of massive galaxies across redshift remain unclear.
A subsequent study by \cite{delarosa} makes use of catalogs based on the Sloan Digital Sky Survey (SDSS) with bulge+disk (B+D) decompositions without restricting the morphology of the host galaxies. This study tests the hypothesis that the massive compact high-redshift galaxies were already quiescent in terms of star formation \citep[termed \textit{red nuggets} in][]{damjanov} and are hidden in the cores of present-day galaxies. In that study the central regions of both spheroids and disks are treated as independent entities, being evaluated according to their compactness. The authors show that the cores of local massive galaxies are structurally equivalent to red nuggets at $z \sim 1.5$.

Studying the structural components of massive galaxies can yield to a better understanding of the different processes behind the assembly of this galaxy population. Several works on photometric decomposition of galaxies contributed to the discovery of many scaling relations \citep[e.g.][]{faberjackson,kormendy77,djorgovski87}, and particularly for the case of massive galaxies, it is established that they tightly follow the mass-size relation in the local Universe \citep[e.g.][]{shen}.
One should bear in mind, however, that, while single Sérsic profiles are frequently used as a standard model to automatically fit large samples \citep[e.g.][]{simard02,vdw,griffith}, galaxies are more complex and their structural characterization typically requires several components.
Traditionally, massive galaxies were described as the outcome of the superposition of a central spheroidal bulge described by the \cite{devaucouleurs} law and a more extended exponential disk \citep{freeman}. However, the increasing quality and lower limiting surface brightness ($\mu$) of imaging data has brought to light new features within bulges, such as: faint spiral-like patterns \citep[e.g.][]{kehrig12,gomes}; surface brightness profiles (SBPs) with a Sérsic index $n \simeq 1$ to 2, being generally attributed to a pseudo-bulge \citep[cf. e.g.][]{kormendy}; or a significant fraction of stellar mass located within the haloes of massive ETGs that likely has accumulated through multiple minor merging events \citep{kaviraj,buitrago17}.

A direct consequence of the findings referred above is the necessity of moving from integrated properties to resolved quantities within galaxies, requiring two components fitting as a way to provide a deeper insight into the different processes at place to assemble the main constituents of galaxies.
Although there are already many computationally expensive works involving B+D decomposition of large samples at low-redshift \citep[e.g.][to say but a few]{allen06,simard11,lackner12,mendel14,meert15,lange16,dimauro18}, a critical amount of post-processing is required to evaluate the quality of the fits. However, this task requires a large amount of interactive work for inspection of the fitting results to assure physical solutions of the models. While such a detailed inspection of the modeling output is both feasible and mandatory for individual galaxies, it is generally out of reach in the case of quasi-automated studies of large galaxy samples.

The present work is dedicated to the structural analysis of 17 low-redshift ($z < 0.5$, for cosmological dimming not to be an issue since its steep increase with redshift) massive galaxies from CANDELS using multi-band surface photometry and B+D profile decomposition. The availability of data from the Hubble Space Telescope (HST) permits a superb (in terms of spatial resolution) investigation of the galaxy structure of our target galaxies, while the Cosmic Assembly Near-infrared Deep Extragalactic Legacy Survey \cite[CANDELS\footnote{\url{http://arcoiris.ucolick.org/candels/data_access/Latest_Release.html}},][]{grogin,koekemoer} ``Deep'' segment of the survey allows us to obtain limiting magnitudes previously unreachable. These HST images at low-z ($z < 0.5$) therefore provide us with a sample of relatively well resolved galaxies from which we can extract precious information especially in the low surface brightness regime.

In Sect.~\ref{sec:data} we describe the data and sample selection. The methodology used for the profile fitting, taking advantage of state-of-the-art source extraction and surface photometry packages is specified in Sect.~\ref{sec:methodology}. The various derived structural properties of the galaxy sample (e.g. Sérsic indices, bulge-to-total luminosity ratios) and the analysis with respect to integral properties (integrated magnitude and total stellar mass) are presented and discussed in Sect.~\ref{sec:results}. Summary and conclusions follow in Sect.~\ref{sec:conclusions}. Simulations performed to quantify the accuracy of our structural parameters measures are included in Appendix~\ref{appen:simulations}. 

Throughout the paper we adopt a cosmology with $\Omega_m = 0.3$, $\Omega_{\Lambda}=0.7$ and $H_0 = 70 \ kms^{-1}Mpc^{-1}$. Magnitudes are provided in the AB system \citep{oke}.

\section{Data and Sample Selection}
\label{sec:data}

\begin{figure*}[ht]
  \centering
     \includegraphics[width=0.95\textwidth] {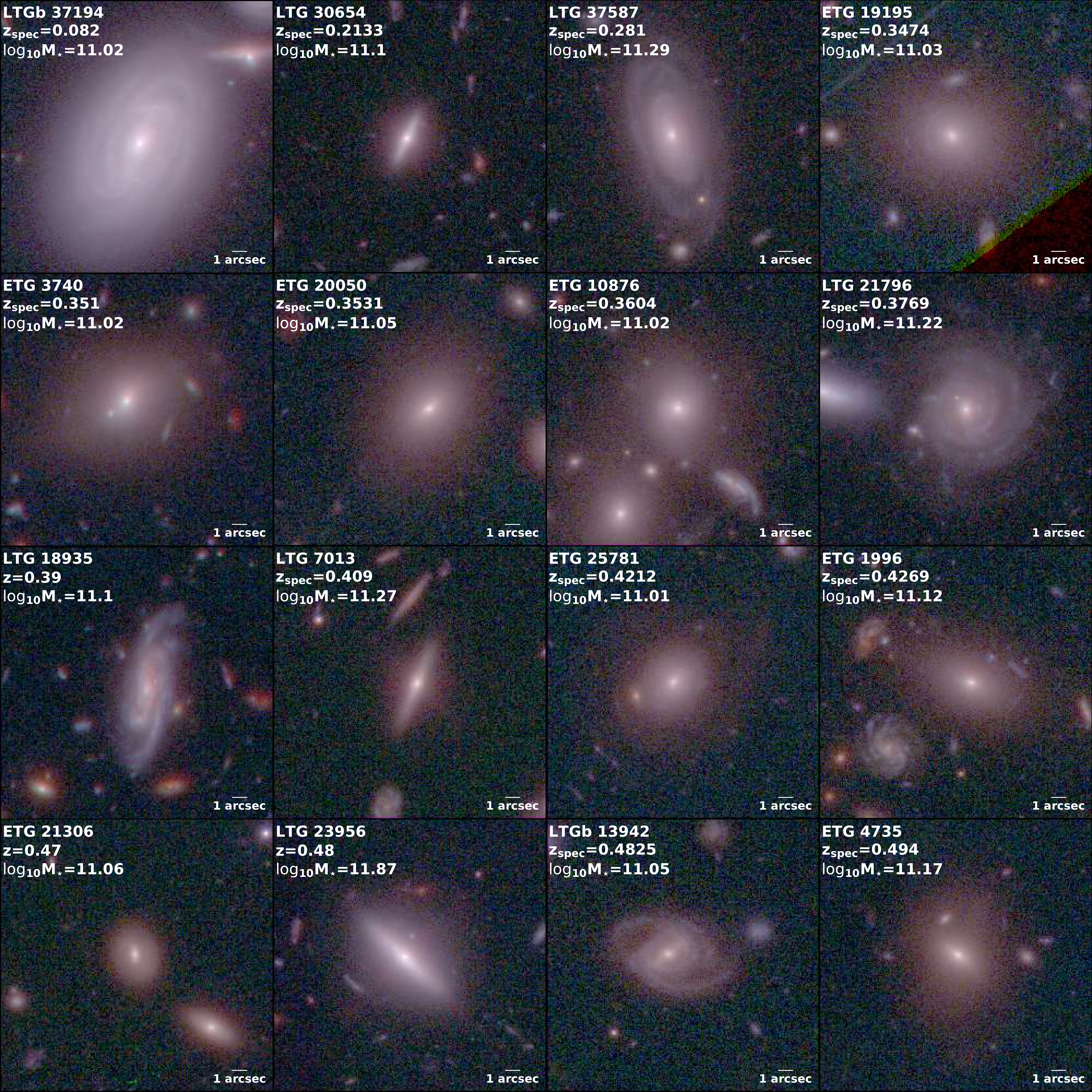}
  \caption{False-color RGB images of the galaxies within our sample, constructed using the filters F160W, F814W, and F606W in the red, green, and blue image channel, respectively. The galaxy with ID \#21604 (Figure~\ref{fig:21604}) is not shown here due to lack of data in the F814W and F606W filters. Each galaxy stamp contains information about our visual classification preceding the object ID, redshift and stellar mass.
This image compilation illustrates the morphological heterogeneity of our galaxy sample, comprising both ETGs and LTGs, with some of the latter containing a bar.}
  \label{fig:mosaic}
\end{figure*}

The HST imaging data were retrieved from the CANDELS survey \citep{grogin,koekemoer} while our sample was selected from the 3D-HST catalog \citep{skelton}, since this catalog includes stellar masses and spectroscopic redshifts for the galaxies within all the five CANDELS fields. CANDELS covers a total of $\sim$800 square arcminutes widely distributed, and contains a 'Deep' and a 'Wide' component. The Deep portion of the survey comprises about 125 arcmin$^2$ to $\sim$10-orbit depth within the GOODS-North and GOODS-South fields. The remaining area includes the shallower Wide component, distributed over all the five fields, to $\sim$2-HST orbit depth. CANDELS is a survey optimized for the detection of galaxies at high-redshifts, and as it is very deep and wide, it allows us to retrieve nearby massive galaxies with a very high signal-to-noise (S/N).

We have used for our analysis two filters from CANDELS: the F160W filter (WFC3 $H$-band) with a pixel scale of $0.06''/\texttt{pixel}$ and pivot wavelength equal to 1536.9 nm corresponding to the reddest HST filter (the most representative of the total stellar component), and the F814W filter (ACS $I$-band) which has pixel scale of $0.03''/\texttt{pixel}$ and pivot wavelength of 805.7 nm and is the most representative of the optical rest-frame in the redshift range considered.
From a total of 207967 sources in all five CANDELS/3D-HST fields, and with the simple selection criteria of stellar masses larger than $\rm 10^{11}$ M$_{\odot}$ and (spectroscopic, when available) redshifts $\rm \leq 0.5$, we extracted 68 sources. However, since most of these sources are stars or objects not observed with the WFC3 F160W filter, the final sample consists of 17 massive galaxies (one object not observed in the ACS $I$-band). The upper limit for the redshift was chosen to assure that we obtain the galaxies with an adequate S/N. All objects in our sample are located in the Wide imaging of CANDELS, reaching 3$\sigma$ limiting magnitudes for extended sources of $H_{F160W}\sim$28.3 mag.arcsec$^2$ in the COSMOS field, $H_{F160W}\sim$28.5 mag.arcsec$^2$ in the GOODS-N field, and $H_{F160W}\sim$28.6 mag.arcsec$^2$ in the UDS and EGS fields.

\begin{figure}[h]
    \centering
    \includegraphics[width=0.5\textwidth, trim={0 30 0 30},clip]{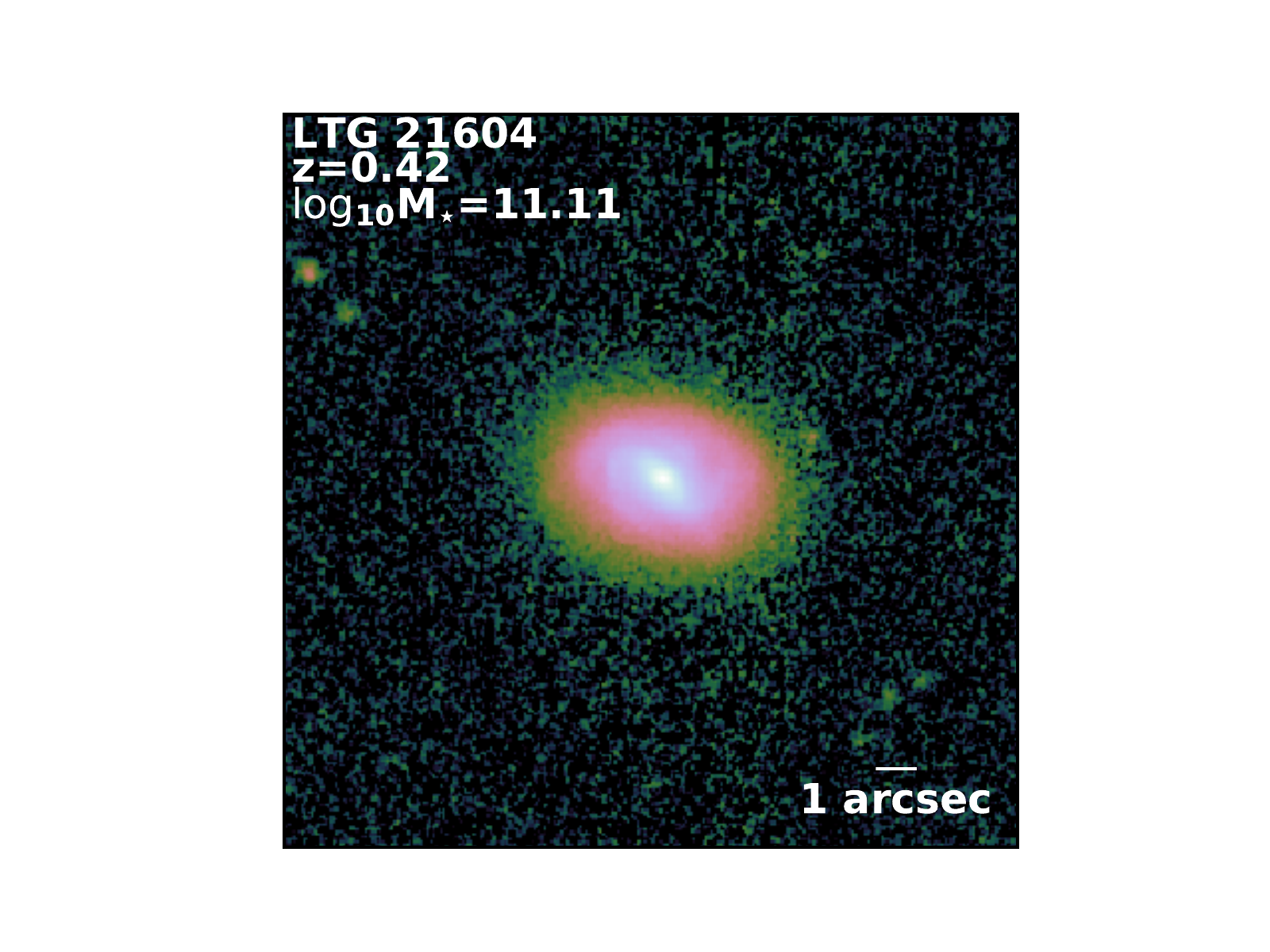}
    \caption{Galaxy stamp of the late-type object \#21604 as observed in the F160W filter.}
    \label{fig:21604}
\end{figure}

By applying a visual morphological classification, our final sample contains nine LTGs (disk-dominated, having two of them a bar) and eight ETGs (spheroid-dominated), with half of the sample being in the redshift range $0.4 \leq z < 0.5 $. Table~\ref{table:sample} lists the data related to the final sample. Figure~\ref{fig:mosaic} displays each source in composite RGB image using $H$-, $I$- and $V$- bands (ACS F606W filter). As mentioned above, one object in our sample (LTG 21604) does not have ACS coverage, thus being shown separately in Figure~\ref{fig:21604}.

\begin{table}[h]
\caption{Summary of the properties for our sample of galaxies, as taken from the 3D-HST catalog \citep{skelton}. Each column shows, from left to right: (1) galaxy IDs matching the identifiers from the 3D-HST master catalog, (2) CANDELS field, (3) Right Ascension (degs), (4) Declination (degs), (5) redshift, (6) stellar mass in units of log(M$_{\odot}$), (7) visual morphology represented as early-type (ETG), late-type (LTG) and late-type with a bar (LTGb).}    
\label{table:sample}      
\centering                          
\resizebox{0.5\textwidth}{!}{
\begin{tabular}{c c c c c c c}      
\hline   
\multirow{ 2}{*}{3D-HST ID} &  \multirow{ 2}{*}{Field}	&	    RA        &	      Dec	    &	   \multirow{ 2}{*}{$z$}	& $\log$M$_\text{stellar}$ & \multirow{ 2}{*}{visual type}\\
          &         & [J2000]         & [J2000]        &            &     [M$_{\odot}]$        &            \\
\hline
37194     &	AEGIS	&	214.7928      &    52.8643	   & 0.0820   	&	       11.02	       & LTGb       \\
30654     &	GOODS-N	&	189.4151	  &	   62.2974	   & 0.2133	&	       11.10	       & LTG        \\
37587     &	GOODS-N	&	189.3515	  &	   62.3660	   & 0.2810 &	       11.29	       & LTG        \\
19195     &	COSMOS	&	150.0581      &	   2.38042	   & 0.3474 &	       11.03           & ETG        \\
3740      & AEGIS	&	215.1804	  &	   52.9960	   & 0.3510	    &	       11.02	       & ETG        \\
20050     &	COSMOS	&	150.0806	  &	   2.39026	   & 0.3531	&	       11.05           & ETG        \\
10876     &	COSMOS	&	150.0951      &	   2.30050	   & 0.3604	&	       11.02           & ETG        \\
21796     &	GOODS-N	&	189.4206	  &	   62.2548	   & 0.3769	&	       11.22           & LTG        \\
18935     &	AEGIS	&	215.2186	  &	   53.0791	   & 0.39$^*$	    &	       11.10           & LTG        \\
7013      & GOODS-N	&	188.9807	  &	   62.1809	   & 0.4090   	&	       11.27	       & LTG        \\
21604     &	UDS	    &	34.55835      &	   -5.2059	   & 0.42$^*$       &	       11.11           & LTG        \\
25781     &	COSMOS	&	150.1822	  &	   2.45136	   & 0.4212	&	       11.01           & ETG        \\
1996      &	COSMOS	&	150.0907      &	   2.20566	   & 0.4269	&	       11.12           & ETG        \\
21306     &	GOODS-N	&	189.4720 	  &	   62.2483	   & 0.47$^*$	    &	       11.06           & ETG        \\
23956     &	GOODS-N	&	189.0092	  &	   62.2638	   & 0.47$^*$  	    &	       11.87	       & LTG        \\
13942     &	COSMOS	&	150.1893      &	   2.32564	   & 0.4825	&	       11.05           & LTGb       \\
4735      & UDS	    &	34.43567	  &	   -5.2562	   & 0.4940	    &	       11.17           & ETG        \\
\hline
\end{tabular}}
\vspace{-0.4cm}
\flushleft{\scriptsize{$^*$ photometric redshifts}}
\end{table}

\section{Methodology}
\label{sec:methodology}

In order to prepare the images for Sérsic model fitting, square postage stamp images of 300$\times$300 pixels for the $H$-band (18\arcsec$\times$18\arcsec), and 600$\times$600 pixels for the $I$-band (same physical size with half pixel scale) were produced from the survey mosaic, updating the headers with the \textsc{Python} package \textsc{Montage}-wrapper\footnote{\url{http://montage.ipac.caltech.edu/}}. The size of the stamps was chosen to be big enough to account for sufficient sky pixels to give a good fit of the main galaxy, while being small enough to be fit in reasonable amounts of computing time. Corresponding noise maps of the same size were also cut.

The next step was to run \textsc{SExtractor} \citep{bertin} in order to identify neighboring objects and to create catalogs with values of apparent magnitude, size, axis ratio, and position angle. These were used later as initial guesses in fitting the galaxies' surface brightness profiles with \textsc{GALFIT} \citep{peng02,peng10}. By using \textsc{SExtractor} we minimize the computing time and ease convergence of \textsc{GALFIT} models to the global $\chi^2$ minimum.
It is worth noting the need of an aperture large enough to capture most of the galaxy light to provide \textsc{GALFIT} with a representative initial guess for the galaxy magnitude. Since our small sample involves closer and well resolved objects, we made several runs with different circular apertures in order to test the influence of such apertures in the estimation of the parameters. We find that for our entire sample we retrieve a more physical set of output parameters when using an aperture of 4 arcsec for the main galaxy, and a smaller one (1 arcsec) for neighboring objects.

Before running \textsc{GALFIT}, masks were created for neighboring sources using the segmentation maps produced with \textsc{SExtractor}. Nearby bright objects were chosen to be fit simultaneously with the main galaxy, in order to remove any light contamination. \textsc{GALFIT} is a state-of-the-art software package which convolves Sérsic 2D models with the point-spread function (PSF), and uses the Levenberg-Marquardt algorithm for error-weighted non-linear fitting, whereby the $\chi^2$
between the PSF-convolved best-fitting model and the observed 2D surface brightness distribution of a galaxy in a given passband are minimized.
As the galaxies in our sample are quite extended (up to 16 arcsec in diameter at $H$-band surface brightness of 26 mag arcsec$^{-2}$), we need to use large PSFs \citep[see][]{sandin14,sandin15,buitrago17}. 
The PSFs of each band were created using the \texttt{TinyTim} HST PSF Modeling tool \citep{krist}, with the maximum sizes available. These PSFs were then rebinned to the current pixel size of our data (0.06\arcsec and 0.03\arcsec for $H$ and $I$-bands respectively), resulting in final sizes of $\sim20\times20$ arcsec$^2$ and $\sim30\times30$ arcsec$^2$.
PSF blurring effects are more relevant for galaxy cores, i.e. in the highest-intensity regions, and also for redder bands where the PSF FWHM is broader \citep{buitrago17}. In this publication, the authors found that correcting the profiles for the PSF is essential for the study of both the bright cores and fainter components of galaxies at intermediate distances.

We performed single Sérsic fits in order to compare our results with published works in the literature. A one-dimensional Sérsic function \citep{sersic} has the form:
\begin{equation}
I(r) = I_e \exp \left\lbrace -b_n \left[ \left( \frac{r}{r_e} \right)^{1/n} -1 \right] \right\rbrace
\label{sersiceq}
\end{equation}
with $b_n$ being a parameter coupled to the Sérsic index $n$ which satisfies the expression $\Gamma (2n) = 2\gamma (2n,b_n)$, where $\Gamma$ and $\gamma$ are the gamma function and the incomplete gamma function, respectively. A Sérsic index of $n=1$ gives an exponential profile, which is commonly assumed to describe perfect exponential galactic disks, whereas $n=4$ yields the de Vaucouleurs’s law. For a 2D case (i.e. an astronomical image) the axis ratio and position angle should also be taken into account.

Moreover, B+D decompositions were also performed, by fitting two Sersic functions, fixing the Sersic index of the disk component to 1 (an exponential). Our aim was not only to describe each galaxy's surface brightness profile as accurately as possible, but to disentangle the bulge and disk in the case of a LTG, or to check if an ETG is purely elliptical or contains other components (e.g., a disk or bar). Since \textsc{GALFIT} is a least-squares fitting algorithm, the use of bad priors can affect the output parameter values settling the solution into a local $\chi^2$ minimum. In the case of single Sérsic fits, the initial parameters do not have a major effect on the fit unless they are considerably off of the actual values. On the other hand, on extensive tests and experience we know that for B+D decompositions the initial parameters given to \textsc{GALFIT} (specially the magnitudes) are much more important to retrieve the models that provide the best residuals. For this reason we adopted different initial parameters for our \textsc{GALFIT} models, according to our visual morphological classification: for ETGs the disk component is set to be 20\% smaller and 0.5 mag fainter than the spheroid, while in LTGs is the opposite. For the two galaxies with visual evidence of a bar an extra component was added to the fit, setting the initial guesses of effective radius and magnitude of the bar with values between the ones of the bulge and of the disk, the starting Sérsic index value to 0.5 for a flat inner and a steep outer profile, and adding an extra diskiness/boxiness parameter ($C0$) fixed to 0.5 ensuring a boxy shape. As demonstrated in e.g. \cite{gadotti2008} \citep[see also][]{2008A&A...478..353M,breda18}, disregarding a bar could contribute to considerable uncertainties in the derived bulge parameters and an overestimation of the bulge-to-total ($B/T$) ratio.

\section{Results and Discussion}
\label{sec:results}

In this section we present the main results from our photometric study, both for single Sérsic and multi-component Sérsic fits, for the HST $H$ and $I$-bands. Note that galaxies with IDs 13942 and 37194 feature a bar component which was accounted for (both in the single and the double component case) through an extra Sérsic component in the fitting procedure. We proceed to compare our results with others from the bibliography to check the limitations and advantages of our adopted approach for image decomposition. However, note that these results from the literature do not take into account the bar component in the fit, whereas we do. 

\subsection{On the reliability of the inferred structural parameters}

\begin{figure}[h]
  \centering
     \includegraphics[page=1,width=0.49\textwidth]{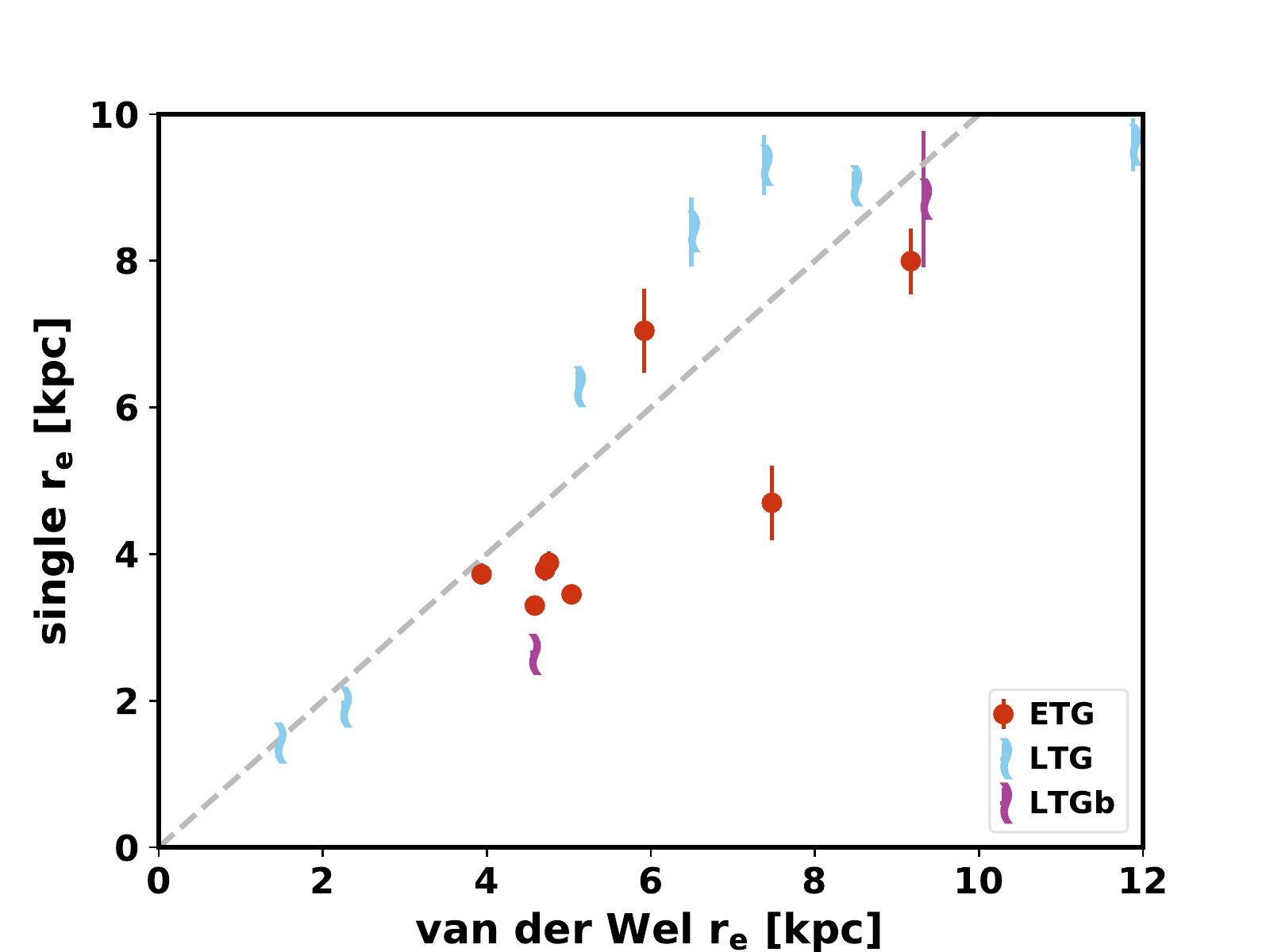}
  \caption{Comparison between our $H$-band single fit and \cite{vdw} effective radius. Our sample is color coded according to the galaxy visual morphology: red for ETGs, blue for LTGs, and violet for late-type barred galaxies. Error bars come from the simulations described in Appendix~\ref{appen:simulations}. The dashed line represents the 1-to-1 relation. Generally our results are in agreement ($\sigma_{\textnormal{NMAD}}$ = 0.13) with \cite{vdw}.
  }
  \label{fig:single_re_vdw}
\end{figure}

\cite{vdw} provides structural parameters from single Sérsic fits of 109533 galaxies within CANDELS in the HST ultra-deep WFC3 $H$-band, which offers a useful database for comparison with our structural analysis. Since our sample is small, we are able to carry out a much more detailed analysis of these low-redshift (and thus comparatively large in terms of angular extent) galaxies. In Figure~\ref{fig:single_re_vdw} we compare the effective radius from our single Sérsic fits with that obtained by \cite{vdw}.
To quantitatively assess our comparisons we adopt the widely-used normalized median absolute deviation ($\sigma_{\textnormal{NMAD}}$), which is equivalent to the standard deviation for a gaussian distribution with the advantage of being less sensitive to outliers \citep[e.g.][]{ilbert06}. The mathematical expression for $\sigma_{\textnormal{NMAD}}$ is the following:
\begin{equation}
    \sigma_{\textnormal{NMAD}} = 1.48 \times \textnormal{median} \ \Big| \ X - \textnormal{median}(X) \ \Big| 
\end{equation}
where $X = \Big| \frac{x_i - \hat{x}_i}{x_i} \Big|$,
$\hat{x}_i$ is our value, and $x_i$ is the value with which we are comparing our results. We find good agreement in general, obtaining
$\sigma_{\textnormal{NMAD}}$ = 0.13. The error bars shown for our determinations come from the simulations described in Appendix~\ref{appen:simulations}. 

\begin{figure}[h]
  \centering
     \includegraphics[page=2,width=0.49\textwidth]{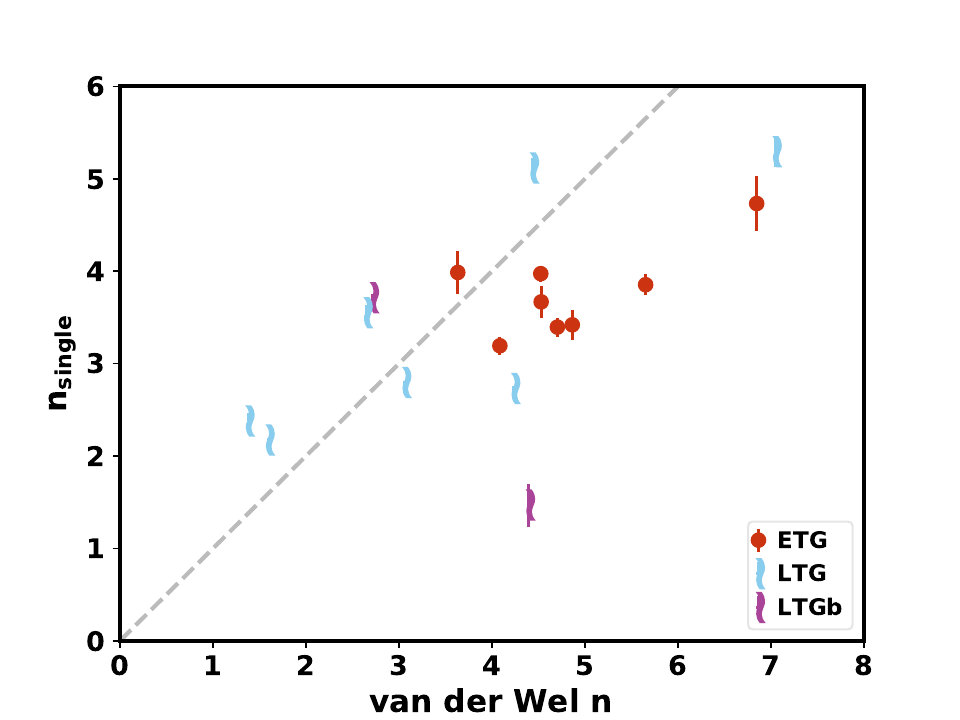}
  \caption{Comparison between our $H$-band single Sérsic index with that obtained by \cite{vdw}. Our sample is color coded as in Figure~\ref{fig:single_re_vdw}. Error bars come from the simulations described in Appendix~\ref{appen:simulations}. 
  The divergence in the case of ETGs can be related with their concentrated nature. In the case of LTGs, the discrepancy may be due to the fact that the luminosity profiles of these galaxies cannot be properly described with a simple Sérsic function.} 
  \label{fig:single_n_vdw}
\end{figure}

Figure~\ref{fig:single_n_vdw} presents the comparison between our values of Sérsic index for a single fit and the values obtained by \cite{vdw}, for which we retrieve $\sigma_{\textnormal{NMAD}}$ = 0.10.
In the light of the simulations we present in the Appendix and previous studies in the literature, the discrepancy on some of the recovered Sérsic index values is unsurprising, especially in the case of ETGs.
For example, \cite{haussler07} tested the performance of two fitting codes (\textsc{GALFIT} and \textsc{GIM2D}) for fitting single Sérsic models for several thousands objects, finding less accurate recovered parameters for ETGs.
This is attributed to the fact that these are very concentrated objects and, as such, very hard to analyze the inner parts of their surface brightness profiles.
As for LTGs, these discrepancies can be explained by the fact that the SBPs of these galaxies cannot be properly described with a simple Sérsic function, given their structural complexity (e.g., additional bulge and bar, spiral arms, and down-bending disk in some cases).

Another possible explanation for the differences between our results with the ones from \cite{vdw}, is the PSF we adopted. While we use for all fields the same pure \texttt{TinyTim} PSF model, \cite{vdw} constructed smaller hybrid PSF models for the different fields, combining stacked stars and \texttt{TinyTim} models. These models were made available in the public release, thus providing us with the possibility to properly compare both results. When using their hybrid PSF we retrieve a much better agreement with \cite{vdw} structural parameters, confirming that our procedure works as expected at least for a single Sérsic fit. However, given the brightness and extent of the galaxies in our sample, we use the larger \texttt{TinyTim} PSFs as it will allow us to recover more precise light profiles. Clearly, the sensitivity of the best-fitting solution from \textsc{GALFIT} on the adopted PSF model is a fact that needs to be taken into account when comparing results from different studies.  

 \begin{figure}[h]
  \centering
     \includegraphics[page=1,width=0.49\textwidth]{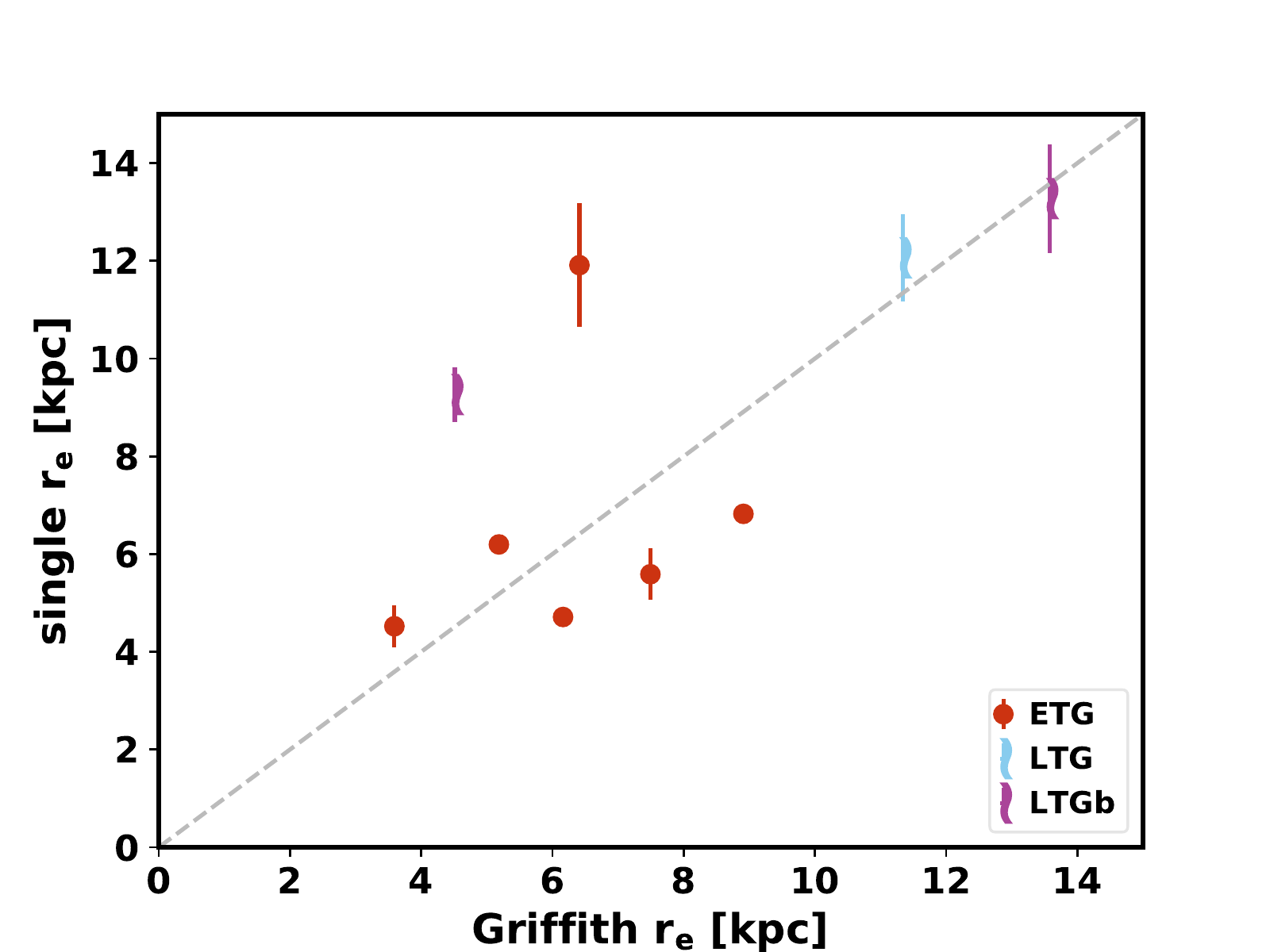}
  \caption{Comparison between our $I$-band effective radius from single Sérsic fits and \cite{griffith} effective radius, in kpc. The two objects deviating from the 1-to-1 line relation are very challenging galaxies to fit in the $I$-band: LTGb 37194 exhibits prominent spiral arms and a nuclear bar, whereas ETG 25781 also appears to have some spiral features evident in the residuals.
}
  \label{fig:single_re_griffith}
 \end{figure}
 
\begin{figure}[h]
  \centering
     \includegraphics[page=2,width=0.49\textwidth]{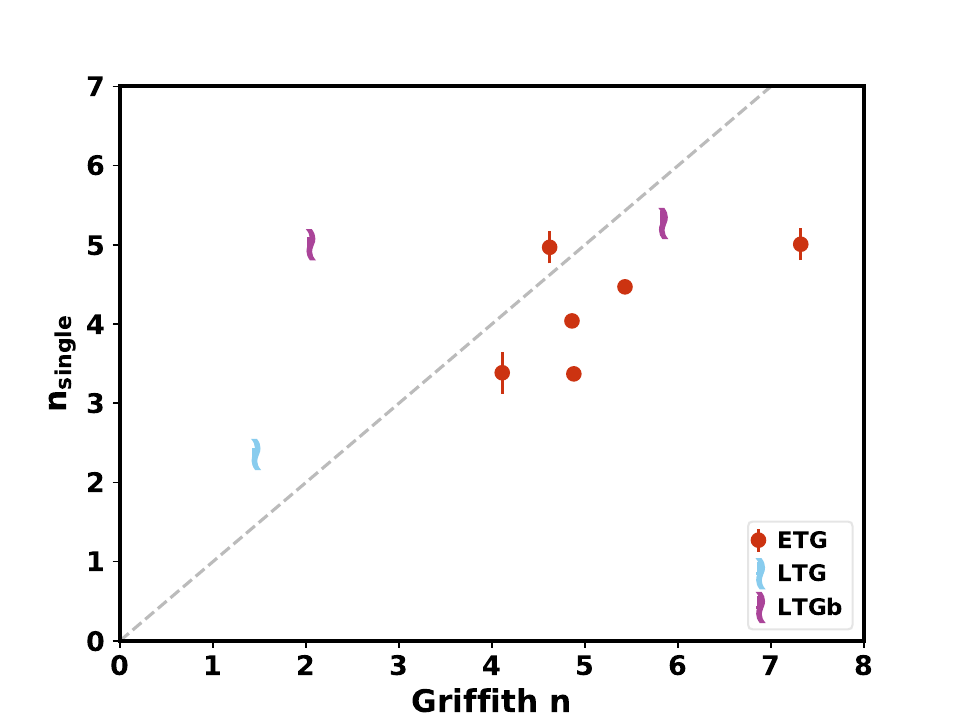}
  \caption{Comparison between our $I$-band Sérsic index from single Sérsic fits and \cite{griffith} Sérsic index.
  }
  \label{fig:single_n_griffith}
\end{figure}

For the ACS $I$-band one can use the \cite{griffith} catalog, in which the authors also fit single Sérsic profiles using \textsc{GALFIT}. Figures~\ref{fig:single_re_griffith} and~\ref{fig:single_n_griffith} show that our results slightly differ from those in this work, obtaining $\sigma_{\textnormal{NMAD}}$ = 0.06 and $\sigma_{\textnormal{NMAD}}$ = 0.15, for effective radii and Sérsic indices respectively.
One possible explanation for the disagreement is that their initial conditions for sizes were derived from the formula $r_e= 0.162R_{\rm flux}^{1.87}$, where $R_{\rm flux}$ denotes the \textsc{SExtractor} \texttt{FLUX\_RADIUS}, and also the initial condition for their Sérsic index was 2.5. In both cases, their choices are at variance with ours. Furthermore, their study was made automatically for half a million galaxies, making their procedure more susceptible to large uncertainties (due to, e.g., an imperfect rejection of overlapping stars or galaxies) that could be eliminated 
in the case of our far smaller sample. These facts, together with only having nine galaxies in common, may explain the somewhat larger disagreement between our results and \cite{griffith} as opposed to a better agreement with \cite{vdw}.
 
Our \textsc{GALFIT} derived structural parameters for the two photometric bands (F160W and F814W, $H$- and $I$-band, respectively) are presented in Appendix~\ref{tables} Tables~\ref{table:singlehband} and~\ref{table:singleiband}.

\subsection{Sérsic index values}

\begin{figure}[!]
\centering
\includegraphics[width=0.49\textwidth]{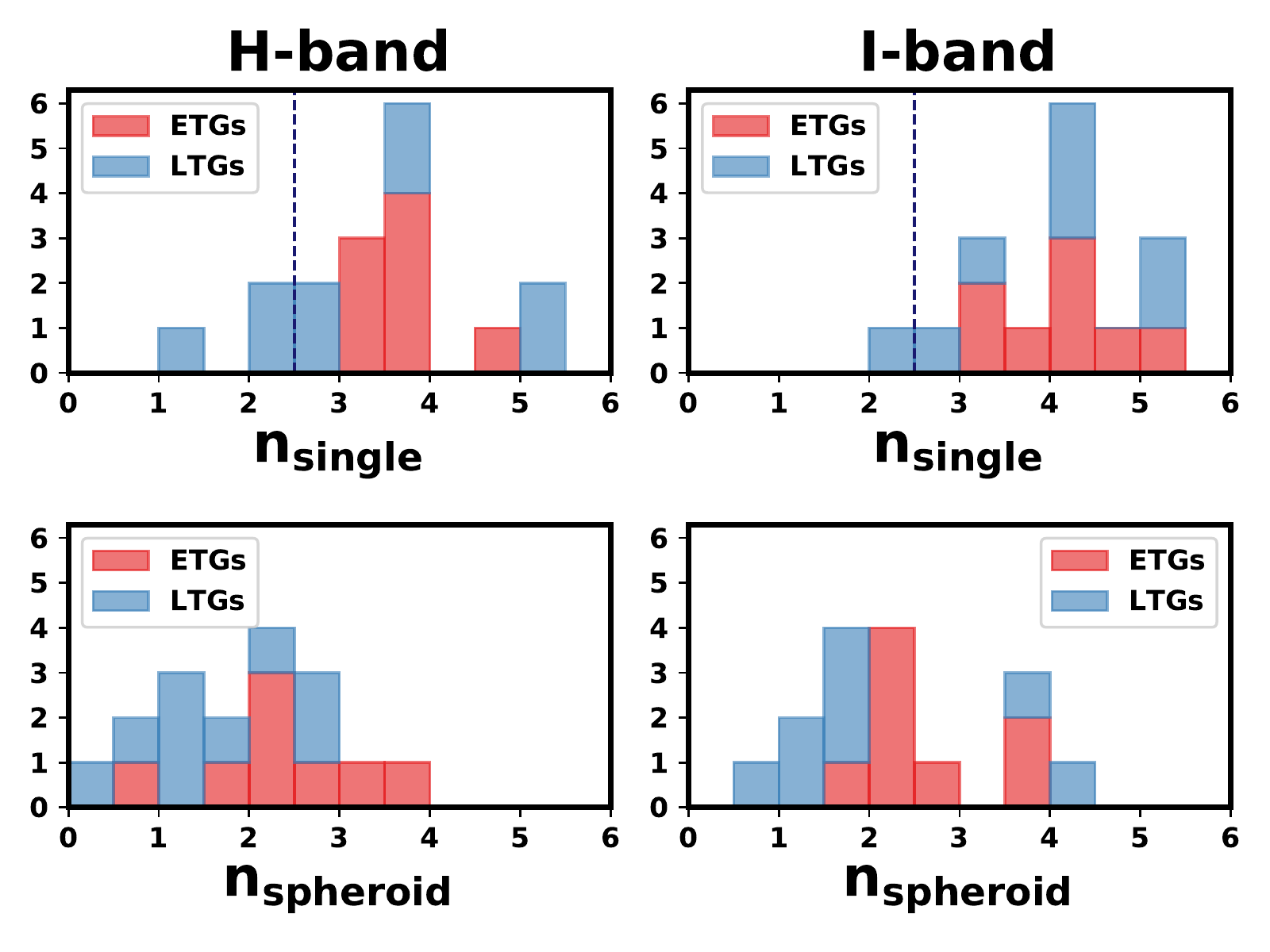}
\caption{Sérsic index histograms for single Sérsic fits values (top panels) and for the spheroid component (bottom panels) in the B+D decomposition, both for the $H$- (left panels) and the $I$-band (right panels). The dashed gray line represents the division line between disk-dominated ($n<2.5$) and spheroid-dominated ($n>2.5$) objects. It is remarkable that, although our sample contains nine visually classified LTGs, single Sérsic fits in the $H$-band yield only three galaxies being classifiable as disk-like on the basis of $n$. For the bottom-panel histograms, it is also noteworthy that the majority of our sample contains a spheroidal component with a low Sérsic index ($n_{\rm spheroid} < 3$) in both bands, pointing to bulges of moderate light concentration.}
\label{fig:n_hist}
\end{figure}

Figure~\ref{fig:n_hist} shows the Sérsic index histograms for the $H$- and $I$-band, in the case of single- (top panels) and multi-component (bottom panels) fits. For this latter case, we show the Sérsic index values of the spheroidal component ($n_{\rm spheroid}$), since the disk component Sérsic index is fixed to 1, i.e. that corresponding to a pure exponential profile. Following \cite{shen} \citep[see also e.g.][]{barden05,trujillo06,buitrago13} we use $n = 2.5$ as a division line between bulge-dominated and disk-dominated systems, shown in the figure by the dashed gray line. A classification solely based on Sérsic index would result in only three galaxies classified as disk-dominated ($n < 2.5$) in the $H$-band, and only one in the $I$-band. However, by visual inspection in the $H$-band we classified morphologically nine galaxies as LTGs/disk-dominated (shown in the histogram with blue color, opposed to ETGs/bulge-dominated in red). The bottom panels of Figure~\ref{fig:n_hist} presents the Sérsic index for the spheroidal component in the B+D decomposition, the majority of the sample having low $n_{\rm spheroid}$ values in both bands, which indicates that its light shows a moderate degree of central concentration.

These results indicate that visual morphologies and Sérsic morphologies are poorly correlated for our galaxy sample.

\subsection{Bulge-to-total light ratio}

\begin{figure}[h]
 \centering
    \includegraphics[page=3,width=0.49\textwidth]{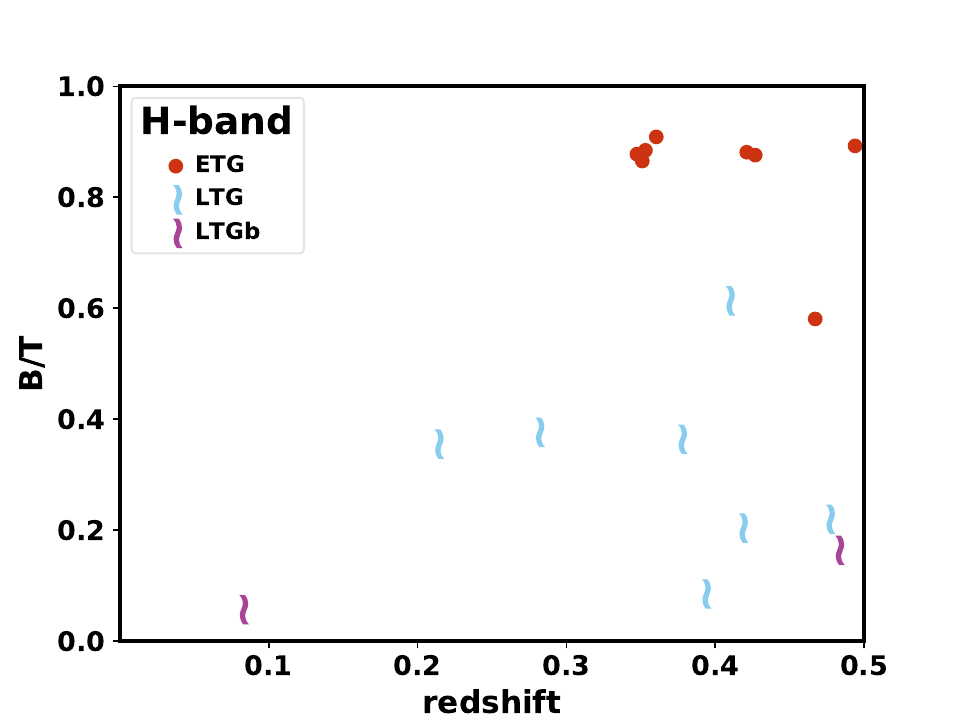}
    \includegraphics[page=3,width=0.49\textwidth]{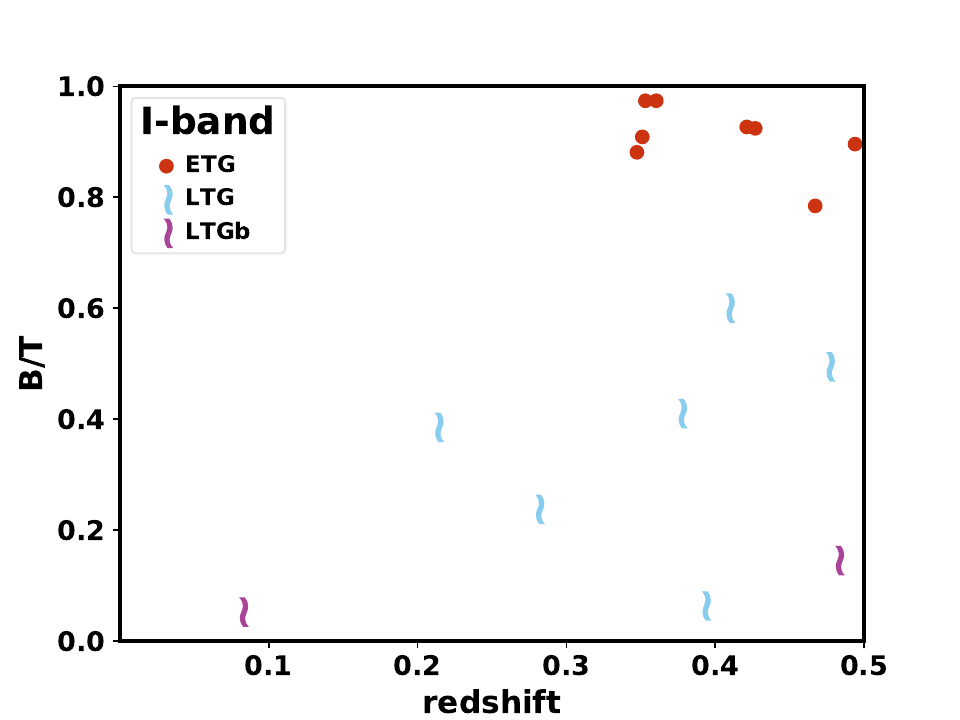}
 \caption{Bulge-to-total ($B/T$) light ratio relation with redshift for $H$-band (top panel) and $I$-band (bottom panel). The $B/T$ values do not change significantly between the $H$-band (NIR rest-frame) and the $I$-band (optical rest-frame) for LTGs, being mostly disk dominated ($B/T$ $\leq$ 0.5), with exception of LTG 7013 which is an edge-on galaxy and visually seems to be a S0 galaxy. In contrast, for ETGs the bulge component seems to dominate ($B/T$ > 0.5) in a similar way in both bands, being the optical rest-frame slightly higher in all cases but particularly for ETG 21306.}
 \label{fig:bt_redshift}
\end{figure}

It is well established that the bulge-to-total ($B/T$) light ratio of a galaxy correlates with its Hubble-type, increasing from LTGs towards ETGs \citep[e.g.][]{hubble26,hubble36,simien,bluck14}. In Figure~\ref{fig:bt_redshift} one can see how this quantity behaves with redshift in the two bands at study. Comparing the optical rest-frame (bottom panel) and the NIR rest-frame (top panel), both seem to display similar values for each population of galaxies. In the case of LTGs the spheroidal component is not dominant for the majority of these objects ($B/T \leq$ 0.5), except for LTG 7013 which is edge-on and might be a S0 galaxy. In contrast, all ETGs have a dominant spheroidal component ($B/T$ > 0.5) for both bands, although the optical rest-frame $B/T$ value seems to be slightly higher in all cases, in particular for ETG 21306. 

Three galaxies in our sample have in both bands a $B/T < 0.2$, which is frequently linked with pseudo-bulges \citep{kormendy,kormendy13}. This is thought to reflect assembly in the course of secular galaxy evolution through \textit{in-situ} star formation fed by gas inflow from their disk, inward migration of star-forming clumps or minor mergers \citep[see, e.g.,][see also \cite{breda18} for a recent spectral modeling study of the assembly history of LTG bulges across $\sim$3 dex in stellar mass]{scannapieco}. Perhaps it is not surprising that the two barred galaxies in our sample are among these small $B/T$ galaxies. \cite{brook}, based on cosmological hydrodynamical simulations, were able to create a disk-like bulge through a bar. However, \cite{kormendy} caution that a very small $B/T$ alone does not ensure that the galaxy contains a disk-like bulge, needing several additional characteristics to be present, namely: a nuclear bar, boxy shape, Sérsic index between 1 to 2, more rotation-dominated than classical bulges, low-sigma outlier in the \cite{faberjackson} correlation, and dominated by young stars, gas and dust but with no evidence for on-going mergers.

For our small albeit meaningful sample, we see no trend between the $B/T$ ratio and redshift.

\subsection{Size estimation}

\begin{figure}[h]
 \centering
    \includegraphics[page=4,width=0.49\textwidth]{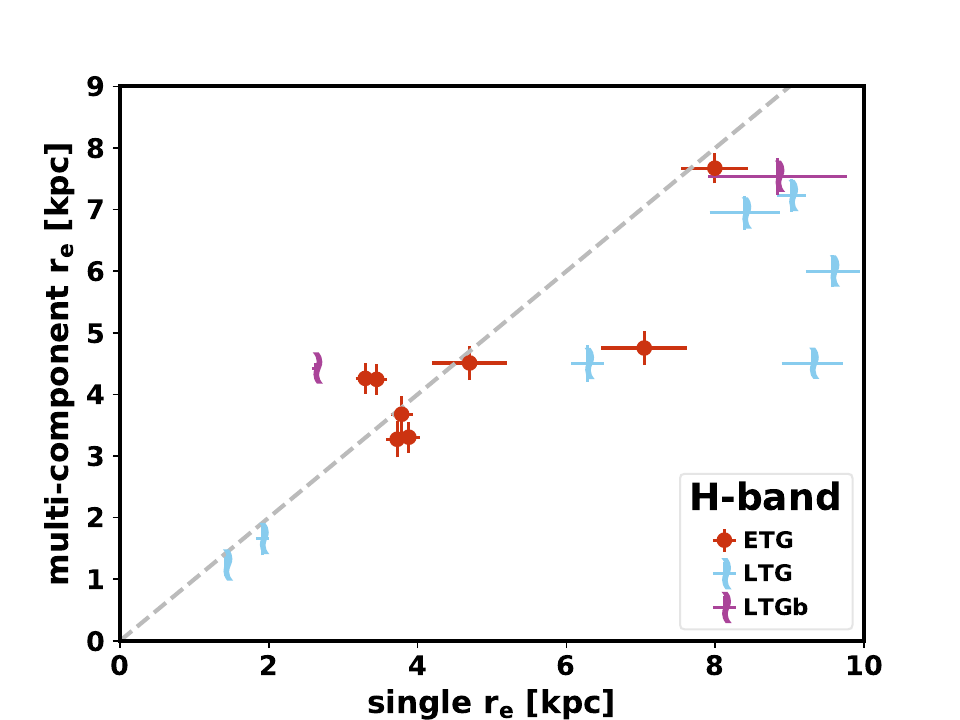}
    \includegraphics[page=4,width=0.49\textwidth]{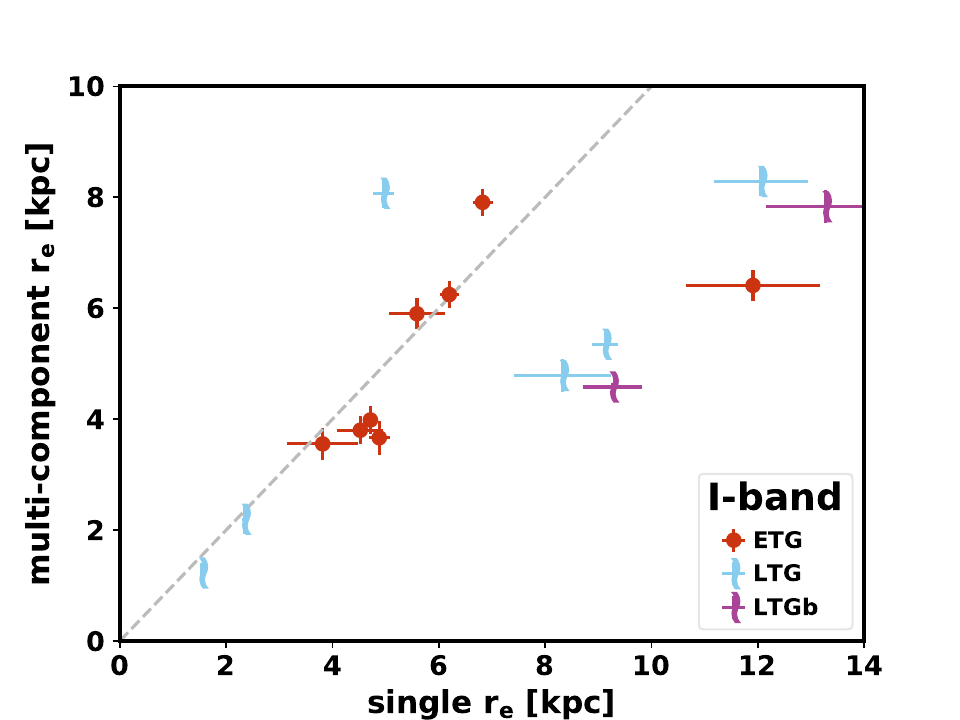}
 \caption{Relation between single and multi-component effective radius in kpc, for the $H$-band (top panel) and for the $I$-band (bottom panel).
 Most offset objects are disk-like, hence these objects demand to use at least a two-component fit. Thinking a two-component analysis is always a better representation of a galaxy surface brightness profile, this plot indicates that high single Sérsic effective radii are not accurate proxies of galaxy sizes, and caution needs to be taken when interpreting effective radii greater than 10 kpc from single Sérsic fits.}
 \label{fig:multi_re}
\end{figure}

Since for our high signal-to-noise ratio galaxy sample a two-component fit provides a better match to the 2D surface brightness profiles than a single Sérsic model, this double-Sérsic model is expected to yield more accurate effective radii. 
Following the definition of the effective radius \reff\, we computed curves of growth for the best-fitting two-component \textsc{GALFIT} models by integrating the flux within concentric elliptical apertures until we reach half of the galaxy total flux. The axis ratio and position angle of these ellipses is fixed to the values of each galaxy's single Sérsic models. In order to check the accuracy of our algorithm, the same scheme was applied to single Sérsic models, yielding a satisfactory agreement between the \reff\ obtained with the curves of growth and the \reff\ obtained directly from the \textsc{GALFIT} output.

In Figure~\ref{fig:multi_re} we show the relation between single and the computed multi-component effective radius in the two bands.
Most of the deviating objects with single Sérsic models implying a higher \reff\ than multi-component Sérsic fits are LTGs. This trend can be explained by the fact that these objects demand to use a combination of at least two models of light (disk + spheroid) for an adequate study of their structure. In the case of ETGs, this effect could also be related with imposing an exponential disk in a multi-component Sérsic fit of a galaxy which does not require such a component. To test this we performed a new multi-component analysis of each early-type galaxy, letting the two Sérsic indices values free. 
The results from this test did not appreciably change the values of effective radius, leaving us with the conclusion that the assumption of a disk ($n$=1) instead of an additional higher-$n$ Sérsic component does not appreciably alter \reff\ determinations for ETGs with \textsc{GALFIT}.

\begin{figure}[h]
 \centering
    \includegraphics[width=0.49\textwidth]{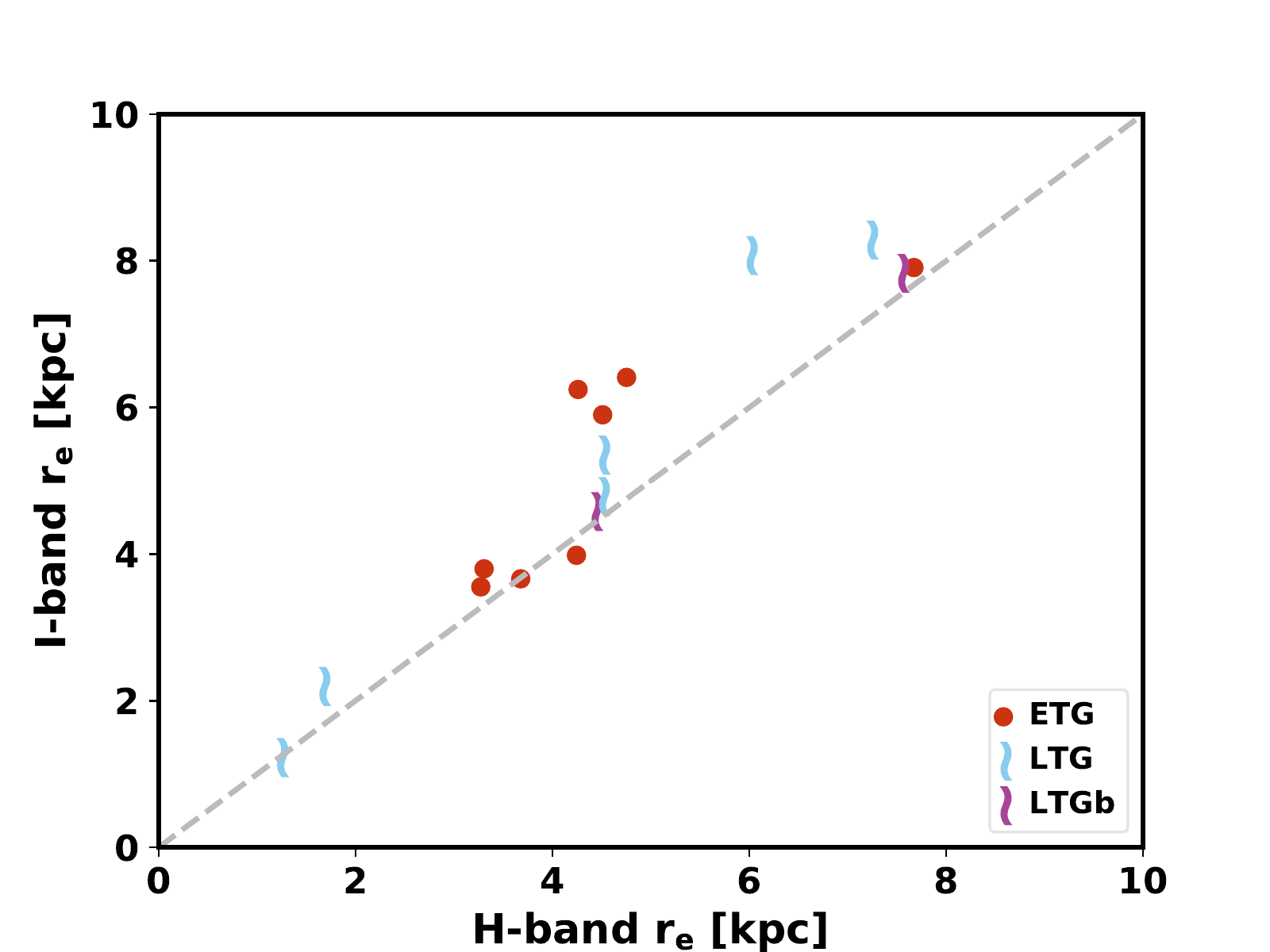}
 \caption{Comparison between the size estimates (computed multi-component effective radii) in units of kpc, using the WFC3 $H$-band versus ACS $I$-band filters, for 16 galaxies in our sample. There is a good correlation between the computed sizes in both bands, albeit a tendency for slightly higher values for the $I$-band.}
 \label{fig:sizes}
\end{figure}

Figure~\ref{fig:sizes} shows our computed multi-component effective radii for both bands. We retrieve a good correlation (values close to the 1-to-1 line relation, together with $\sigma_{\textnormal{NMAD}}$ = 0.12) between the computed sizes in the different bands, yet systematically obtaining slightly higher values for the $I$-band.

\subsection{Mass-size relation}

The relation between galaxy size (usually quantified as \reff) and mass evolves differently for spheroid-like and disk-like massive galaxies, with spheroids having a stronger evolution in size than disk-like objects \citep[e.g.][]{trujillo07,buitrago08,vdw14}.

\begin{figure*}[ht]
 \centering
    \includegraphics[page=5,width=1\textwidth]{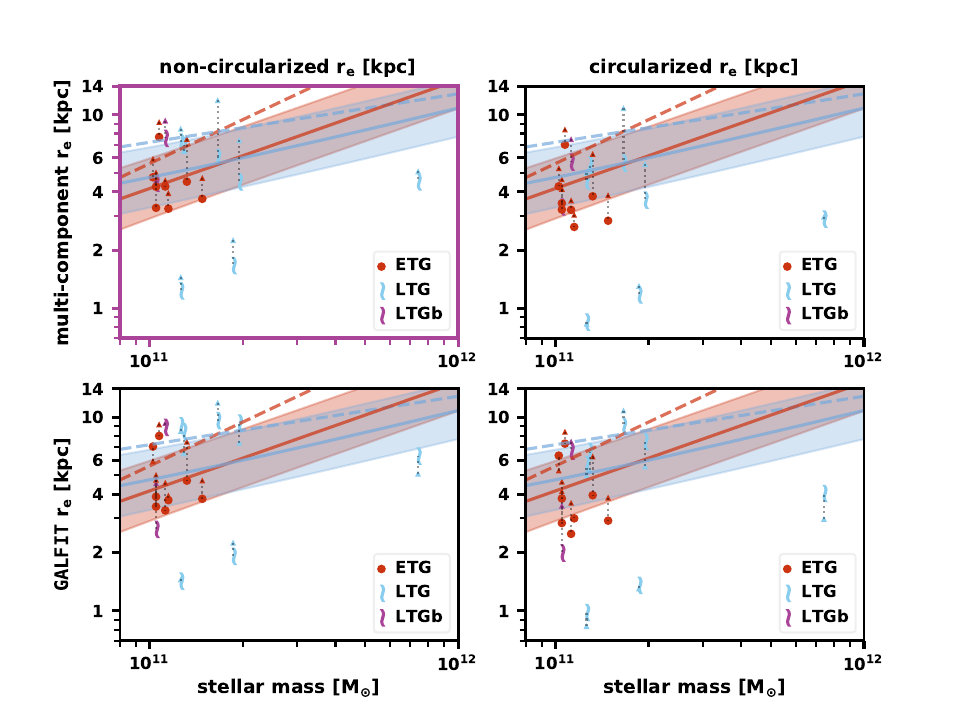}
 \caption{Size-stellar mass distribution of our $H$-band data, linked to \cite{vdw} results represented by triangles. Left panels show non-circularized effective radius, while in the right panels are displayed the circularized ones ($r_{e, \rm circ}$). Top panels correspond to our computed multi-component effective radii, whereas bottom panels show the \textsc{GALFIT} single Sérsic effective radii. \cite{shen} local relation for ETGs and LTGs is represented by the solid red and blue lines, respectively, with the corresponding scatter being the shaded red and blue regions. The colored dashed lines correspond to \cite{vdw14} mass-size relations at $z = 0.25$, as obtained from non-circularized $r_e$ determinations. Our results lay within the scatter of \cite{shen} local relation. However, there are two disk-like galaxies significantly smaller (by $\sim2\sigma$) than the rest of objects. We choose to highlight with purple color the axis of our preferred plot, since it shows the values of the most representative effective radius (multi-component $r_e$) without circularization (as in the works that we are comparing our results with).}
 \label{fig:mass_size}
\end{figure*}

\cite{shen} investigated the size distribution of $\sim$140000 galaxies from the SDSS and its dependence on luminosity, stellar mass and morphology, separating spheroid-dominated and disk-dominated systems accordingly with the indicators $n=2.5$ and $c=2.86$, which refer to Sérsic index and concentration index \citep[see][]{shimasaku2001}, respectively.
In another work, \cite{vdw14} took advantage of the slitless spectroscopy from the 3D-HST survey combined with CANDELS photometry to study the evolution of the galaxy size-mass distribution since $z = 3$. The authors separate the two classes of galaxies on the basis of star formation activity, using rest-frame $U$-$V$ and $V$-$J$ color distributions of galaxies with stellar mass above $10^{10}M_\odot$.
The combination of these two works is the reference we utilize to compare with our results.

\begin{figure*}[ht]
 \centering
    \includegraphics[page=5,width=1\textwidth]{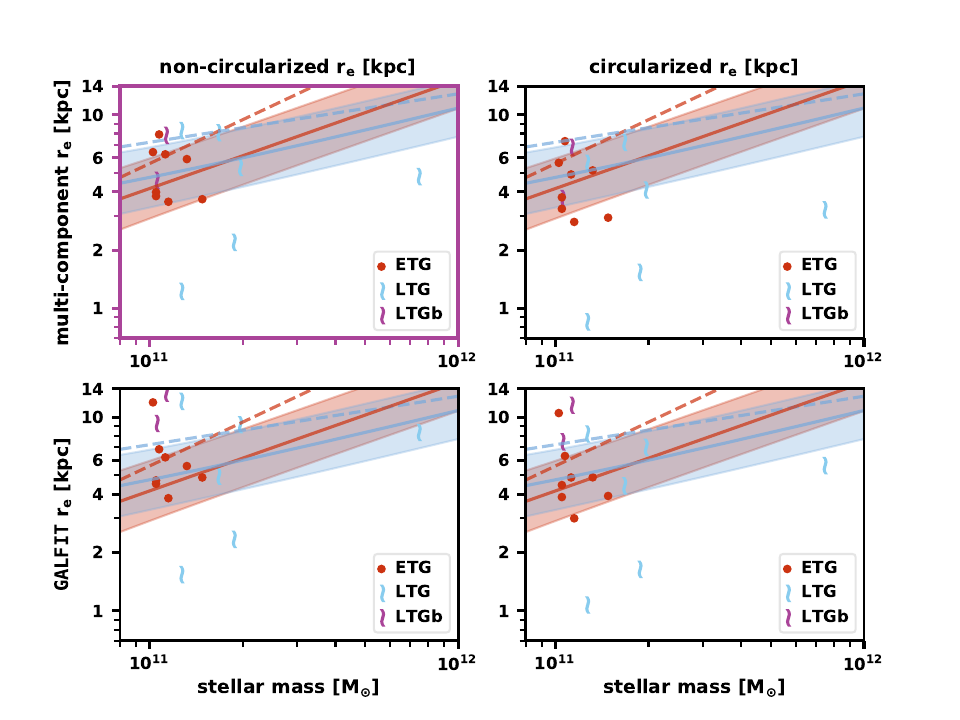}
 \caption{Size-stellar mass distribution of our $I$-band data. See the caption of Figure~\ref{fig:mass_size} for detailed description of this image. Similar to the results of the $H$-band, most of our sample lies in the local relation of \cite{shen}, with exception for the same two LTGs, that are around $\sim$2$\sigma$ smaller than expected.}
 \label{fig:mass_size_i}
\end{figure*}

We construct the mass-size relation for our $H$ and $I$-bands data in Figs.~\ref{fig:mass_size} and ~\ref{fig:mass_size_i}, and compare them with the results from \cite{shen}, and \cite{vdw14} for their lowest redshift bin centered at $z = 0.25$. We highlight with purple axis the plot we consider more relevant, i.e. the one considering that it has the most representative effective radius (multi-component $r_e$, and not circularized as in the works we use as reference to compare). The circularized effective radius ($r_{e,circ} = r_e \sqrt{b/a}$) is commonly used in the literature in order to correct somehow for the line-of-sight projection of triaxial ETGs, and thus we display our results both with and without circularization (right and left panels, respectively).
Early-type galaxies are represented with red dotted points, the late-type population is represented with blue spirals, and barred LTGs by purple spirals.
We show in the y-axis of the top panels the computed multi-component effective radii and, in the bottom panels, we present the \textsc{GALFIT} single Sérsic effective radii. The results from \cite{vdw} for each galaxy in our sample are represented by triangles and connected with a line to our data. \cite{shen} local relation is represented by red (for ETGs) and blue (for LTGs) solid lines, with their respective scatter represented by the shaded regions. The dashed lines correspond to \cite{vdw14} at $z = 0.25$ (again, red for ETGs and blue for LTGs). We recall that \reff\ values from \cite{vdw,vdw14} and \cite{shen} were inferred from fitting single Sérsic functions.
It is noteworthy, even though the median redshift of our sample is 0.39, that our results are consistent with those found by \cite{shen}, having almost the entire sample falling inside their scatter. This similarity between both mass-size relations indicates that these galaxies do not experience a big evolution in sizes. Furthermore we find in both bands two LTGs that are outliers (with sizes $\sim$2$\sigma$ below \cite{shen} relation), implying that ETGs seem to be at place already early-on.

\subsection{Surface brightness profiles and color profiles}

To check how the observed surface brightness profiles of our sample galaxies compare with 
those implied by the best-fitting models with \textsc{GALFIT} we computed from the latter  1D SBPs for both $H$ and $I$ band. The profiles for the entire sample can be found in Appendix~\ref{appen:sbs}.

In Figure~\ref{fig:SBs} we present SBPs for three representative galaxies from our sample that were visually classified as ETGs (top panels), barred LTGs (middle panels), and LTGs (bottom panels). The profiles in the $H$-band and $I$ band are displayed in the left and right panel, respectively. The inset in each profile contains the corresponding galaxy stamp in units of surface brightness (mag arcsec$^{-2}$; see the corresponding color bar) with shadowed areas matching the masks used to recover the observed light profile, the galaxy ID on top right, and a scale bar corresponding to 10 kpc on top left. In the profile itself, black points show the observed SBP, and violet and green solid lines correspond to the results from single and multi-component Sérsic models after convolution with the PSF. For multi-component fits, the dashed colored lines illustrate the bulge (red), disk (blue) and bar (orange). The dotted vertical lines depict the different effective radii: violet color for the single Sérsic model, green color for our computed from the multi-component model, and gray color for single Sérsic results in \cite{vdw}.

\begin{figure*}[!]
 \centering
    \vspace{-10pt}
    \includegraphics[page=4,width=0.45\textwidth]{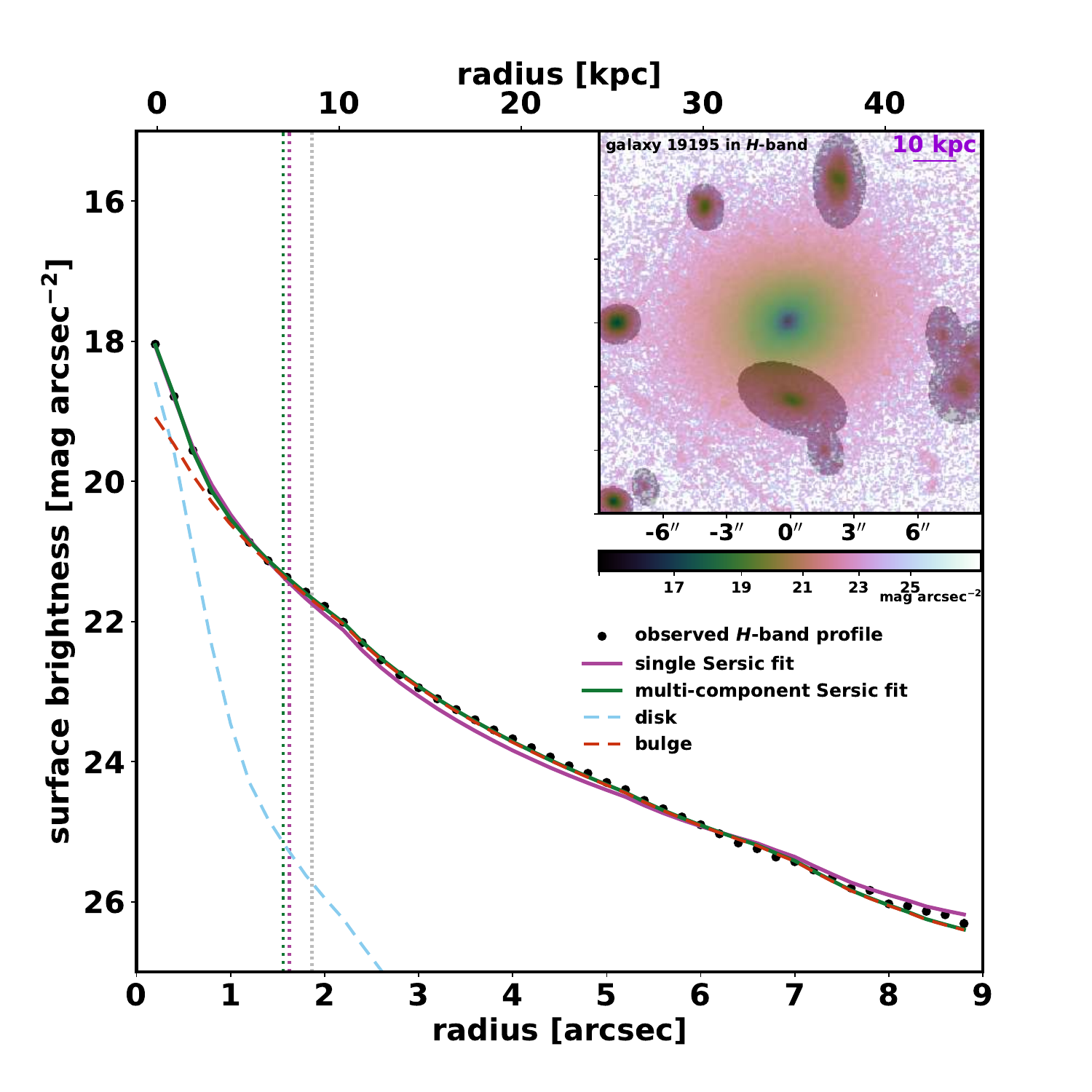}
    	\vspace{-10pt}
    \includegraphics[page=4,width=0.45\textwidth]{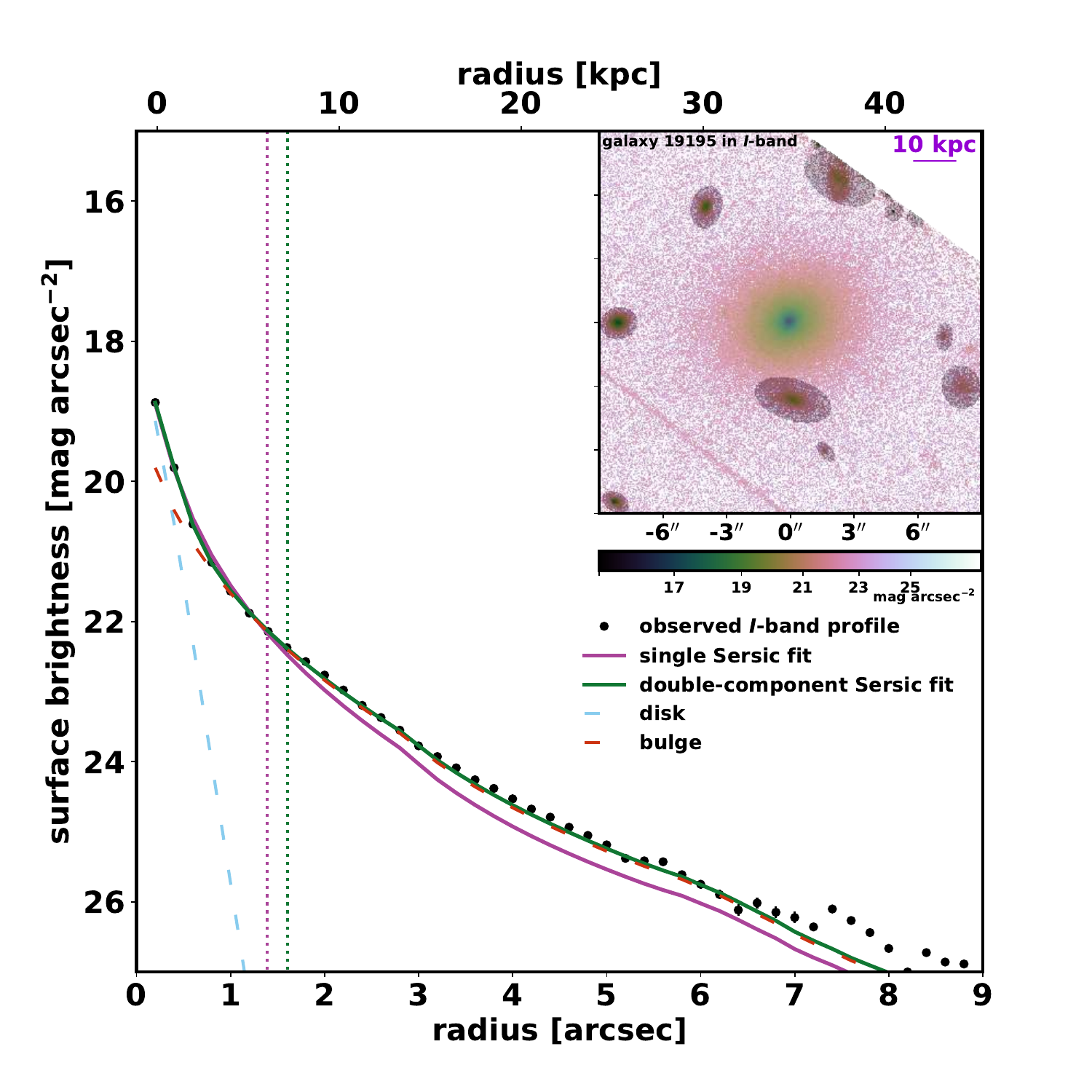}
    \vspace{-10pt}
    \includegraphics[page=2,width=0.45\textwidth]{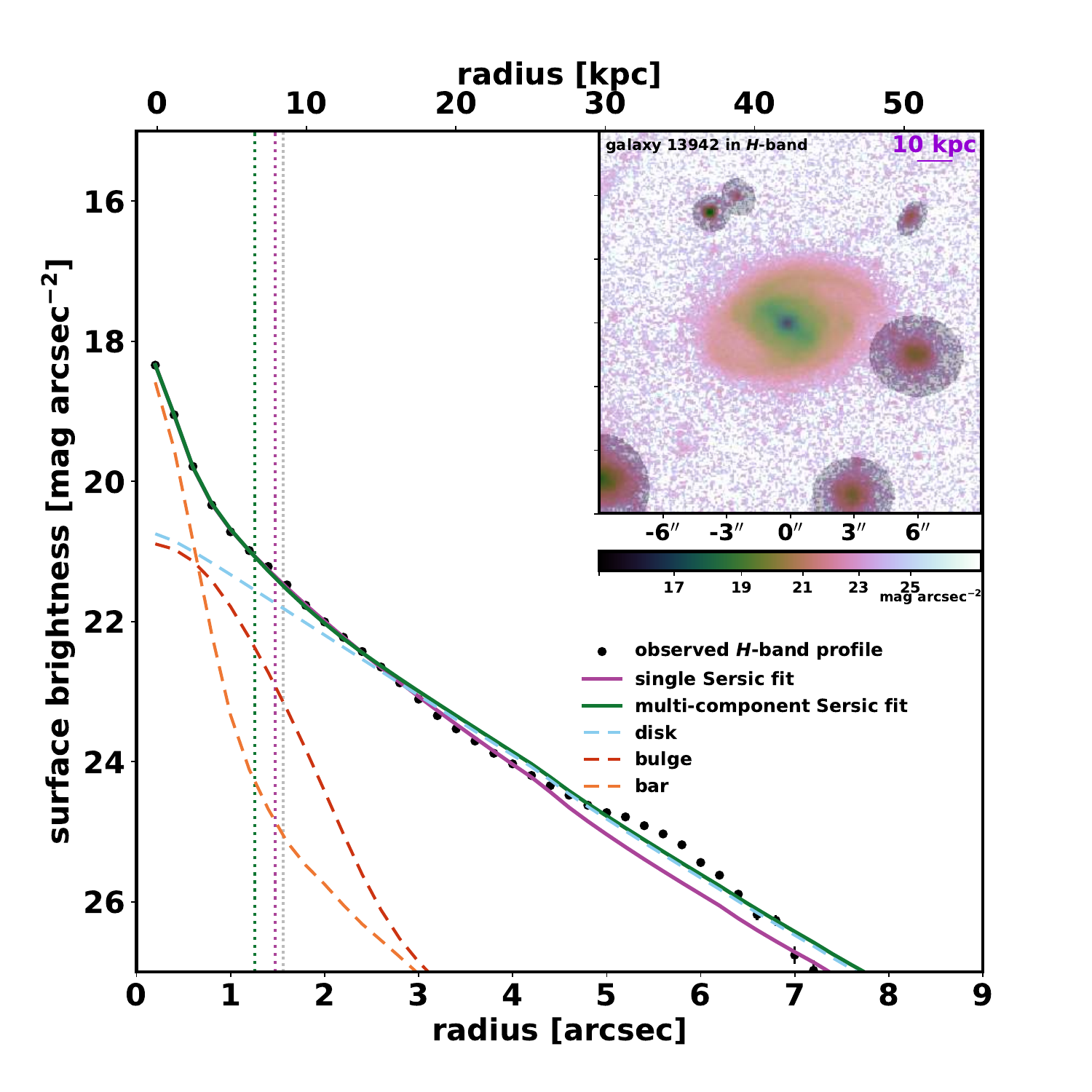}
    \includegraphics[page=2,width=0.45\textwidth]{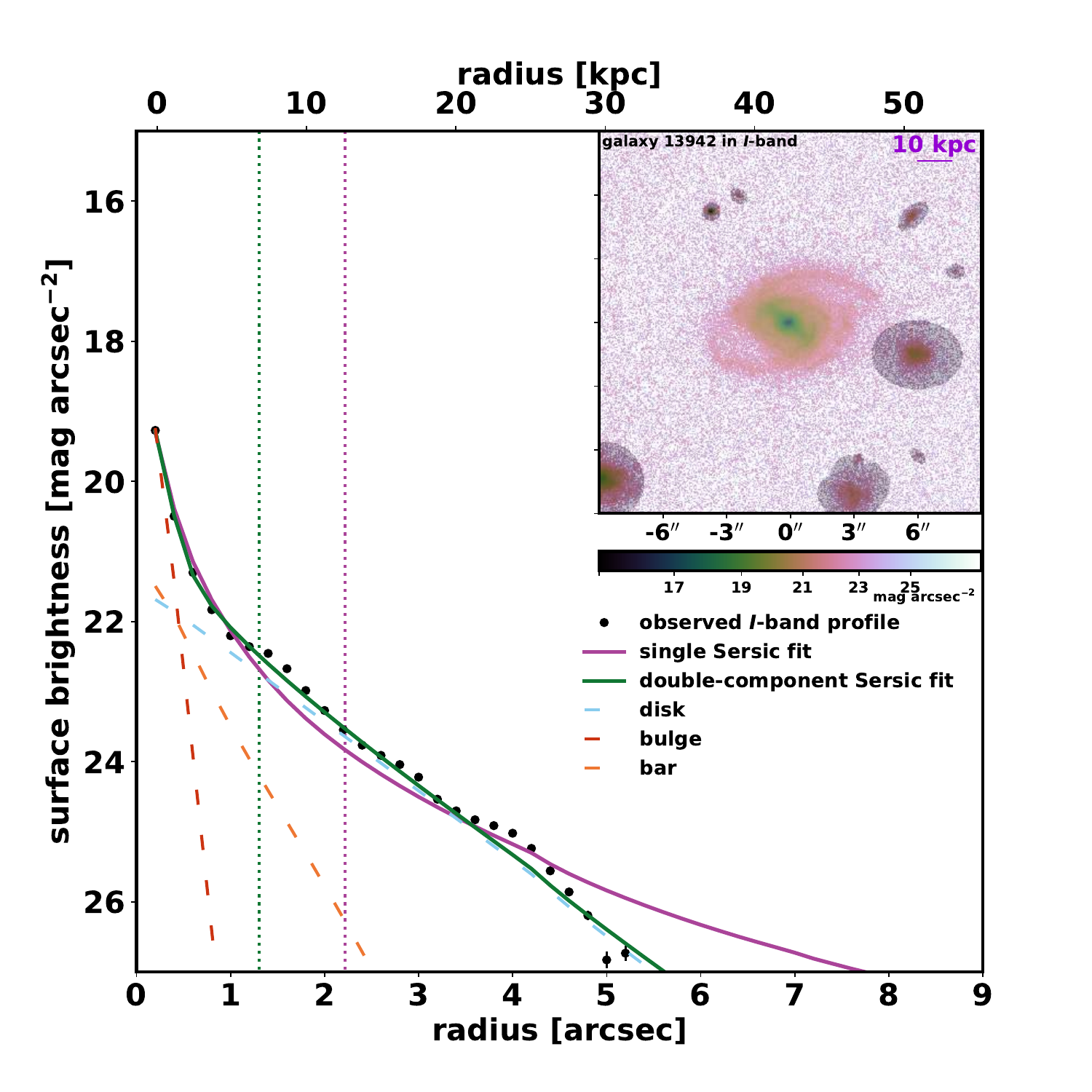}
	\vspace{-10pt}    
    \includegraphics[page=9,width=0.45\textwidth]{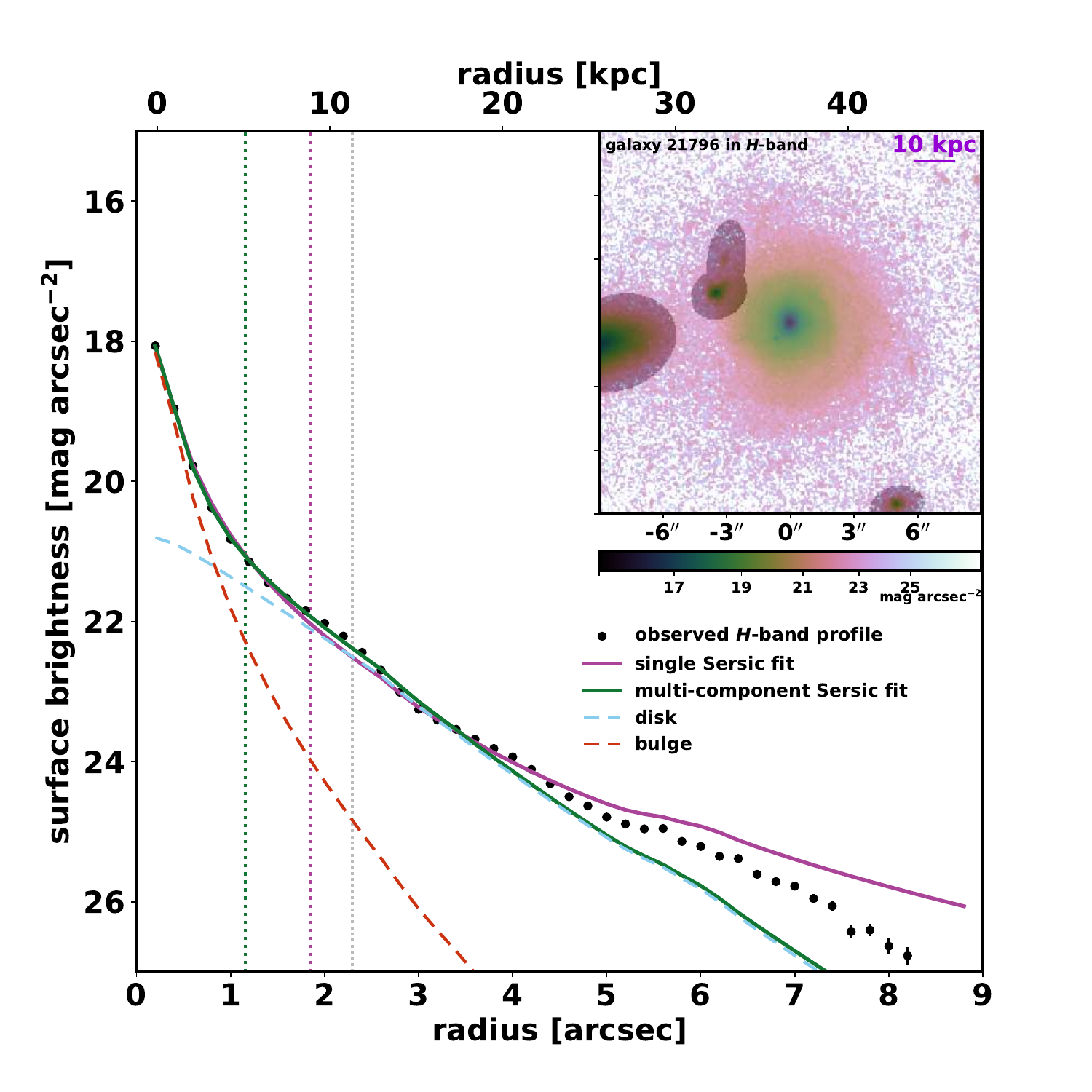}
    \includegraphics[page=8,width=0.45\textwidth]{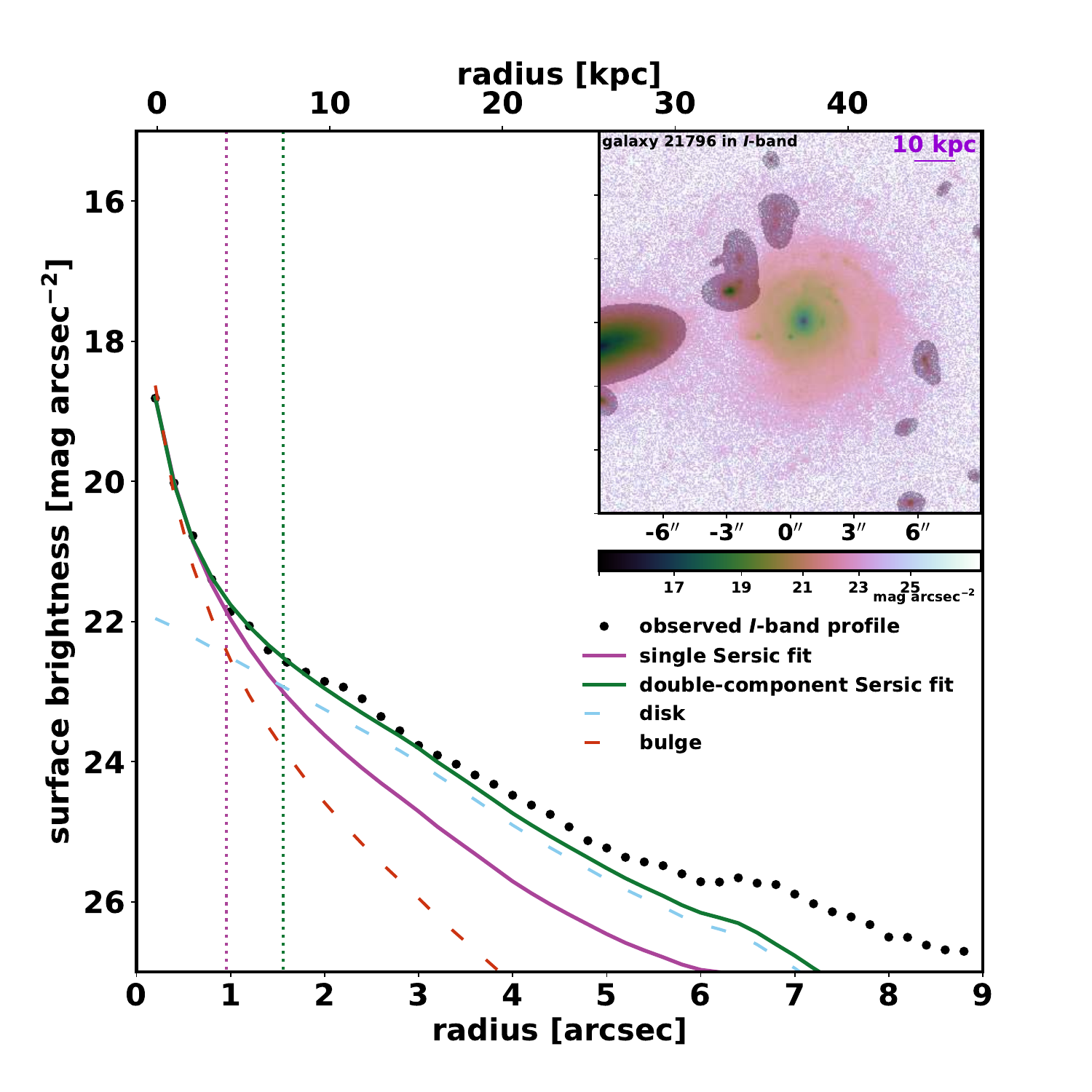}
 \caption{Surface brightness profiles in the $H$-band (left panels) and in the $I$-band (right panels) for three systems illustrating the different galaxy morphologies in our sample: the early-type galaxy ID 19195 (top panels), the barred late-type galaxy ID 13942 (middle panels), and the late-type galaxy ID 21796 (bottom panels). The violet and green colors correspond, respectively, to the results from single- and multi-component Sérsic models. Black points represent the observed surface brightness profile; solid lines show the models convolved with the PSF; dashed lines stand for the decomposition of the multi-component model into bulge (red), disk (blue), and bar (orange); dotted vertical lines represent the effective radius; and the dotted gray vertical line is the \cite{vdw} (single Sérsic) effective radius. We display the galaxy stamp in units of surface brightness (mag arcsec$^{-2}$; cf. the horizontal color bar beneath each map) with shadowed areas matching the masks used to obtain the observed light profile, the galaxy ID on top left and a scale bar on top right corresponding to 10 kpc. 
 }
\label{fig:SBs}
\end{figure*}

All galaxies in our sample are more luminous in the $H$-band. For almost half of our sample an over-prediction of the light in the galaxy outskirts is obtained when performing a single Sérsic fit (solid purple line). For the rest of the objects the light is under-estimated in the outskirts, both for the single Sérsic fit and the multi-component fit (solid green line). As expected, multi-component Sérsic fits yield a better match to the observed SBPs (black points).

Despite our efforts with B+D decompositions, many objects show small-scale patterns in their residuals in both bands after the removal of the principal galaxy components (bulge + disk). These residuals hold clues on galaxy substructures that cannot be described by axis-symmetric Sérsic functions, although their physical interpretation is not always clear. These subtle features are listed below and included as a figure in Appendix~\ref{appen:residuals}: 
\begin{itemize}
    \item LTGb 37194 shows large over-subtractions surrounding the entire galaxy, but also between the spiral arms.
    \item LTG 30654 has symmetric over-subtractions on both sides from the center of the disk plane, perpendicular to the minor axis of the galaxy, standing out a nuclear stellar disk.
    \item LTG 37587 exposes its spiral arms with some over-subtractions around them.
    \item ETG 3740 displays asymmetric over- and under-subtractions in the galactic center, suggesting the presence of a faint halo enclosing the galaxy, and a nuclear disk or bar.
    \item ETG 19195 remains with faint circumnuclear rings.
    \item ETG 20050 shows asymmetric over-subtractions along the galaxy plane, suggesting the presence of a nuclear bar with low-level dust patterns.
    \item ETG 10876 reveals nuclear spiral arms with asymmetric over-subtractions, and a nuclear point source.
    \item LTG 21796 exhibits asymmetric over- and under-subtractions, emerging some spiral patterns and two point sources on each side of the nucleus.
    \item LTG 18935 has considerable over- and under-subtractions, standing out in its spiral arms and clumps.
    \item LTG 21604 remains with an interesting asymmetric luminous central substructure with spiral-like features. 
    \item ETG 25781 exposes very faint circular rings.
    \item ETG 1996 displays a small under-subtraction in the galactic center. 
    \item ETG 21306 presents an asymmetric under-subtraction in the outskirts of the galaxy, suggesting the presence of an outer asymmetric envelope. 
    \item ETG 4735 has symmetric over-subtractions perpendicular to the major axis of the galaxy, on both sides from the center of the disk plane, remaining a disk likely to be an artifact created by a model mismatch.
    \item LTG 23956 displays asymmetric over-subtractions in the central part of the galaxy, standing out a nuclear substructure with dust patterns, and a faint halo surrounding the outer parts of the galaxy.
    \item LTG 7013 exhibits symmetric cross-shaped over-subtractions.
    \item LTGb 13942 exposes spiral arms in the outer parts of the disk, a circumnuclear ring, and another ring around the bar component.
\end{itemize}

Whereas visual inspection of the residuals after subtraction of the best-fitting 2D model from an image clearly reveals large contiguous regions where the fit over/under-estimates the galaxy surface brightness, this comparison does not allow for a quantitative inference on these residuals in units of mag. Having this information would be important, however, since among other things it is also needed for estimating the goodness of the color maps implied by subtraction of the best-fitting 2D model with {\sc GALFIT} in two different bands. 
An inference on such deviations in mag can, however, be more easily obtained from comparison of synthetic and observed SBPs in Figure~\ref{fig:colors} and in the figures of Appendix~\ref{appen:colors}. It can be seen that, especially in the periphery of galaxies, the fit in many cases deviates by 0.5-1 mag from the observed SBP, both for single- and double-Sérsic fits. This documents a global failure of the adopted fitting scheme with {\sc GALFIT} to give a proper approximation to the surface brightness (and color) of the low-surface brightness outskirts of the galaxies under study. Therefore, whereas the extensive set of simulations on mock galaxy images (Appendix~\ref{appen:simulations}) demonstrates that, on the statistical average, {\sc GALFIT} can retrieve Sérsic model parameters with a satisfactory accuracy, individual fits can yield systematic and substantial deviations from the true 1D/2D surface brightness distribution of even for a relatively 'simple' galaxy, as an ETG. This may be partly attributed to the inherent uncertainties in non-linear fitting which, as empirically quantified in the Appendix~\ref{appen:simulations}, increase with increasing Sérsic exponent and decreasing luminosity (i.e. signal-to-noise).   

\begin{figure*}
 \centering
    \includegraphics[page=4,width=0.32\textwidth]{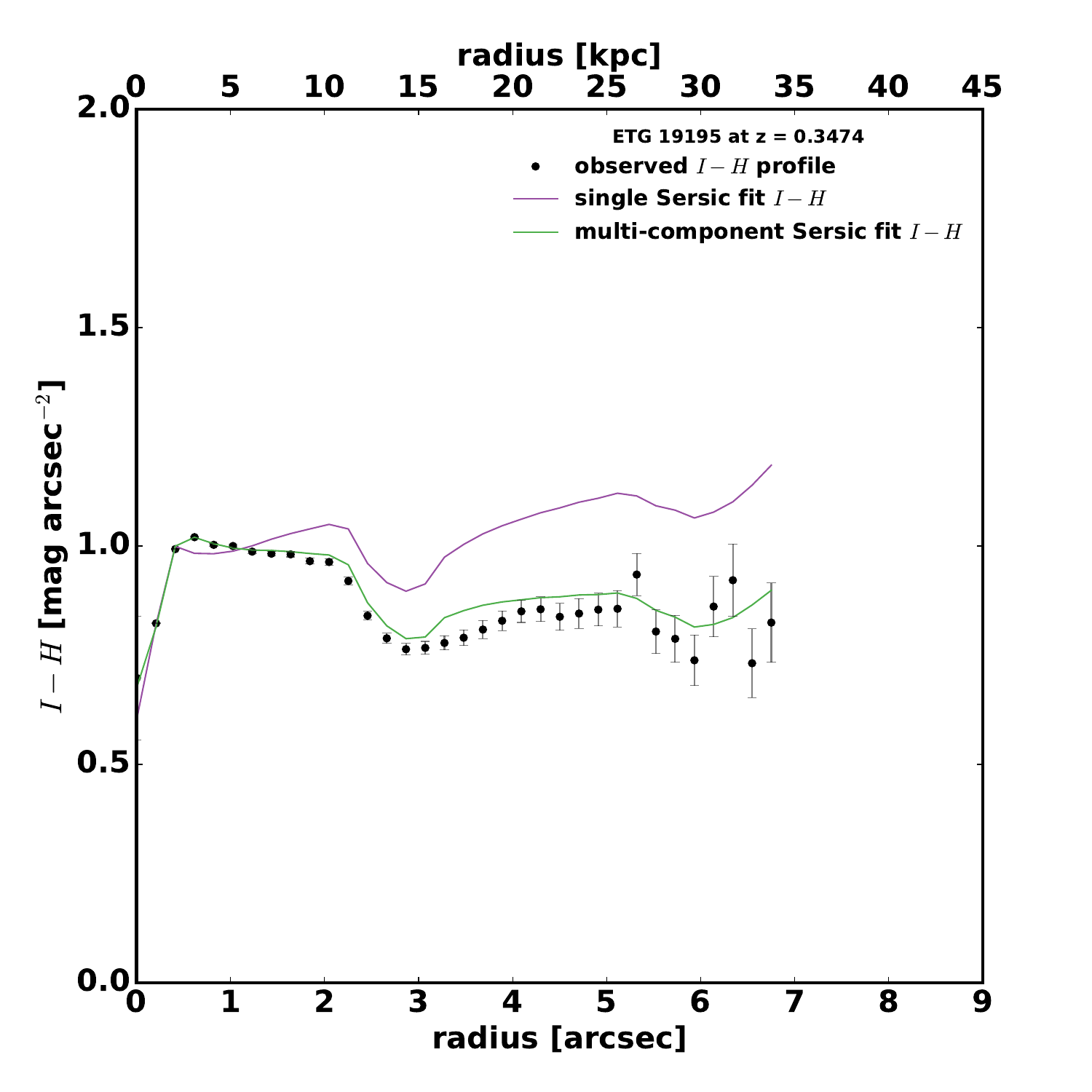}
    \includegraphics[page=2,width=0.32\textwidth]{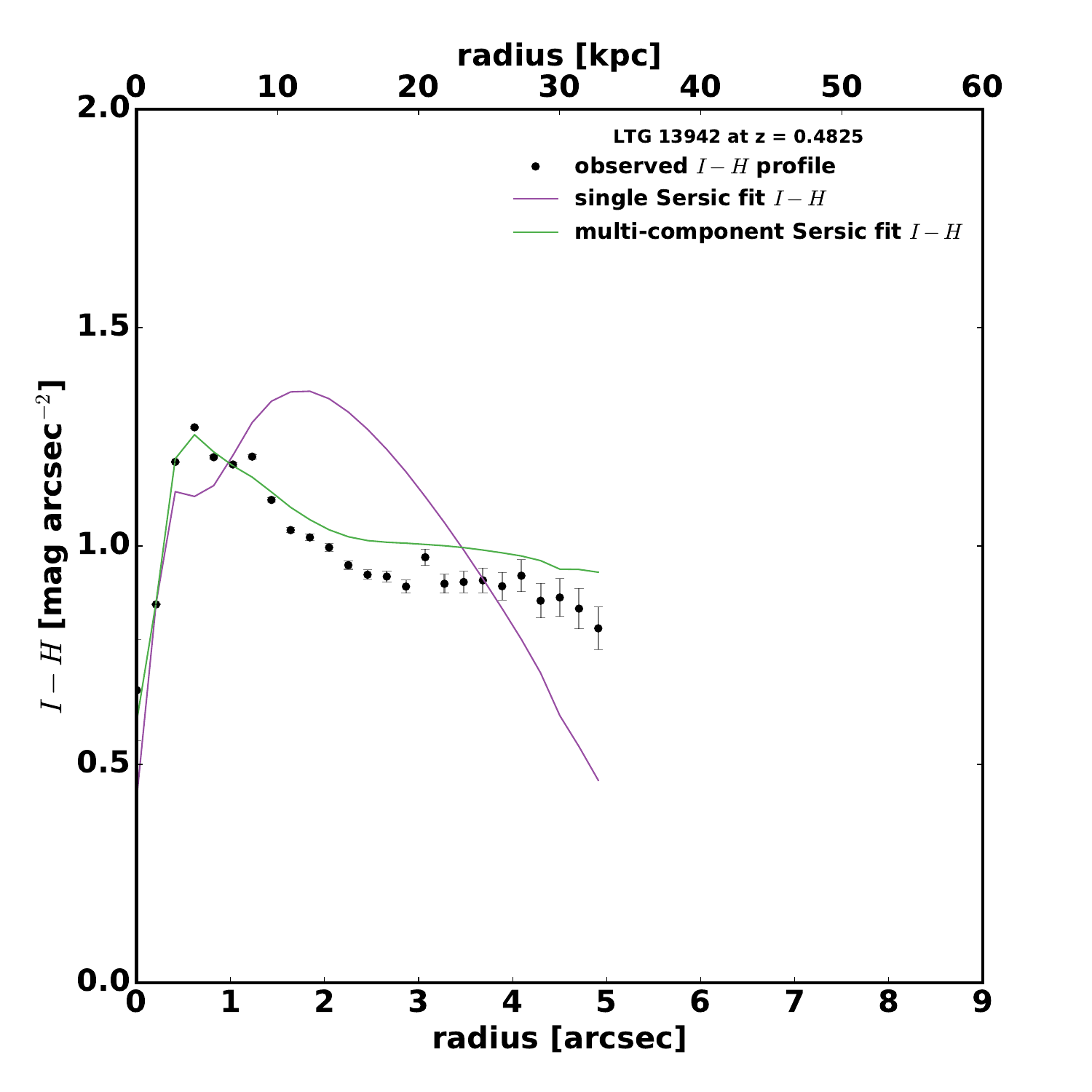}
    \includegraphics[page=8,width=0.32\textwidth]{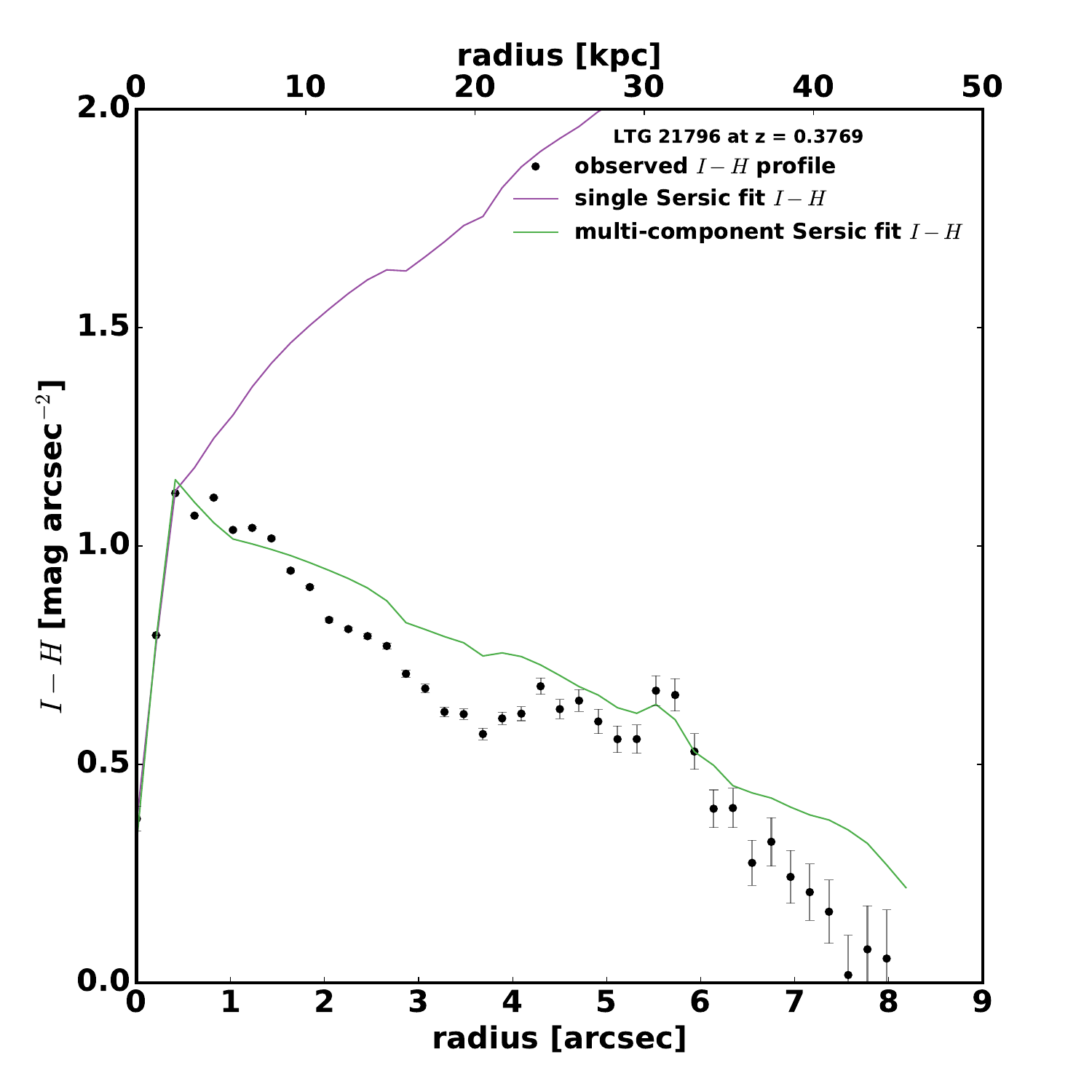}
 \caption{$I$--$H$ color profiles for ETG 19195, LTGb 13942 and LTG 21796. Data points with larger error bars (> 0.3 mag) are not shown. }
 \label{fig:colors}
\end{figure*}

Furthermore, the specifics of the fitting procedure could also strongly influence Sérsic model solutions. For example, \cite{papaderos96a} used the Levenberg-Marquardt non-linear fitting algorithm to decompose SBPs of blue compact dwarf galaxies into a Sérsic and a Gaussian component, accounting for the luminosity contribution of the star-forming component, plus an exponential component approximating the more extended underlying stellar host. 
They pointed out that the solution significantly depends on whether or not SBP data points are weighted by photometric uncertainties $\sigma\mu_{\rm i}$ (because central points have lower uncertainties): doing so, generally leads the innermost (highest surface brightness and thus lowest-$\sigma\mu_{\rm i}$ points) dictating the solution, which typically leads to a 'compactification' of the exponential host in the sense of an overestimated central surface brightness and underestimated exponential scale length\footnote{For this reason, these authors considered in their analysis both weighted and non-weighted non-linear fitting solutions.}.
This may also reflect on the rest of structural parameters \citep[see also][]{ribeiro16}.

The dependence of the fit on the central data points makes a precise correction for PSF convolution effects critically important and eventually partly accounts for the systematic deviations between the best-fitting Sérsic model and the observed SBP in the low-surface brightness periphery of galaxies. Furthermore, at this depth and as far as ETGs are concerned, we may be looking at stellar haloes being both structurally and evolutionary distinct from the galaxy main body. Such deviations are apparent from several \textsc{GALFIT} models, and more noticeable when producing the $I$--$H$ color profiles implied by subtraction of \textsc{GALFIT} models in the two different bands (see Figure~\ref{fig:colors} and Appendix~\ref{appen:colors}). This is an elementary consistency check for any 1D or 2D parametric fitting approach, namely whether the color profiles implied by fits in individual bands are in reasonable agreement with the observed ones and make sense from the astrophysical point of view. The fulfillment of the latter condition is obviously crucial for ensuring that the adopted 1D/2D parametric fitting procedure leads to meaningful observational constraints on the age distribution of stellar populations in galaxies. As apparent from Figure~\ref{fig:colors_appendix}, this is not the case for most galaxies under study, given that deviations by >0.3 mag between observed and modeled color profiles are apparent over extended zones, both for multi- and single-component Sérsic fits (7 and 3 of the 13 profiles, respectively).
In the case of the ETGs in our sample, the reason behind the extra light at larger galactocentric distances is most probably their haloes. Nevertheless, CANDELS is not optimized to study the low surface brightness Universe, which means that our images could be potentially affected by the way their data reduction was performed \citep[see][]{buitrago17,borlaff19}.

\section{Summary and conclusions}
\label{sec:conclusions}

We accomplished a photometric study of the 17 low-redshift ($z < 0.5$, with 13 spectroscopic redshifts and a median redshift of 0.39) massive galaxies ($M_{stellar} \geq 10^{11} M_\odot$) in the CANDELS fields. Using superb-quality imaging data from a survey intended for high-redshift science provides an unprecedented photometric depth and spatial resolution. This information was complemented with the slitless spectroscopy data in the 3D-HST survey.

We obtained single Sérsic fits to compare with previous works in the literature, namely \cite{vdw} for the $H$-band and \cite{griffith} for the $I$-band. In general we find good agreement with their results, suggesting that a detailed modeling of individual galaxies with \textsc{GALFIT} yields a rather satisfactory agreement with results obtained from automated application of the code to large galaxy samples.

We perform B+D decompositions on our entire sample in the $H$- and $I$-bands, deriving multi-component effective radii for each galaxy, their surface brightness profiles, and construct a mass-size relation for comparison with standard published references.

Our results indicate morphology classifications based on single Sérsic fitting are poorly correlated with the visual morphology of our galaxy sample population. In particular, by using the division line of $n=2.5$ to differentiate spheroid-dominated from disk-dominated systems in the $H$-band, we would have only three galaxies classified as disk-like, whereas nine LTGs were identified in our sample on the basis of visual classification.
Regarding galaxy sizes, by computing more accurate multi-component effective radii, we show that they systematically deliver small sizes than single Sérsic effective radii when the values for the latter are greater than 10 kpc.

From the constructed mass-size relation, we located most of the galaxies in our sample within the scatter of the local \cite{shen} mass-size relation, indicating that these galaxies have a small evolution in sizes since $z < 0.5$. Moreover, the two detected outliers in the relation are disk-like galaxies, implying that spheroids are already at place early-on.

Although \textsc{GALFIT} generally gives satisfactory fits (in terms of $\chi^2$ minimization) to galaxy images in individual bands, we find that the best-fitting Sérsic model can substantially and systematically deviate from observed surface brightness profiles at intermediate-to-low surface brightness levels. This issue probably arises from the fact that the global solution in profile fitting and decomposition primarily depends on the innermost (highest S/N yet most affected by the PSF) regions of a galaxy. 
Additionally, in the case of the deep observations we are dealing with, these deviations could be partly due to the extended stellar halo of ETGs.
We also warn about the fact that a data reduction optimized for the preservation of low surface brightness features is mandatory for the next generation surveys to be also able to utilize their imaging for the $z < 1$ Universe.
Regardless of its origin, the failure of parametric 2D decomposition with \textsc{GALFIT} to closely fit the low-surface brightness periphery in more than one half of galaxies in our sample is a significant concern.
One of the implications of it is that color profiles constructed from subtraction of 2D models in two different bands may differ from observed ones by more than one mag. 
A consequence of it is that studies of the inside-out growth of galaxies across redshift that are based on synthetic color profiles from \textsc{GALFIT} could be subject of hardly predictable systematic uncertainties. 

\begin{acknowledgements}
We would like to thank the anonymous referee for valuable comments and suggestions.
We thank the European taxpayer, who in the spirit of solidarity and mutual respect between EU countries provides the Funda\c{c}\~{a}o para a Ci\^{e}ncia e a Tecnologia (FCT) with a substantial fraction of the financial resources that allow it to sustain a research infrastructure in astrophysics in Portugal. Specifically, this work was supported by FCT through national funds and by FEDER through COMPETE by the grants UID/FIS/04434/2013 \& POCI-01-0145-FEDER-007672 and 
PTDC/FIS-AST/3214/2012 \& FCOMP-01-0124-FEDER-029170.
Additionally, we acknowledge support by FCT/MCTES through national funds (PIDDAC) by  grant UID/FIS/04434/2019. We thank all members of the Thematic Line ``The assembly history of galaxies resolved in space and time'' of Instituto de Astrof\'isica e Ci\^encias do Espa\c co (IA)
for a critical review of the manuscript and numerous valuable comments and suggestions. 
FB acknowledges the support by FCT via the postdoctoral fellowship SFRH/BPD/103958/2014, the conversations with the TRACES group of galaxy evolution in the IAC, and also the RAVET programme. 
PP acknowledges support by FCT through Investigador FCT contract IF/01220/2013/CP1191/CT0002, as well as support by FCT/MCTES through national funds (PIDDAC) by PTDC/FIS-AST/29245/2017.
This project has benefited from support by European Community Programme (FP7/2007-2013)
under grant agreement No. PIRSES-GA-2013-612701 (SELGIFS).
I.B. was supported by the FCT PhD::SPACE Doctoral Network (PD/00040/2012) through the fellowship PD/BD/52707/2014 funded by FCT (Portugal) and POPH/FSE (EC) and by the fellowship CAUP-07/2014-BI in the context of the FCT project PTDC/FIS-AST/3214/2012 \& FCOMP-01-0124-FEDER-029170, as well as by the fellowship with reference CIAAUP-19/2019-BIM within the scope of the research unit IA.
J.M.G. is supported by the DL 57/2016/CP1364/CT0003 contract and acknowledges the previous support by the fellowships CIAAUP-04/2016-BPD in the context of the FCT project UID/-FIS/04434/2013 \& POCI-010145-FEDER-007672, and SFRH/BPD/66958/2009 funded by FCT and POPH/FSE(EC).
This work is based on observations taken by the CANDELS Multi-Cycle Treasury Program with the NASA/ESA HST, which is operated by the Association of Universities for Research in Astronomy, Inc., under NASA contract NAS5-26555.
We acknowledge the use of the following software packages: TOPCAT \citep{taylor}, ALADIN \citep{bonnarel}, \textsc{SExtractor} \citep{bertin}, \texttt{TinyTim} HST PSF Modeling \citep{krist}.
This research has been partly supported by the grant AYA-2016-77237-C3-1-P from the Spanish Ministry of Science, Innovation and Universities.
This research made use of \textsc{Montage}. It is funded by the National Science Foundation under Grant Number ACI-1440620, and was previously funded by the National Aeronautics and Space Administration's Earth Science Technology Office, Computation Technologies Project, under Cooperative Agreement Number NCC5-626 between NASA and the California Institute of Technology. It also made use of the following \textsc{Python} packages: NumPy \& SciPy \citep{scipy}, Matplotlib \citep{matplotlib}, Astropy \citep{astropy}, a community-developed core Python package for Astronomy; APLpy \citep{aplpy}, an open-source plotting package.
\end{acknowledgements}

\bibliographystyle{aa}
\bibliography{references.bib}
\begin{appendix}
\section{Simulations to test the estimates of the structural parameters}
\label{appen:simulations}

\begin{figure*}[h]
 \centering
    \includegraphics[page=5,width=0.45\textwidth]{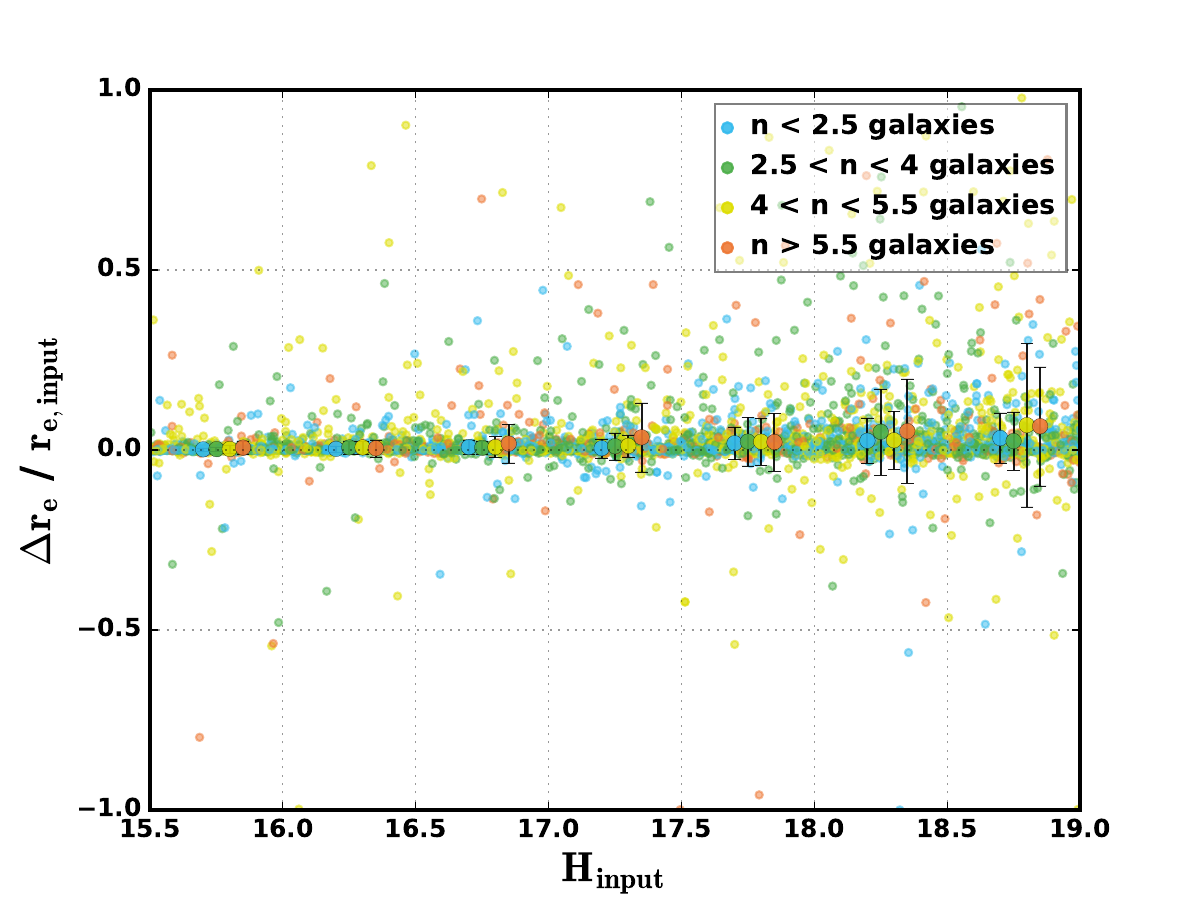}
    \includegraphics[page=2,width=0.45\textwidth]{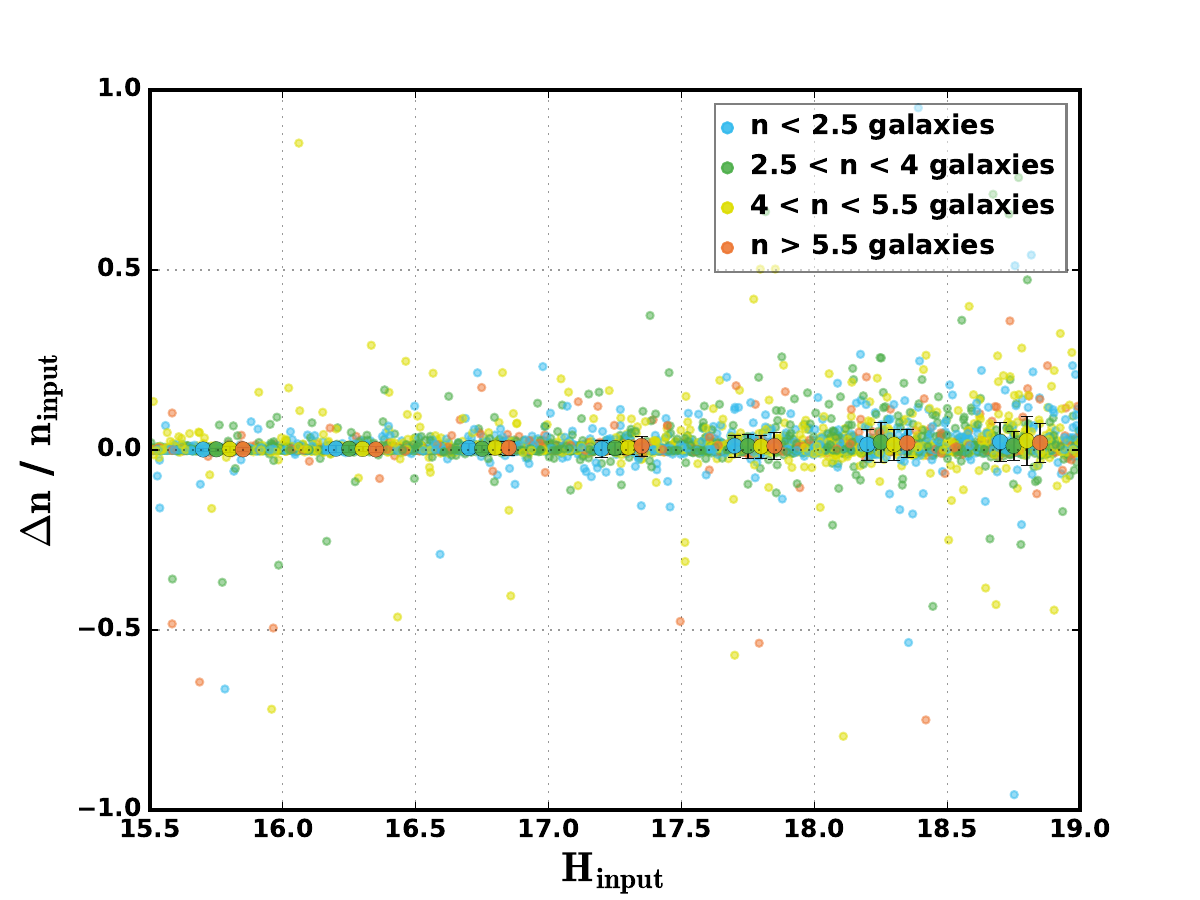}
 \caption{Relative errors – $(output-input)/input$ – of the structural parameters effective radius (left panel) and Sérsic index (right panel) of our 4500 galaxies in the $H$-band. Galaxies are color coded accordingly with the input Sérsic index. The higher is the input Sérsic index the more affected is the recovery of the structural parameters.}
\label{fig:Herrors}
\end{figure*}

We conducted a set of simulations to test the robustness of our measured structural parameters, both for the $H$- and $I$-bands. We used the ranges of the structural parameters from our analysis as a guide to create our artificial galaxies, uniformly randomized within the following ranges:
\begin{center}
$15.5 < H_{AB} [mag] < 19$
\end{center}
\begin{center}
$0.3 < r_{e,H\textnormal{-band}} [arcsec] <3$
\end{center}
\begin{center}
$1 < n_{H\textnormal{-band}} < 6$
\end{center}
\begin{center}
$0.4 < ar < 1$
\end{center}
where $H_{AB}$, $r_{e,H\textnormal{-band}}$, $n_{H\textnormal{-band}}$, and $ar$ stand for derived $H_{AB}$-band magnitude, effective radius, Sérsic index, and axis ratio, respectively. For the $I$-band we used the values of the structural parameters were uniformly randomized within the following ranges:
\begin{center}
$15.5 < I_{AB} [mag] < 20$
\end{center}
\begin{center}
$0.3 < r_{e,I\textnormal{-band}} [arcsec] <3$
\end{center}
\begin{center}
$1 < n_{I\textnormal{-band}} < 7$
\end{center}
\begin{center}
$0.4 < ar < 1$
\end{center}
where $I_{AB}$, $r_{e,I\textnormal{-band}}$, $n_{I\textnormal{-band}}$, and $ar$ stand for derived $I_{AB}$-band magnitude, effective radius, Sérsic index, and axis ratio, respectively. Taking these values into account, we created 4500 mock galaxies for the $H$-band and 6000 for the $I$-band (because of the bigger range of parameters for this band) uniformly distributed along the entire parameter space, placing each galaxy randomly on the correspondent band image and convolving with the respective PSF. We analyzed each artificial galaxy with the same methodology used in our real sample for the single Sérsic fits.

\begin{figure*}
 \centering
	\includegraphics[page=5,width=0.45\textwidth]{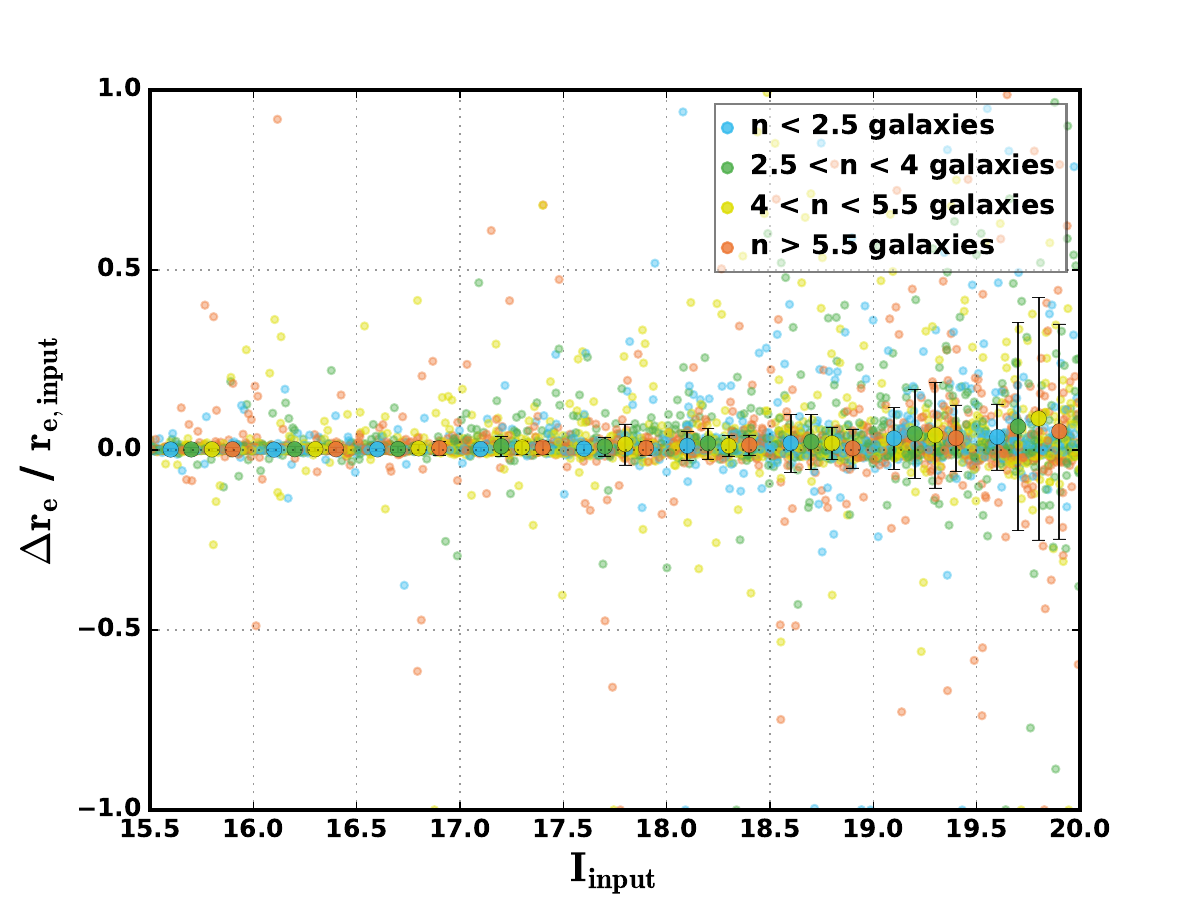}    
    \includegraphics[page=2,width=0.45\textwidth]{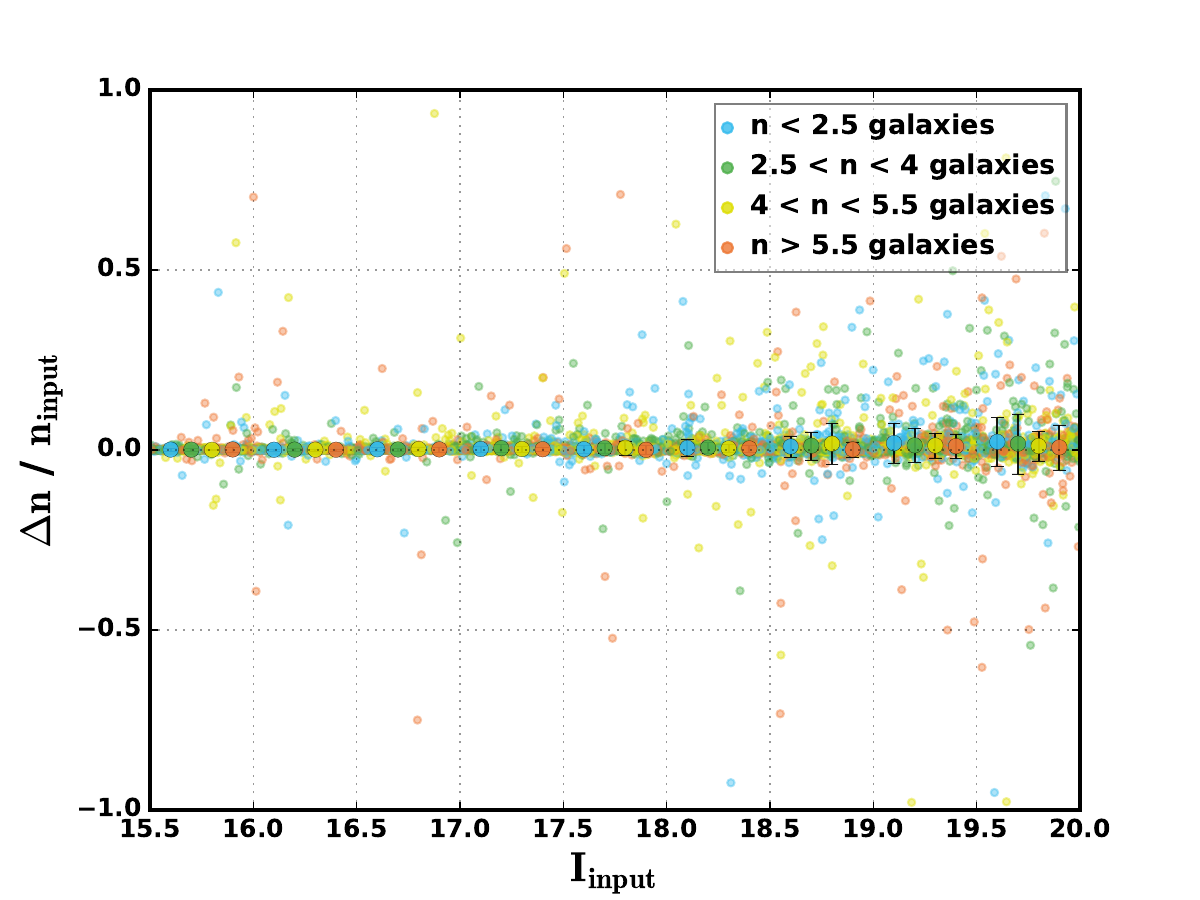}
 \caption{Relative errors – $(output-input)/input$ – of the structural parameters of our 6000 galaxies in the $I$-band. Galaxies are color coded accordingly with the input Sérsic index. Fainter objects are more affected in the recovery of the structural parameters.}
 \label{fig:Ierrors}
\end{figure*}

Figures~\ref{fig:Herrors} and~\ref{fig:Ierrors} show the relative errors of the structural parameters calculated as $(output-input)/input$ versus the input magnitude of the artificial galaxy. The sample is color coded accordingly to the input Sérsic index for the purpose of exploring the associated effects on the structural parameter. From these figures one can see that the recovery of the structural parameters in the $H$-band is more affected at increasing Sérsic index values. Low Sérsic index galaxies (blue points) are properly recovered even at faintest magnitudes.

\begin{figure*}
 \centering
    \includegraphics[page=6,width=0.45\textwidth]{figures/simulations_errorbars_hband.pdf}
    \includegraphics[page=1,width=0.45\textwidth]{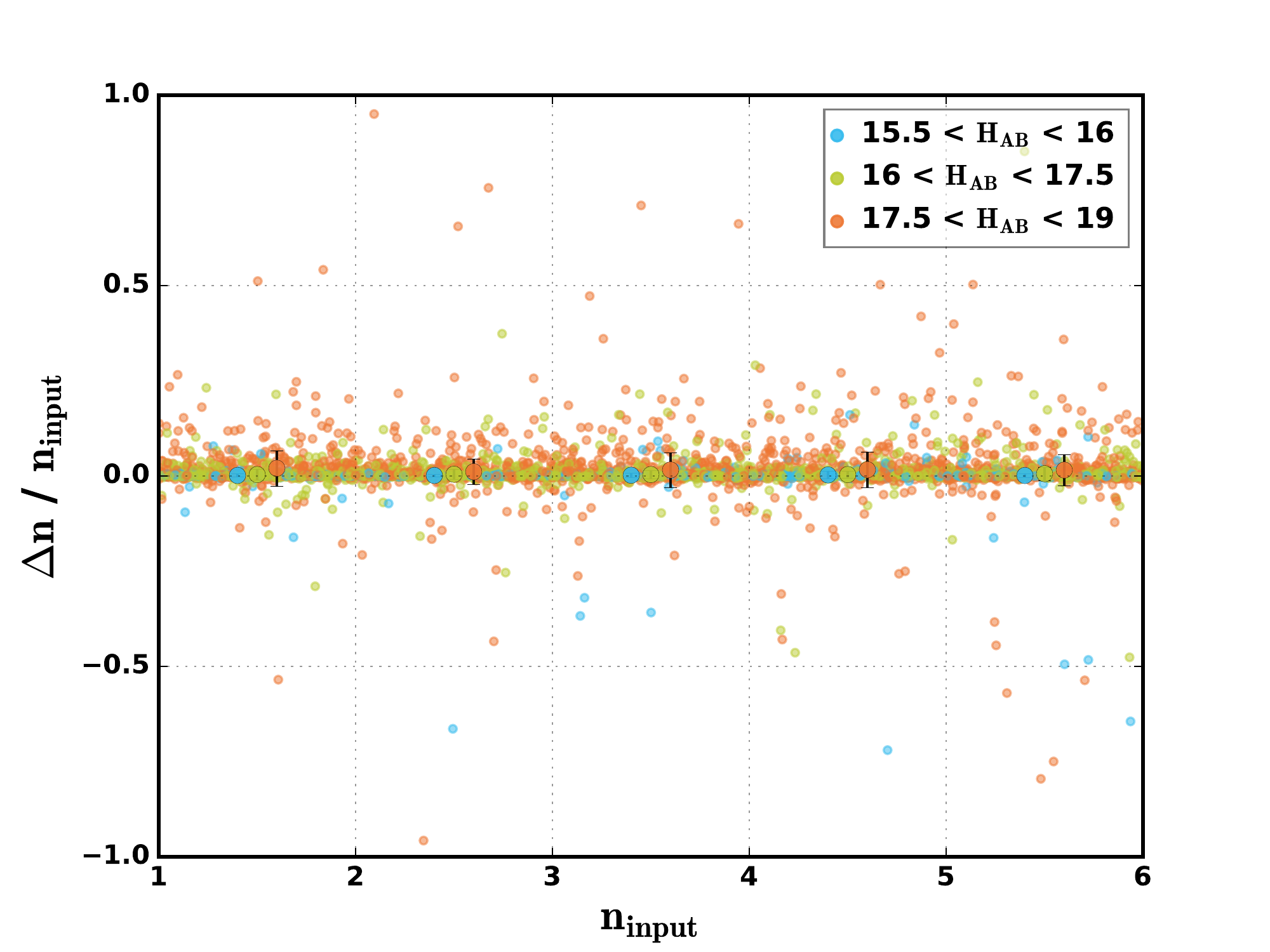}
    \includegraphics[page=4,width=0.45\textwidth]{figures/simulations_errorbars_hband.pdf}
    \includegraphics[page=3,width=0.45\textwidth]{figures/simulations_errorbars_hband.pdf}
 \caption{Relative errors – $(output-input)/input$ – of the effective radius and the Sérsic index as a function of the input Sérsic index and the input effective radius in the $H$-band.}
 \label{fig:err_simulations_H}
\end{figure*}

\begin{figure*}[h]
 \centering
	\includegraphics[page=6,width=0.45\textwidth]{figures/simulations_errorbars_iband.pdf}    
    \includegraphics[page=1,width=0.45\textwidth]{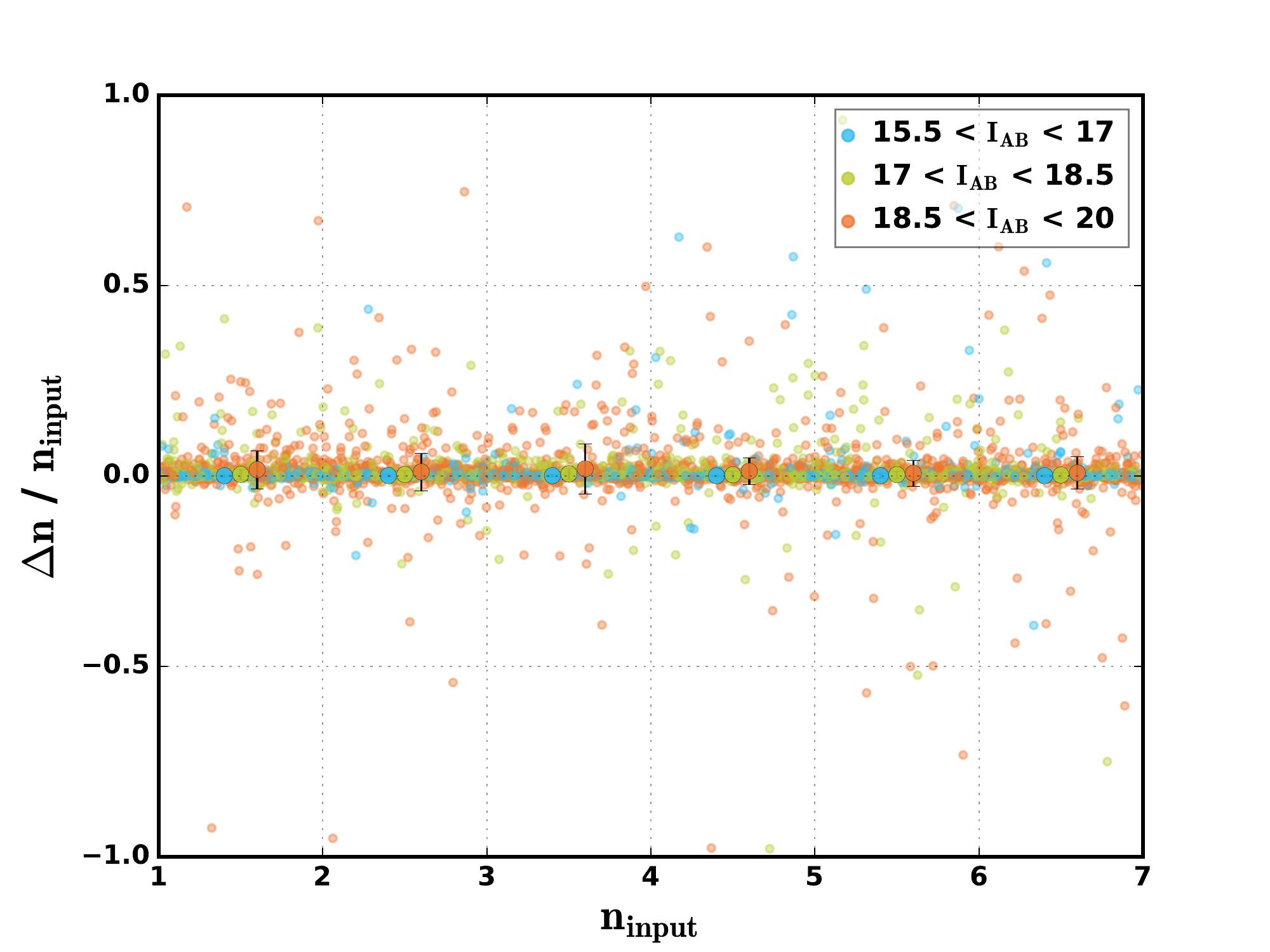}
    \includegraphics[page=4,width=0.45\textwidth]{figures/simulations_errorbars_iband.pdf}
    \includegraphics[page=3,width=0.45\textwidth]{figures/simulations_errorbars_iband.pdf}
 \caption{Relative errors – $(output-input)/input$ – of the effective radius and the Sérsic index as a function of the input Sérsic index and the input effective radius in the $I$-band.}
 \label{fig:err_simulations_I}
\end{figure*}

In Figures~\ref{fig:err_simulations_H} and~\ref{fig:err_simulations_I} the sample is splitted into three groups according to their input magnitude to appreciate the effect of the apparent magnitude additionally with the effect of the input Sérsic index (top panels) and the input effective radii (bottom panels). Fainter galaxies are more affected in recovering both the effective radius and the Sérsic index, independently of their input Sérsic index value. 

Combining the results obtained with these simulations, we find that the apparent magnitude and the Sérsic index are the essential parameters on the good recovery of the structural parameters, having the galaxy size playing a minor role.
\clearpage
\section{Tables of the retrieved structural parameters}\label{tables}

In this Appendix are presented the results of the analysis of our sample with \textsc{GALFIT}. Tables ~\ref{table:singlehband} and ~\ref{table:singleiband} list the single Sérsic fits for the $H$- and $I$-band respectively, containing the structural parameters. Tables ~\ref{table:doublehband} and ~\ref{table:doubleiband} are related to the multi-component Sérsic fits in the $H$- and $I$-bands respectively.

\begin{table*}[h]
\caption{Structural parameters for single Sérsic fits in the WFC3 $H$-band. Columns description: (1) galaxy ID, (2) magnitude, (3) effective radius, (4) Sérsic index, (5) axis ratio, (6) position angle, (7) magnitude of the bar component, (8) effective radius of the bar component, (9) position angle of the bar component.}    
\label{table:singlehband}      
\centering                          
\resizebox{0.9\textwidth}{!}{
\begin{tabular}{c c c c c c c c c}      
\hline\hline                 
\multirow{ 2}{*}{galaxy ID} & \multirow{ 2}{*}{mag $\pm$ $\Delta$mag} & r$_\text{e}$ $\pm$ $\Delta$r$_\text{e}$ & \multirow{ 2}{*}{n $\pm$ $\Delta$n} & \multirow{ 2}{*}{q} & pa & \multirow{ 2}{*}{mag$_\text{bar}$} & r$_\text{e,bar}$ & pa$_\text{bar}$\\
 & & [kpc] & & & [degs] & & [kpc] & [degs] \\ 
\hline                        

37194 & 15.61 $\pm$ 0.01 & 2.63 $\pm$ 0.05 & 3.71 $\pm$ 0.04 & 0.57 & 66.17 & 15.95 & 5.35 & 60.22 \\
30654 & 17.98 $\pm$ 0.01 & 1.42 $\pm$ 0.01 & 2.80 $\pm$ 0.02 & 0.47 & -28.96 & - & - & - \\
37587 & 16.87 $\pm$ 0.03 & 9.30 $\pm$ 0.41 & 5.12 $\pm$ 0.10 & 0.65 & 18.64 & - & - & - \\
19195 & 17.40 $\pm$ 0.05 & 8.00 $\pm$ 0.45 & 3.97 $\pm$ 0.09 & 0.83 & 57.98 & - & - & - \\
3740 & 17.48 $\pm$ 0.03 & 3.45 $\pm$ 0.14 & 3.40 $\pm$ 0.10 & 0.68 & 64.87 & - & - & - \\
20050 & 17.54 $\pm$ 0.03 & 3.30 $\pm$ 0.13 & 3.85 $\pm$ 0.12 & 0.57 & -54.07 & - & - & - \\
10876 & 17.43 $\pm$ 0.03 & 3.88 $\pm$ 0.15 & 3.19 $\pm$ 0.10 & 0.96 & -87.80 & - & - & - \\
21796 & 17.50 $\pm$ 0.03 & 9.58 $\pm$ 0.36 & 5.29 $\pm$ 0.17 & 0.88 & 6.85 & - & - & - \\
18935 & 17.79 $\pm$ 0.02 & 9.03 $\pm$ 0.19 & 2.17 $\pm$ 0.02 & 0.40 & 86.65 & - & - & - \\
7013 & 18.17 $\pm$ 0.01 & 1.91 $\pm$ 0.09 & 2.73 $\pm$ 0.03 & 0.50 & -23.05 & - & - & - \\
21604 & 17.57 $\pm$ 0.07 & 8.40 $\pm$ 0.47 & 2.38 $\pm$ 0.09 & 0.63 & 59.67 & - & - & - \\
25781 & 17.77 $\pm$ 0.06 & 7.05 $\pm$ 0.58 & 3.99 $\pm$ 0.23 & 0.81 & -46.99 & - & - & - \\
1996 & 18.03 $\pm$ 0.08 & 4.70 $\pm$ 0.51 & 4.73 $\pm$ 0.30 & 0.71 & 75.63 & - & - & - \\
21306 & 18.66 $\pm$ 0.03 & 3.73 $\pm$ 0.14 & 3.67 $\pm$ 0.17 & 0.65 & 12.50 & - & - & - \\
23956 & 16.44 $\pm$ 0.02 & 6.28 $\pm$ 0.22 & 3.55 $\pm$ 0.08 & 0.41 & 47.39 & - & - & - \\
13942 & 18.41 $\pm$ 0.28 & 8.84 $\pm$ 0.93 & 1.47 $\pm$ 0.23 & 0.57 & -59.77 & 20.30 & 0.84 & 78.48 \\
4735 & 18.01 $\pm$ 0.03 & 3.79 $\pm$ 0.15 & 3.42 $\pm$ 0.16 & 0.59 & 55.30 & - & - & - \\

\hline                                   
\end{tabular}}
\end{table*}

\begin{table*}[h]
\caption{Structural parameters for single Sérsic fits in the ACS $I$-band. Columns description: (1) galaxy ID, (2) magnitude, (3) effective radius, (4) Sérsic index, (5) axis ratio, (6) position angle, (7) magnitude of the bar component, (8) effective radius of the bar component, (9) position angle of the bar component.} 
\label{table:singleiband}
\centering   
\resizebox{0.9\textwidth}{!}{
\begin{tabular}{c c c c c c c c c}
\hline
\multirow{ 2}{*}{galaxy ID}	&	\multirow{ 2}{*}{mag	$\pm$	$\Delta$mag}	&	r$_\text{e}$	$\pm$	$\Delta$r$_\text{e}$	&	\multirow{ 2}{*}{n	$\pm$	$\Delta$n}	&	\multirow{ 2}{*}{q}	&	pa	&	 \multirow{ 2}{*}{mag$_\text{bar}$} 	&	 r$_\text{e,bar}$	&	 pa$_\text{bar}$\\
 & & [kpc] & & & [degs] & & [kpc] & [degs] \\
\hline

37194 & 15.33 $\pm$ 0.03 & 9.26 $\pm$ 0.56 & 5.00 $\pm$ 0.10 & 0.65 & 63.89 & 19.22 & 4.63 & 69.33 \\
30654 & 18.87 $\pm$ 0.01 & 1.54 $\pm$ 0.02 & 3.20 $\pm$ 0.26 & 0.49 & -27.65 & - & - & - \\
37587 & 17.97 $\pm$ 0.02 & 9.13 $\pm$ 0.25 & 4.45 $\pm$ 0.05 & 0.58 & 15.81 & - & - & - \\
19195 & 18.50 $\pm$ 0.02 & 6.82 $\pm$ 0.19 & 3.37 $\pm$ 0.04 & 0.85 & 48.89 & - & - & - \\
3740 & 18.31 $\pm$ 0.01 & 4.71 $\pm$ 0.07 & 4.04 $\pm$ 0.03 & 0.67 & 64.78 & - & - & - \\
20050 & 18.52 $\pm$ 0.02 & 6.20 $\pm$ 0.18 & 4.47 $\pm$ 0.04 & 0.62 & -47.46 & - & - & - \\
10876 & 18.38 $\pm$ 0.04 & 4.52 $\pm$ 0.43 & 3.38 $\pm$ 0.26 & 0.97 & -30.69 & - & - & - \\
21796 & 19.06 $\pm$ 0.03 & 4.95 $\pm$ 0.20 & 4.20 $\pm$ 0.08 & 0.80 & -0.79 & - & - & - \\
18935 & 18.74 $\pm$ 0.06 & 12.06 $\pm$ 0.89 & 2.35 $\pm$ 0.09 & 0.46 & 84.31 & - & - & - \\
7013 & 19.40 $\pm$ 0.01 & 2.34 $\pm$ 0.04 & 2.71 $\pm$ 0.02 & 0.49 & -20.02 & - & - & - \\
25781 & 18.63 $\pm$ 0.07 & 11.93 $\pm$ 1.27 & 4.97 $\pm$ 0.20 & 0.77 & -37.28 & - & - & - \\
1996 & 19.11 $\pm$ 0.08 & 5.59 $\pm$ 0.53 & 5.01 $\pm$ 0.20 & 0.76 & 78.80 & - & - & - \\
21306 & 19.79 $\pm$ 0.21 & 3.81 $\pm$ 0.67 & 3.78 $\pm$ 0.47 & 0.62 & 9.31 & - & - & - \\
23956 & 17.19 $\pm$ 0.09 & 8.33 $\pm$ 0.92 & 4.20 $\pm$ 0.26 & 0.45 & 46.33 & - & - & - \\
13942 & 19.02 $\pm$ 0.05 & 13.27 $\pm$ 1.11 & 5.27 $\pm$ 0.16 & 0.75 & -79.53 & 22.82 & 4.40 & -50.36 \\
4735 & 19.02 $\pm$ 0.03 & 4.88 $\pm$ 0.20 & 4.15 $\pm$ 0.08 & 0.64 & 53.86 & - & - & - \\

\hline                          
\end{tabular}}
\end{table*}

\begin{table*}[h]
 \caption{Structural parameters for multi-component Sérsic fits in the WFC3 $H$-band. Columns description: (1) galaxy ID, (2) magnitude of the disk component, (3) effective radius of the disk component, (4) exponential Sérsic index, (5) axis ratio of the disk component, (6) position angle of the disk component, (7) magnitude of the bulge component, (8) effective radius of the bulge component, (9) Sérsic index of the bulge component, (10) axis ratio of the bulge component, (11) position angle of the bulge component, (12) magnitude of the bar component, (13) effective radius of the bar component, (14) position angle of the bar component.} 
 \label{table:doublehband} 
 \centering 
 \resizebox{0.9\textwidth}{!}{
  \begin{tabular}{c c c c c c c c c c c c c c} 
  \hline\hline
\multirow{ 2}{*}{galaxy ID}	&	\multirow{ 2}{*}{mag$_\text{disk}$}	&	r$_\text{e,disk}$ &	\multirow{ 2}{*}{n$_\text{disk}$}	&	\multirow{ 2}{*}{q$_\text{disk}$}	&	pa$_\text{disk}$	&	\multirow{ 2}{*}{mag$_\text{bulge}$}	&	r$_\text{e,bulge}$ &	\multirow{ 2}{*}{n$_\text{bulge}$}	&	\multirow{ 2}{*}{q$_\text{bulge}$}	&	pa$_\text{bulge}$	&	\multirow{ 2}{*}{mag$_\text{bar}$}	&	r$_\text{e,bar}$ &	pa$_\text{bar}$	\\
 & & [kpc] & & & [degs] & & [kpc] & & & [degs] & & [kpc] & [degs]\\
\hline

37194 & 15.17 & 5.42 & 1 & 0.54 & 64.95 & 18.10 & 0.23 & 1.21 & 0.70 & 66.25 & 17.48 & 0.82 & 64.39 \\
30654 & 18.51 & 2.62 & 1 & 0.30 & -25.64 & 19.16 & 0.39 & 0.94 & 0.79 & -51.10 & - & - & - \\
37587 & 17.81 & 8.20 & 1 & 0.50 & 16.19 & 18.36 & 1.46 & 2.58 & 0.72 & 21.95 & - & - & - \\
19195 & 19.72 & 1.01 & 1 & 0.77 & 39.61 & 17.57 & 9.49 & 2.26 & 0.83 & 67.56 & - & - & - \\
3740 & 19.60 & 0.65 & 1 & 0.70 & 79.64 & 17.58 & 4.82 & 2.35 & 0.66 & 60.08 & - & - & - \\
20050 & 19.74 & 0.74 & 1 & 0.28 & -52.47 & 17.53 & 5.50 & 3.79 & 0.60 & -51.87 & - & - & - \\
10876 & 20.11 & 0.57 & 1 & 0.78 & 54.67 & 17.61 & 4.01 & 2.07 & 0.98 & -67.44 & - & - & - \\
21796 & 18.27 & 9.90 & 1 & 0.93 & 63.25 & 18.88 & 1.64 & 2.30 & 0.82 & 1.39 & - & - & - \\
18935 & 18.05 & 8.60 & 1 & 0.35 & 86.25 & 20.64 & 0.92 & 1.06 & 0.74 & 84.02 & - & - & - \\
7013 & 19.20 & 6.52 & 1 & 0.22 & -20.12 & 18.70 & 0.99 & 1.67 & 0.64 & -26.39 & - & - & - \\
21604 & 17.96 & 8.03 & 1 & 0.67 & 73.91 & 19.45 & 2.86 & 2.55 & 0.39 & 46.43 & - & - & - \\
25781 & 20.31 & 0.66 & 1 & 0.86 & -49.91 & 18.14 & 5.94 & 1.77 & 0.80 & -44.76 & - & - & - \\
1996 & 20.32 & 0.46 & 1 & 0.85 & 82.31 & 18.20 & 5.82 & 3.00 & 0.71 & 70.76 & - & - & - \\
21306 & 19.71 & 1.06 & 1 & 0.67 & 9.04 & 19.35 & 6.26 & 0.61 & 0.71 & 23.99 & - & - & - \\
23956 & 16.94 & 6.71 & 1 & 0.31 & 46.11 & 18.32 & 0.76 & 1.37 & 0.78 & 70.39 & - & - & - \\
13942 & 18.65 & 10.29 & 1 & 0.77 & -79.67 & 20.16 & 6.35 & 0.35 & 0.27 & -53.23 & 20.05 & 0.93 & 83.19 \\
4735 & 20.42 & 0.64 & 1 & 0.53 & 46.22 & 18.12 & 4.77 & 2.66 & 0.57 & 55.32 & - & - & - \\

  \hline
  \end{tabular}
 }
\end{table*}

\begin{table*}[h]
\caption{Structural parameters for multi-component Sérsic fits in the ACS $I$-band. Columns description: (1) galaxy ID, (2) magnitude of the disk component, (3) effective radius of the disk component, (4) exponential Sérsic index, (5) axis ratio of the disk component, (6) position angle of the disk component, (7) magnitude of the bulge component, (8) effective radius of the bulge component, (9) Sérsic index of the bulge component, (10) axis ratio of the bulge component, (11) position angle of the bulge component, (12) magnitude of the bar component, (13) effective radius of the bar component, (14) position angle of the bar component} 
\label{table:doubleiband} 
\centering 
\resizebox{0.9\textwidth}{!}{
\begin{tabular}{c c c c c c c c c c c c c c} 
\hline\hline
\multirow{ 2}{*}{galaxy ID}	&	\multirow{ 2}{*}{mag$_\text{disk}$}	&	r$_\text{e,disk}$ &	\multirow{ 2}{*}{n$_\text{disk}$}	&	\multirow{ 2}{*}{q$_\text{disk}$}	&	pa$_\text{disk}$	&	\multirow{ 2}{*}{mag$_\text{bulge}$}	&	r$_\text{e,bulge}$ &	\multirow{ 2}{*}{n$_\text{bulge}$}	&	\multirow{ 2}{*}{q$_\text{bulge}$}	&	pa$_\text{bulge}$	&	\multirow{ 2}{*}{mag$_\text{bar}$}	&	r$_\text{e,bar}$ &	pa$_\text{e,bar}$	\\
 & & [kpc] & & & [degs] & & [kpc] & & & [degs] & & [kpc] & [degs] \\
\hline

37194 & 15.84 & 5.80 & 1 & 0.55 & 64.18 & 18.89 & 0.26 & 1.57 & 0.74 & 64.22 & 18.35 & 0.81 & 64.92 \\
30654 & 19.50 & 2.86 & 1 & 0.27 & -25.31 & 20.01 & 0.42 & 1.55 & 0.71 & -34.45 & - & - & - \\
37587 & 18.62 & 7.83 & 1 & 0.51 & 16.62 & 19.89 & 1.03 & 1.45 & 0.61 & 15.29 & - & - & - \\
19195 & 20.73 & 1.06 & 1 & 0.80 & 28.58 & 18.55 & 9.75 & 2.12 & 0.83 & 66.62 & - & - & - \\
3740 & 20.98 & 0.51 & 1 & 0.54 & 77.33 & 18.48 & 4.83 & 2.29 & 0.67 & 56.20 & - & - & - \\
20050 & 22.48 & 0.50 & 1 & 0.22 & -53.12 & 18.54 & 6.53 & 3.99 & 0.63 & -47.79 & - & - & - \\
10876 & 22.43 & 0.42 & 1 & 0.62 & -73.94 & 18.49 & 4.14 & 2.67 & 0.96 & -13.19 & - & - & - \\
21796 & 19.26 & 11.82 & 1 & 0.93 & 46.48 & 19.65 & 2.62 & 3.61 & 0.77 & -0.94 & - & - & - \\
18935 & 19.09 & 9.59 & 1 & 0.40 & 84.67 & 22.02 & 0.95 & 1.24 & 0.70 & 78.89 & - & - & - \\
7013 & 20.37 & 8.24 & 1 & 0.22 & -21.88 & 19.93 & 1.27 & 1.84 & 0.56 & -18.78 & - & - & - \\
25781 & 21.84 & 0.56 & 1 & 0.71 & -30.33 & 19.08 & 7.12 & 2.17 & 0.78 & -40.23 & - & - & - \\
1996 & 21.88 & 0.53 & 1 & 0.58 & -75.26 & 19.16 & 7.10 & 3.78 & 0.72 & 67.15 & - & - & - \\
21306 & 21.50 & 0.70 & 1 & 0.69 & 4.05 & 20.10 & 5.44 & 1.76 & 0.58 & 13.63 & - & - & - \\
23956 & 18.24 & 7.79 & 1 & 0.22 & 45.69 & 18.26 & 2.91 & 4.47 & 0.71 & 48.32 & - & - & - \\
13942 & 19.66 & 10.88 & 1 & 0.77 & -77.98 & 21.37 & 0.89 & 0.95 & 0.68 & 70.48 & 21.30 & 6.20 & -51.69 \\
4735 & 21.61 & 0.44 & 1 & 0.91 & 60.45 & 19.28 & 4.86 & 2.32 & 0.59 & 53.99 & - & - & - \\

\hline
\end{tabular}
}
\end{table*}

\clearpage
\section{Surface Brightness Profiles}
\label{appen:sbs}

  \begin{figure*}[h]
  \centering
      \includegraphics[page=13,width=0.4\textwidth]{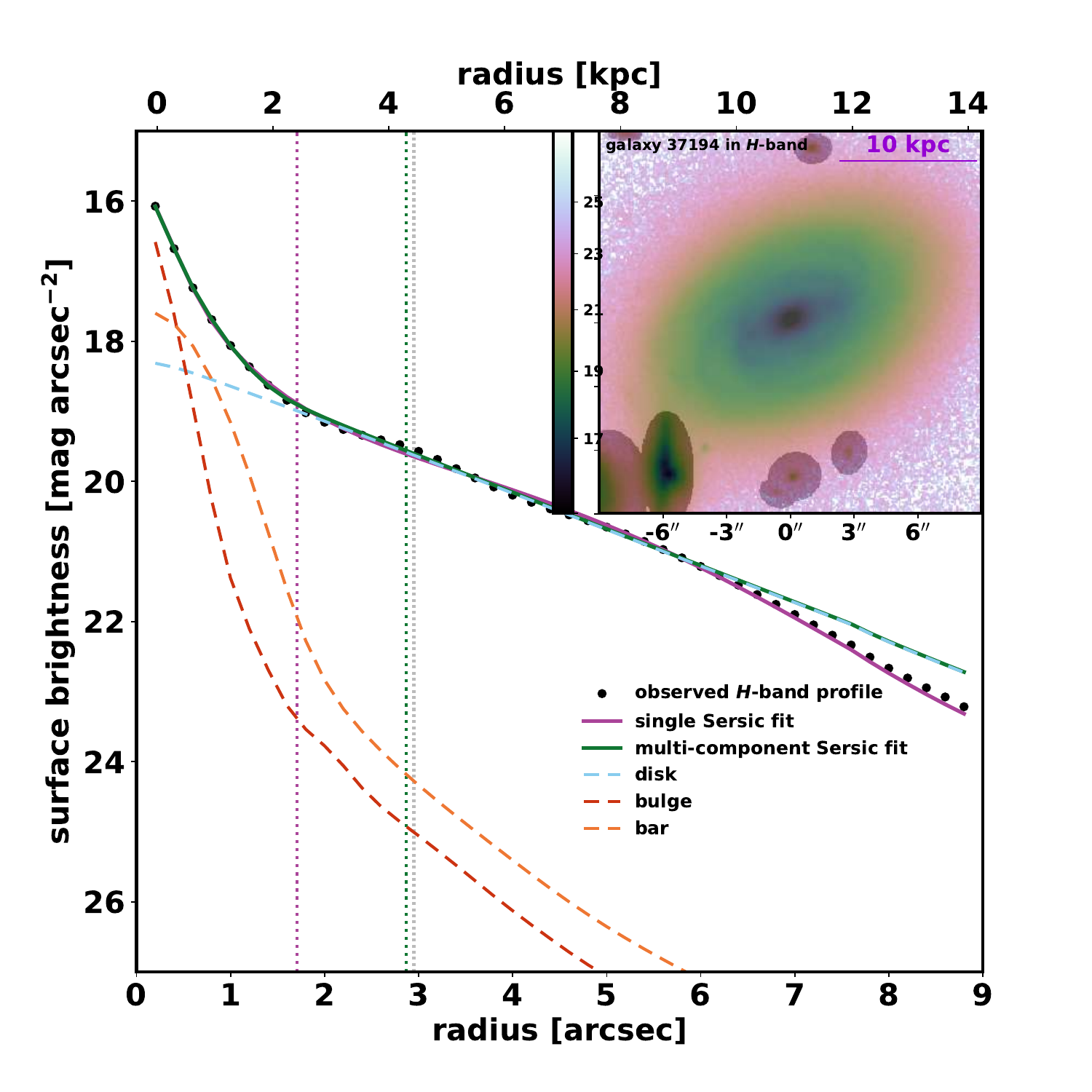}
  	\includegraphics[page=12,width=0.4\textwidth]{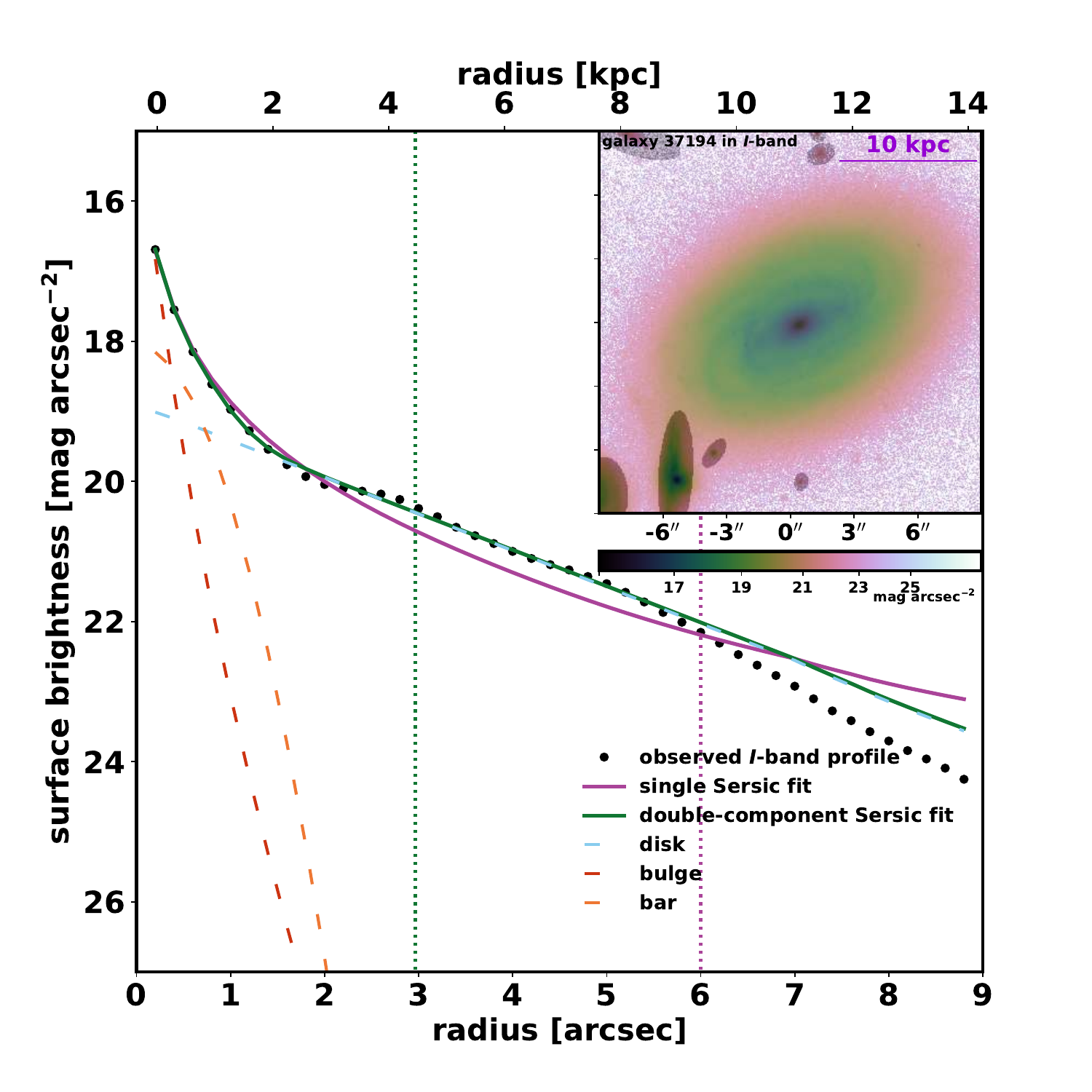}
   \caption{Surface brightness profiles of barred late-type galaxy 37194 in the $H$-band (left panel) and $I$-band (right panel).}
  \vspace{-10pt}
   \label{fig:13942_SB}
  \end{figure*}

\begin{figure*}[h]
 \centering
     \includegraphics[page=12,width=0.4\textwidth]{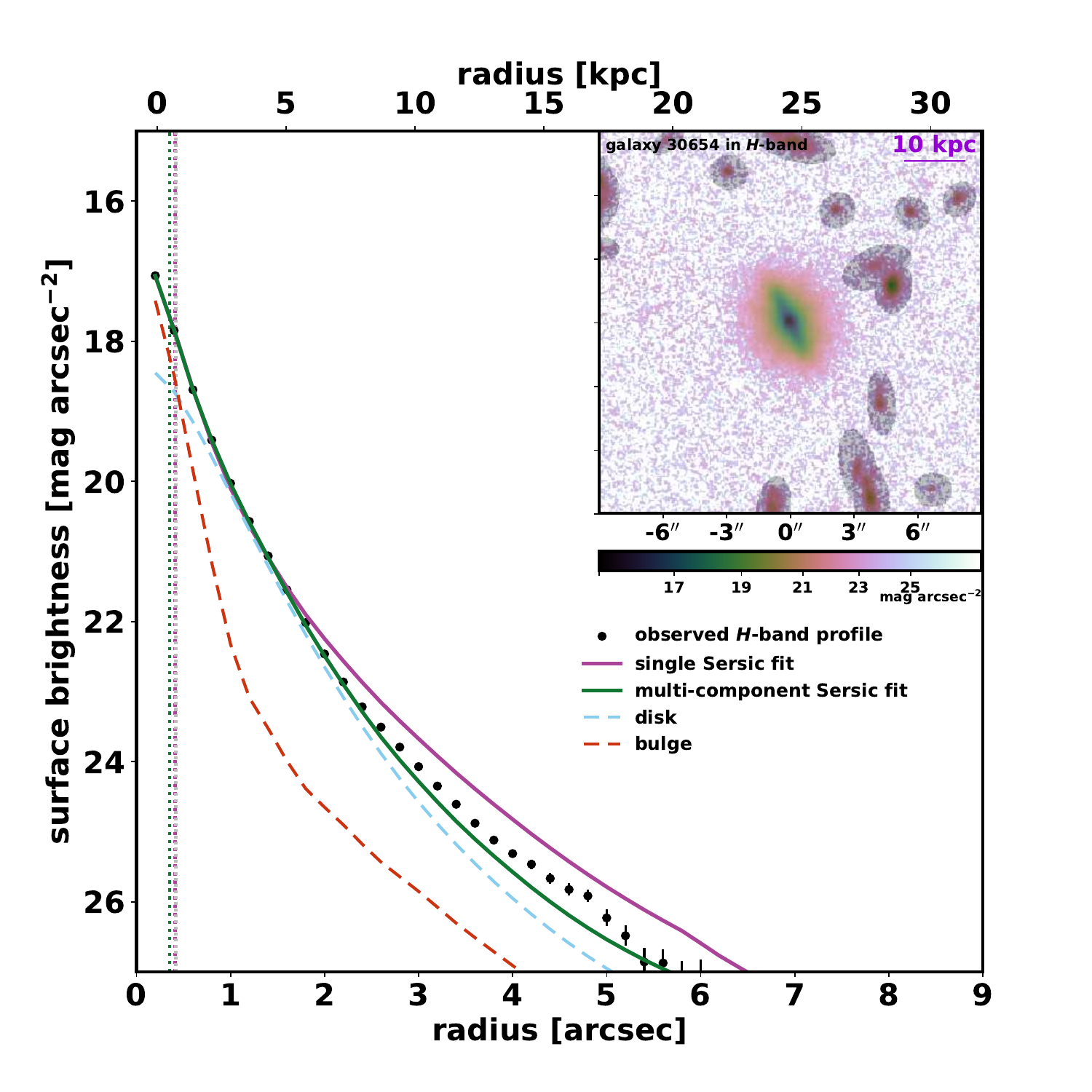}
 	\includegraphics[page=11,width=0.4\textwidth]{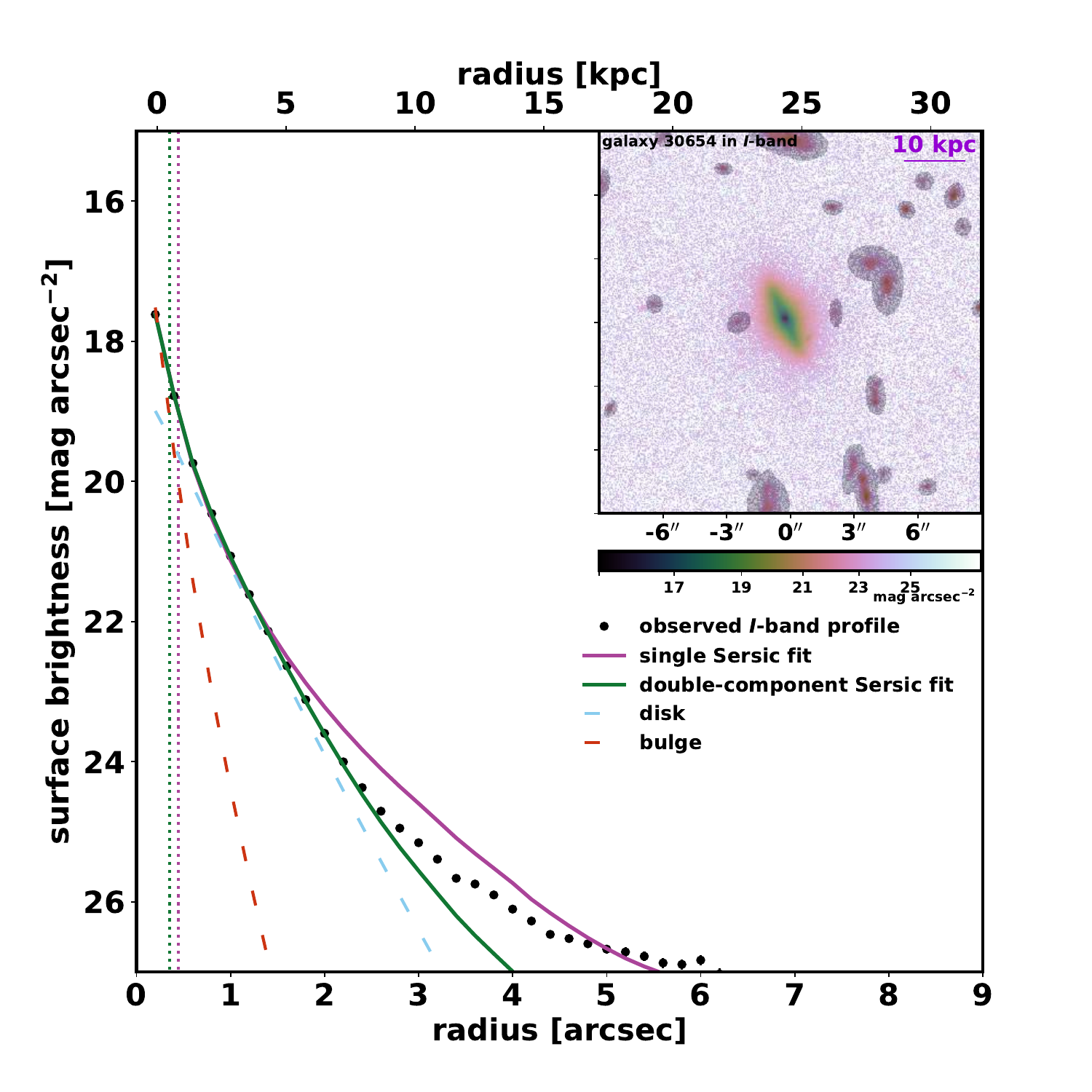}
  \caption{Surface brightness profiles of late-type galaxy 30654 in the $H$-band (left panel) and $I$-band (right panel).}
  \vspace{-10pt}
  \label{fig:30654_SB}
 \end{figure*}

\begin{figure*}[h]
\centering
    \includegraphics[page=15,width=0.4\textwidth]{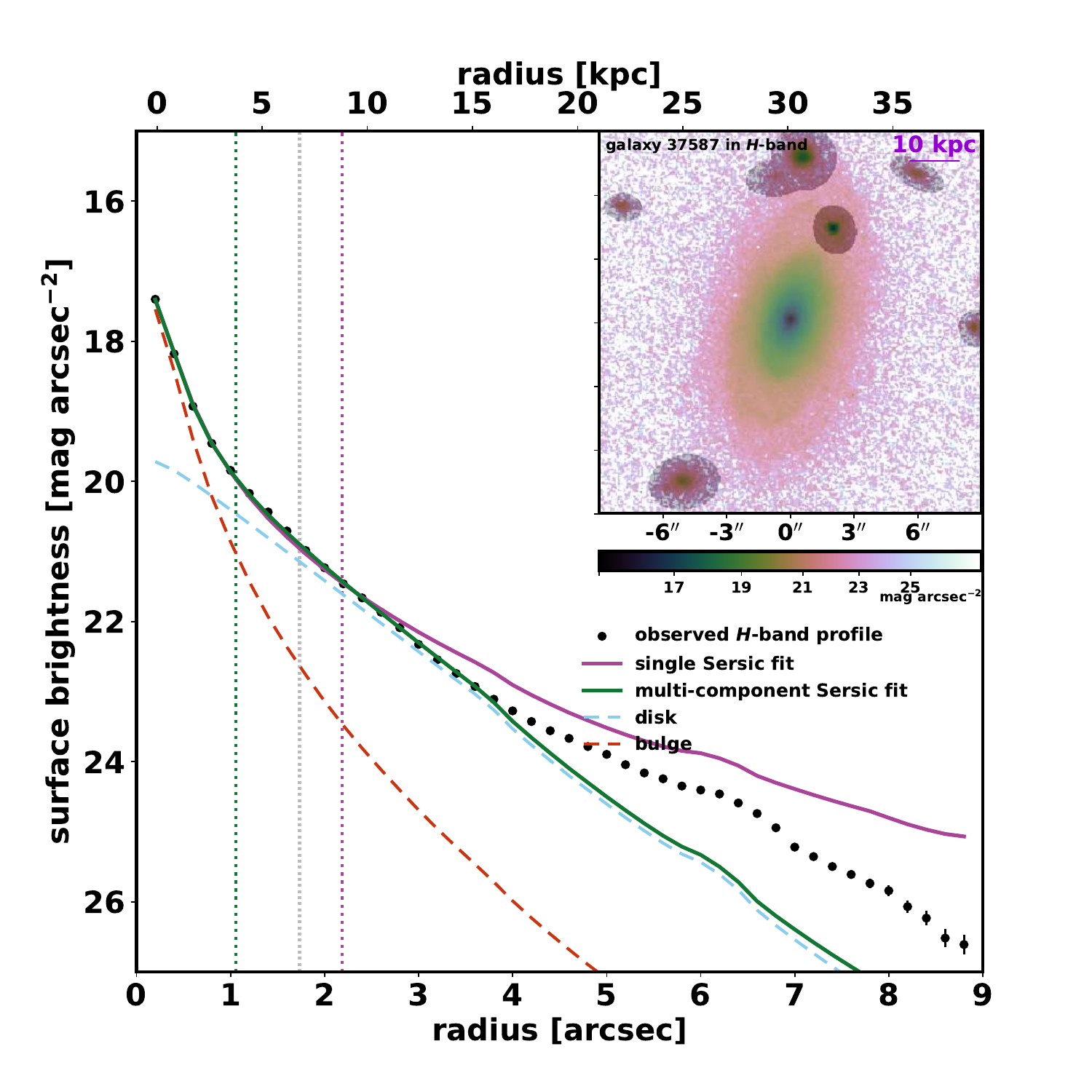}
	\includegraphics[page=14,width=0.4\textwidth]{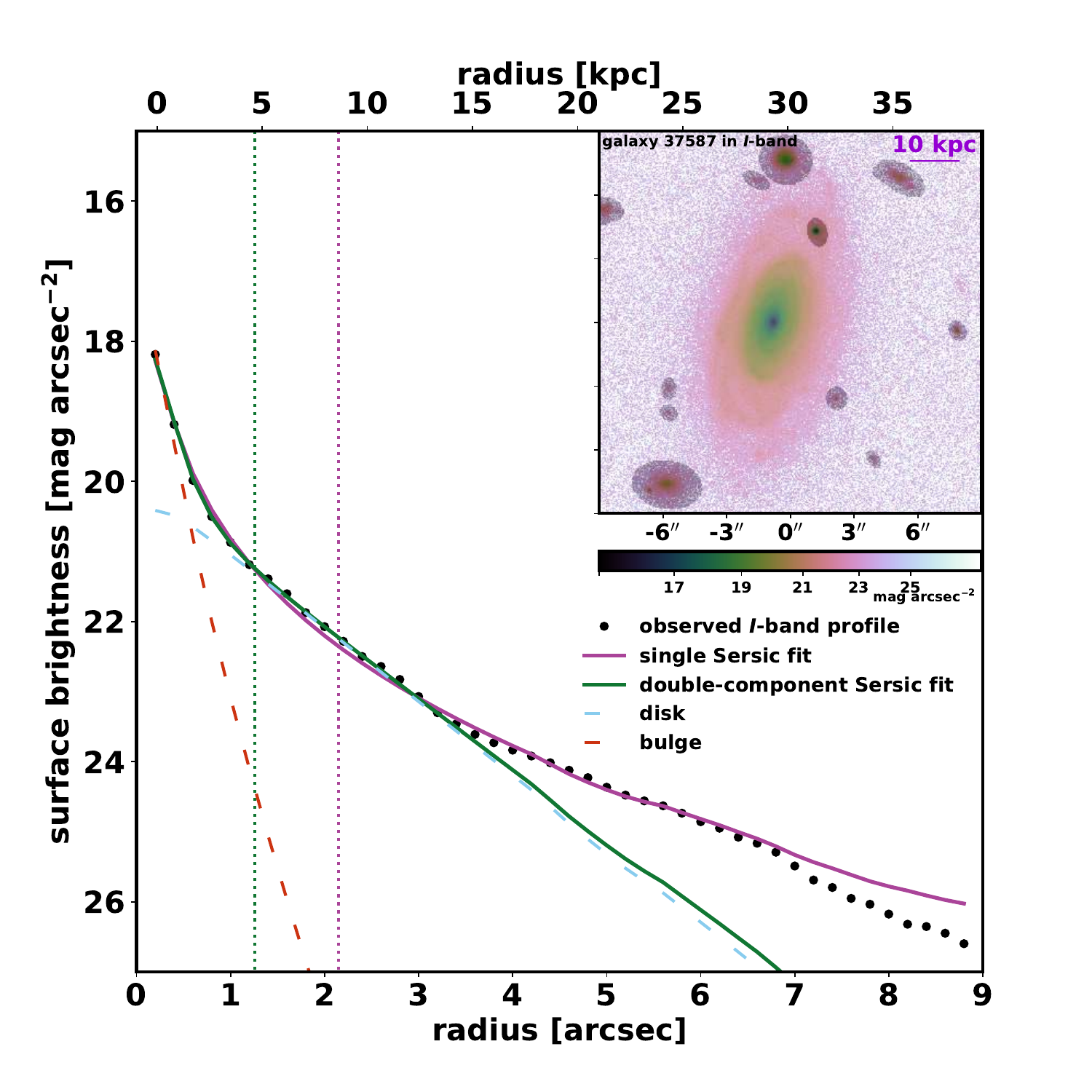}
 \caption{Surface brightness profiles of late-type galaxy \#37587 in the $H$-band (left panel) and $I$-band (right panel).}
 \vspace{-10pt}
 \label{fig:37587_SB}
\end{figure*}

\begin{figure*}[h]
 \centering
     \includegraphics[page=14,width=0.4\textwidth]{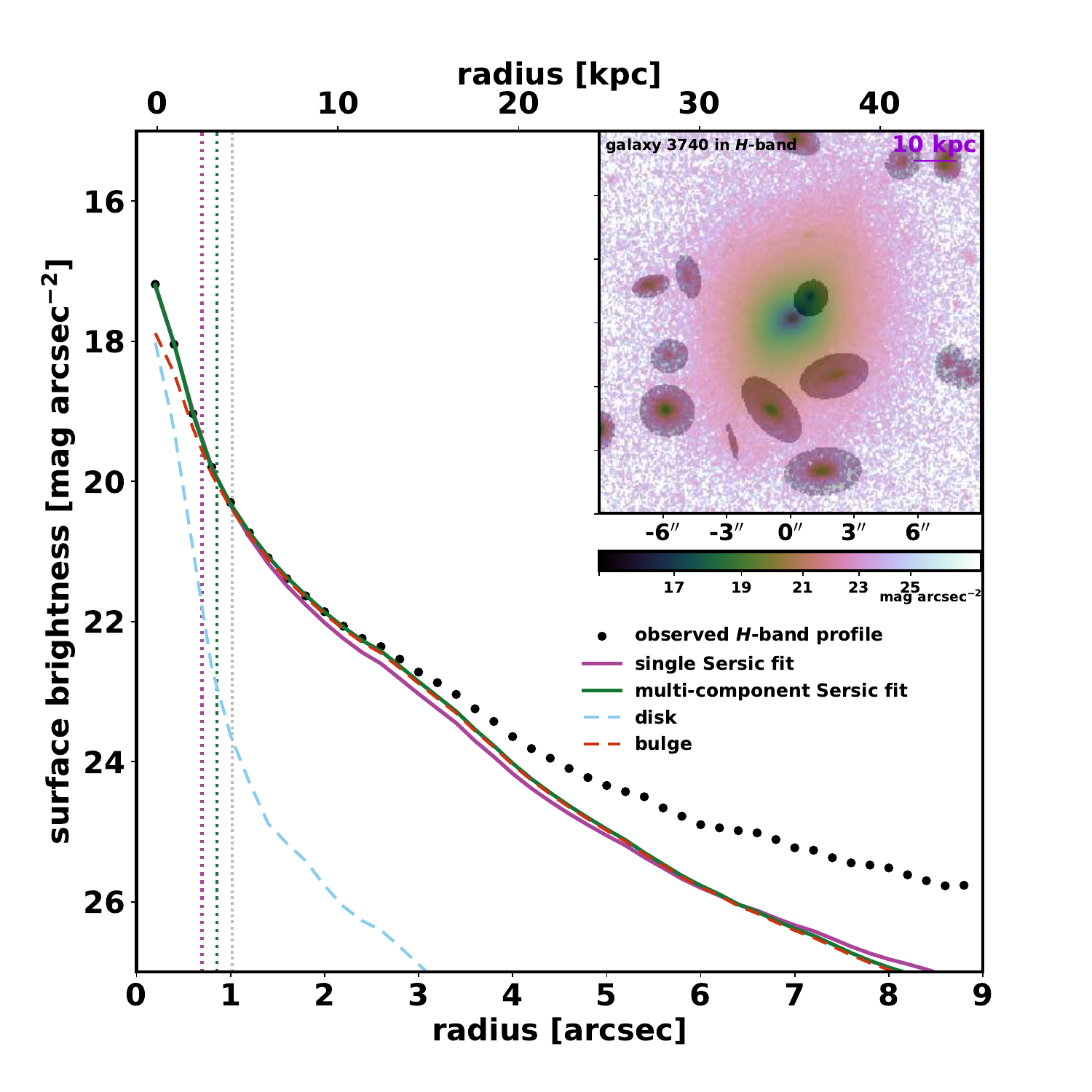}
 	\includegraphics[page=13,width=0.4\textwidth]{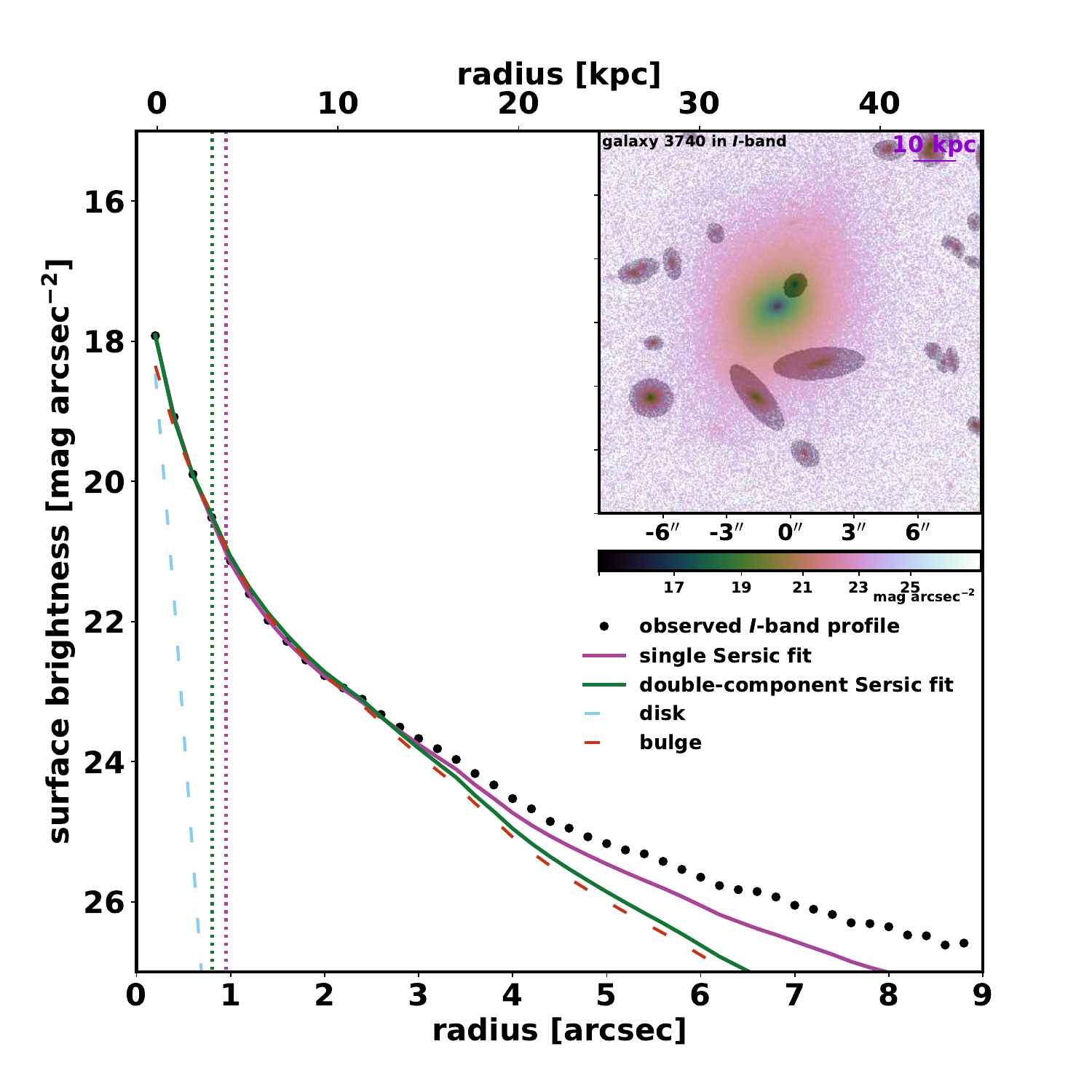}
  \caption{Surface brightness profiles of early-type galaxy \#3740 in the $H$-band (left panel) and $I$-band (right panel).}
  \vspace{-10pt}
  \label{fig:3740_SB}
 \end{figure*}

\begin{figure*}[h]
\centering
    \includegraphics[page=6,width=0.4\textwidth]{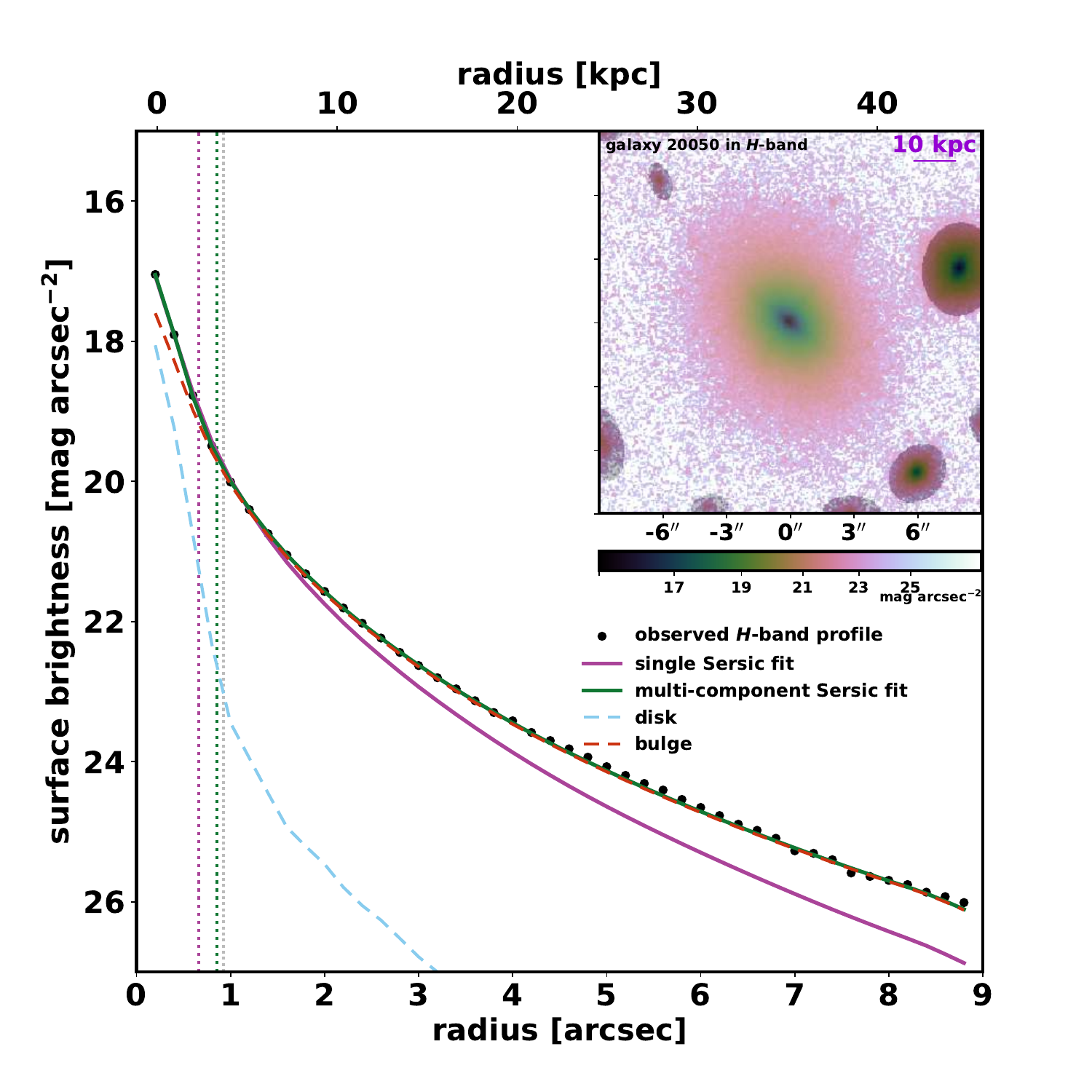}
	\includegraphics[page=6,width=0.4\textwidth]{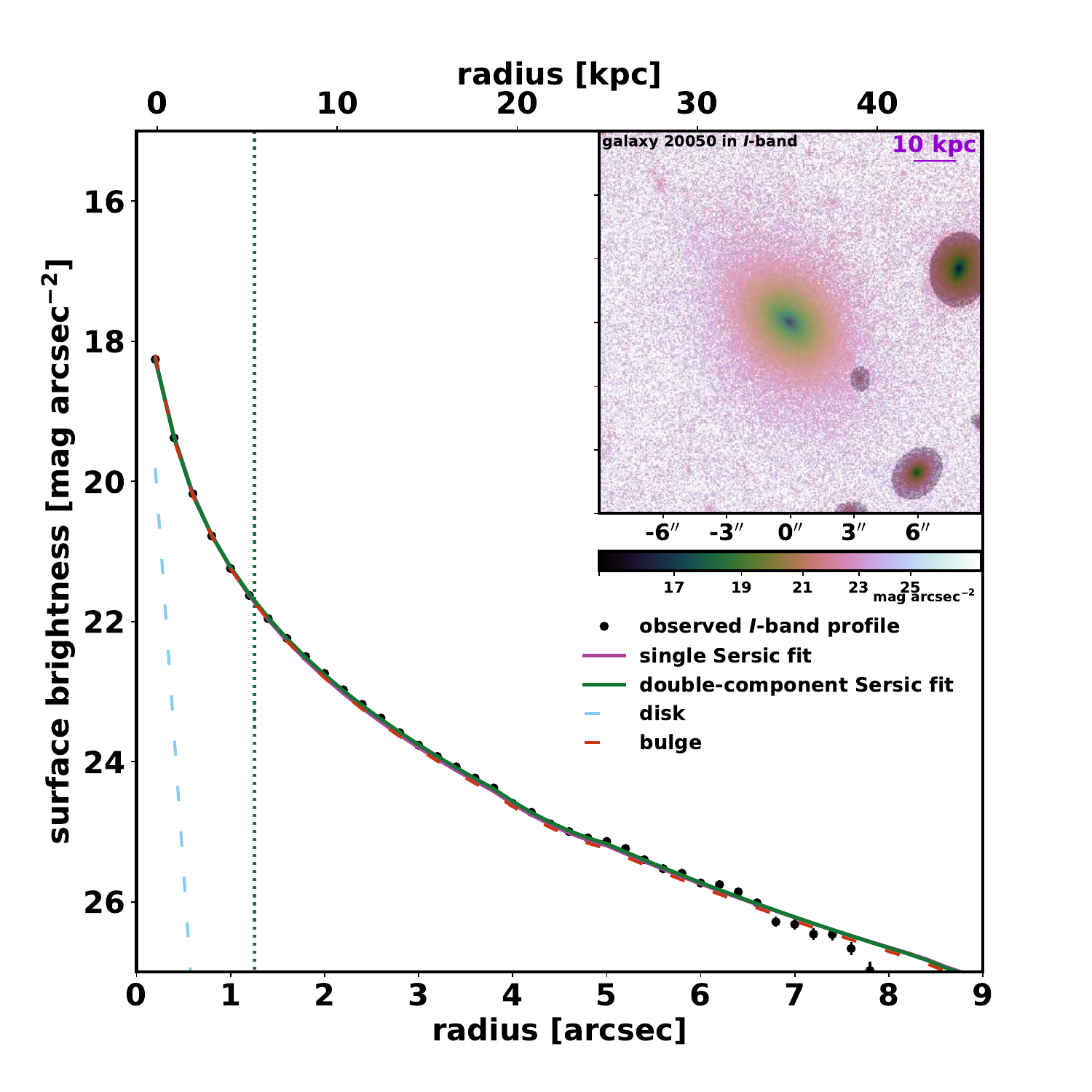}
 \caption{Surface brightness profiles of early-type galaxy \#20050 in the $H$-band (left panel) and $I$-band (right panel).}
 \vspace{-10pt}
 \label{fig:20050_SB}
\end{figure*}

\begin{figure*}[h]
\centering
    \includegraphics[page=1,width=0.4\textwidth]{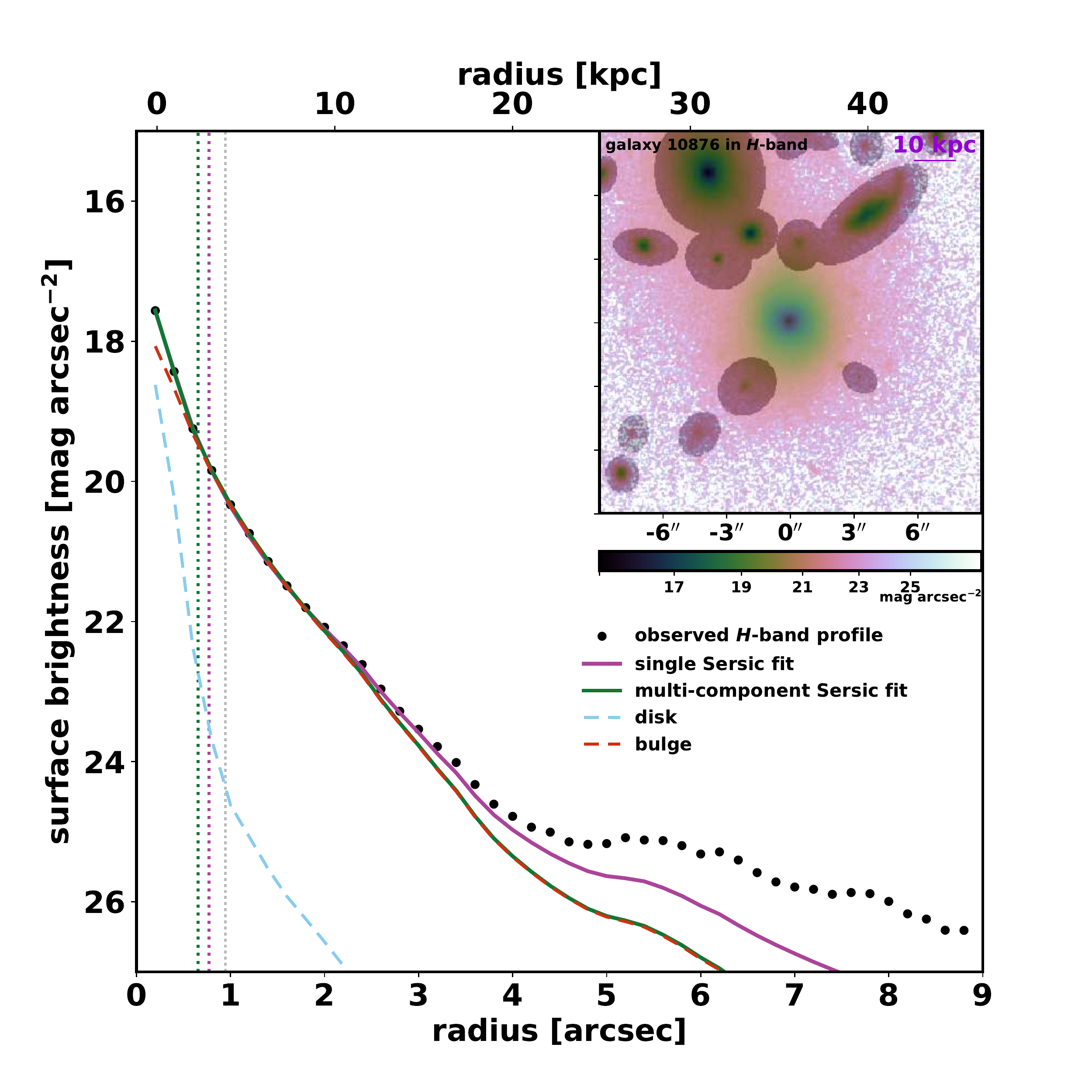}
	\includegraphics[page=1,width=0.4\textwidth]{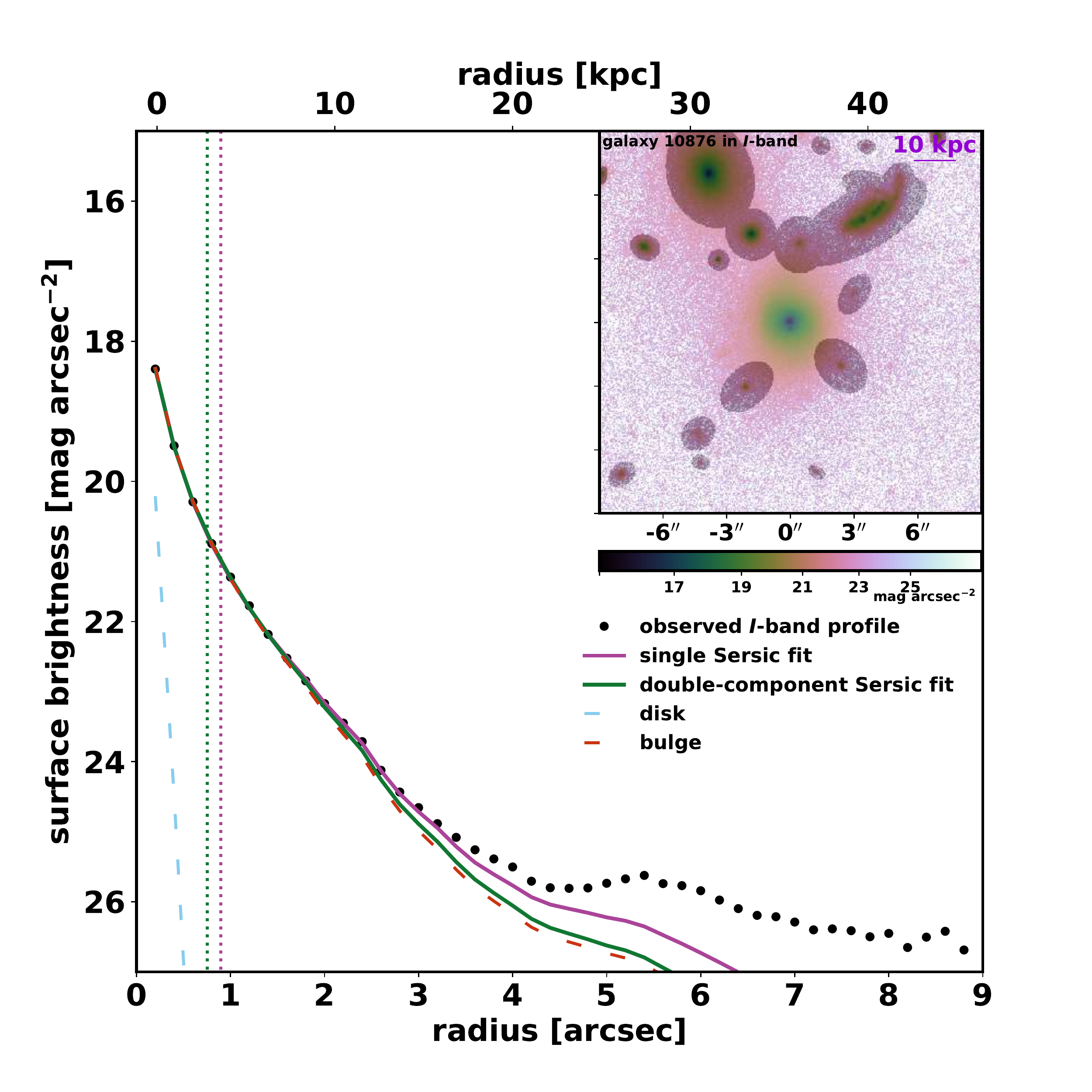}
 \caption{Surface brightness profiles of early-type galaxy \#10876 in the $H$-band (left panel) and $I$-band (right panel).}
 \vspace{-10pt}
 \label{fig:10876_SB}
\end{figure*}

\begin{figure*}[h]
\centering
    \includegraphics[page=3,width=0.4\textwidth]{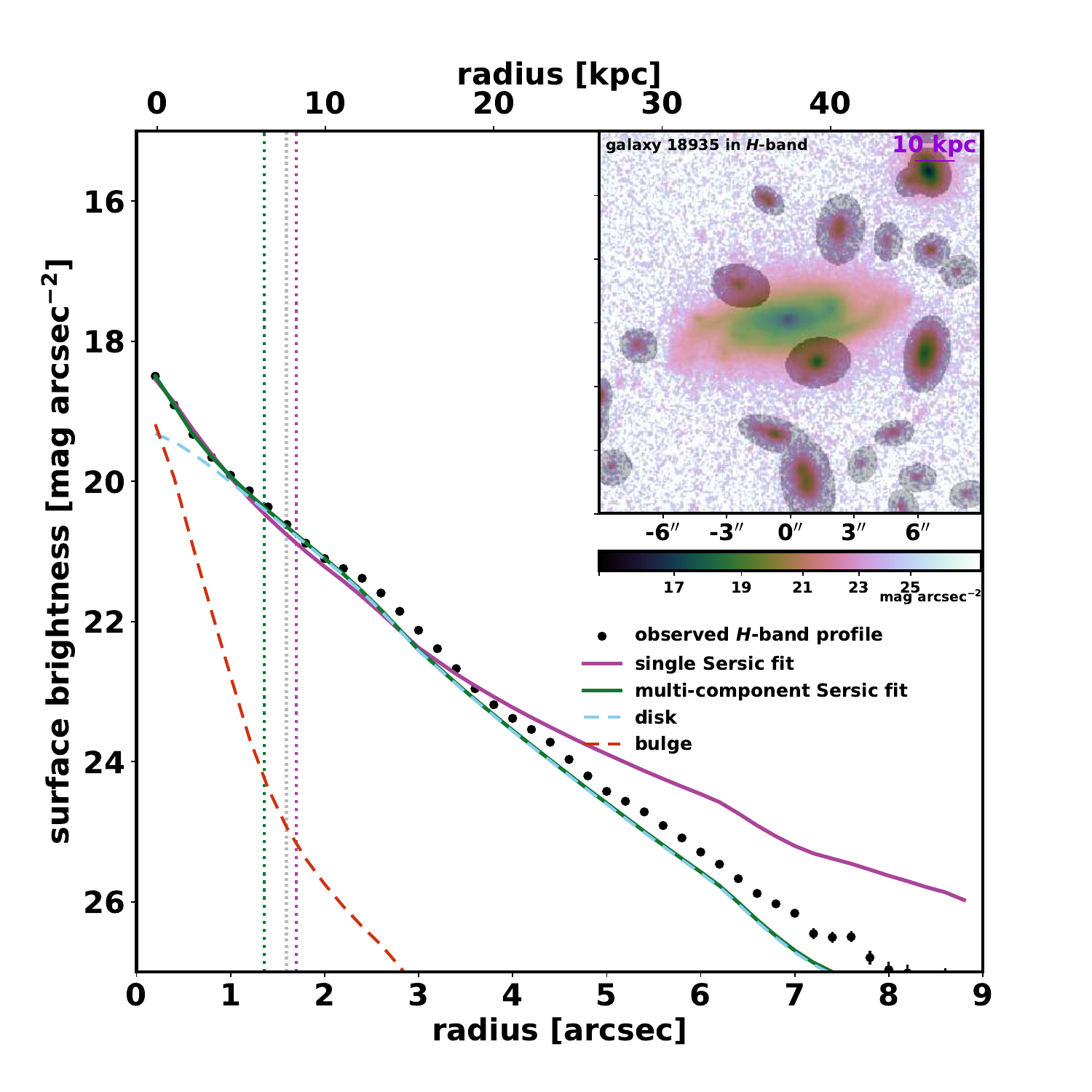}
	\includegraphics[page=3,width=0.4\textwidth]{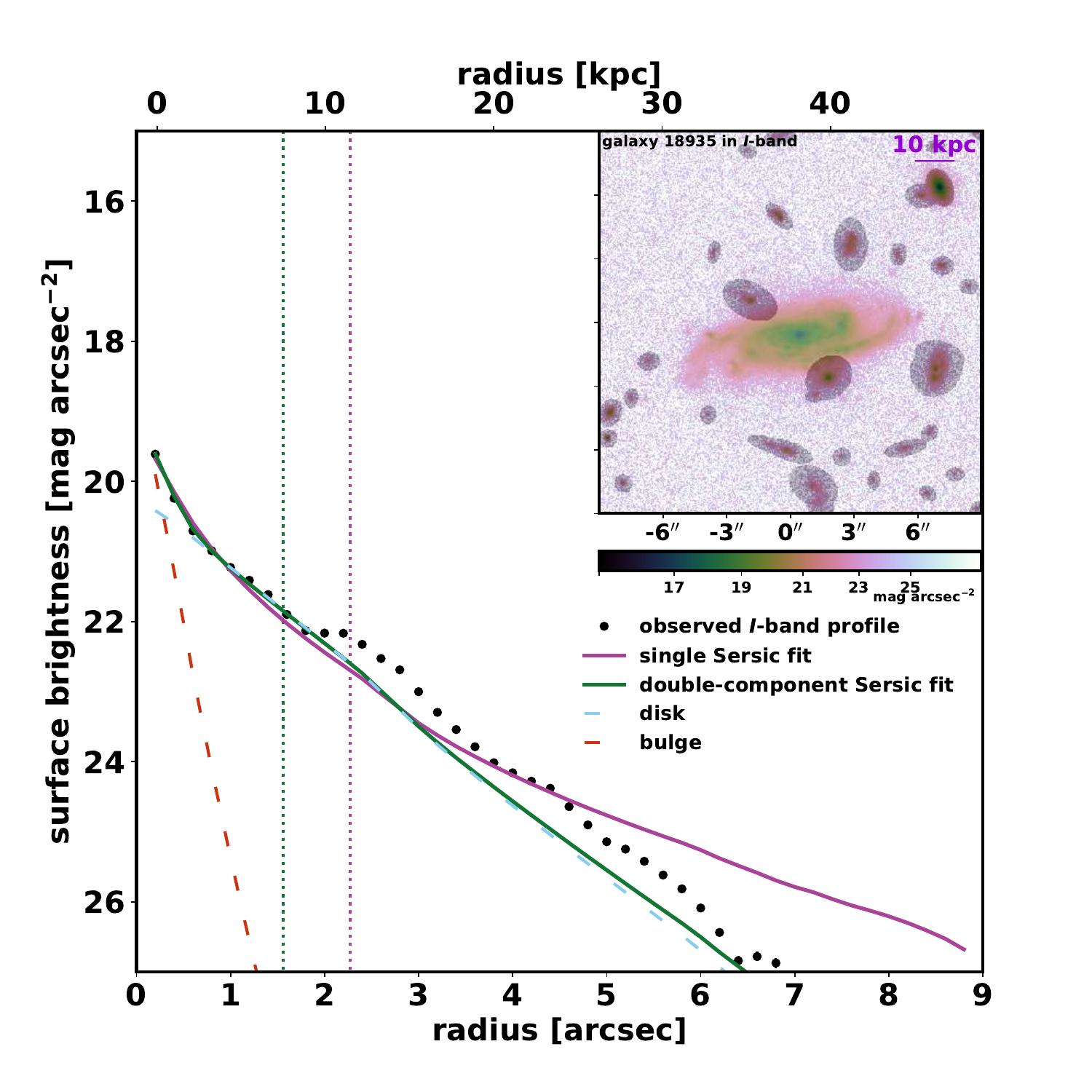}
 \caption{Surface brightness profiles of late-type galaxy \#18935 in the $H$-band (left panel) and $I$-band (right panel).}
 \vspace{-10pt}
 \label{fig:18935_SB}
\end{figure*}

\begin{figure*}[h]
\centering
    \includegraphics[page=8,width=0.45\textwidth]{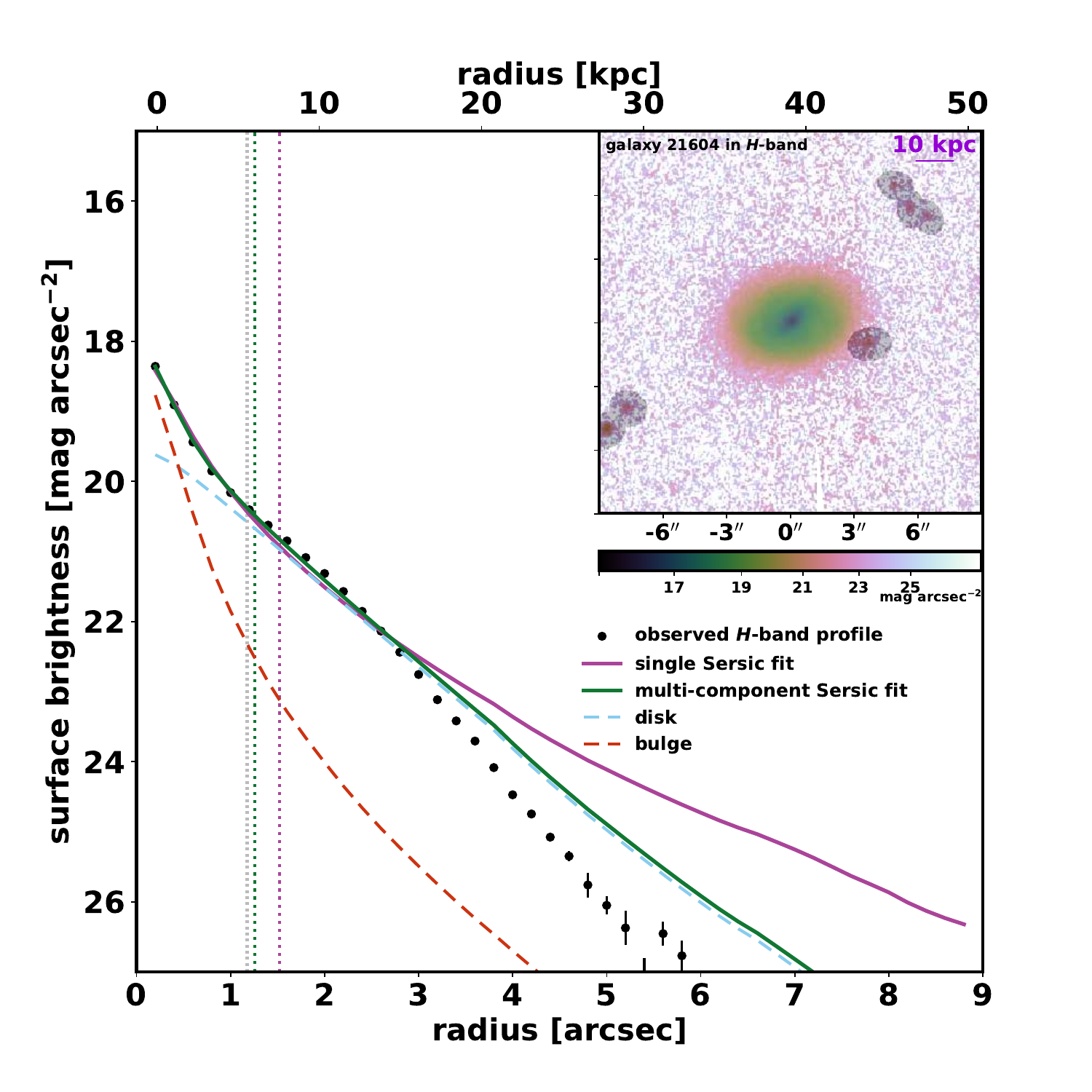}
 \caption{Surface brightness profiles of late-type galaxy \#21604 in the $H$-band.}
 \vspace{-10pt}
 \label{fig:21604_SB}
\end{figure*}

\begin{figure*}[h]
\centering
    \includegraphics[page=11,width=0.4\textwidth]{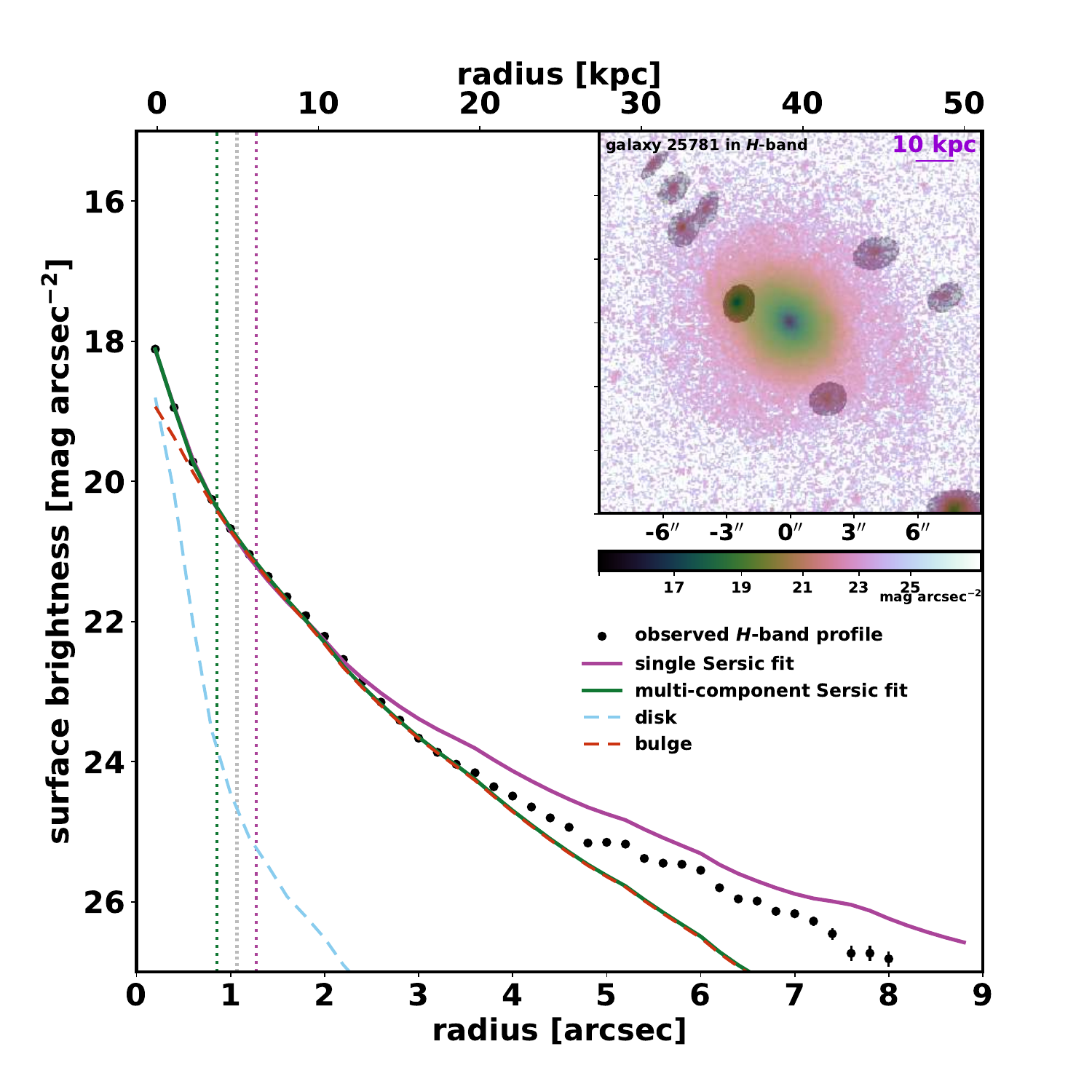}
	\includegraphics[page=10,width=0.4\textwidth]{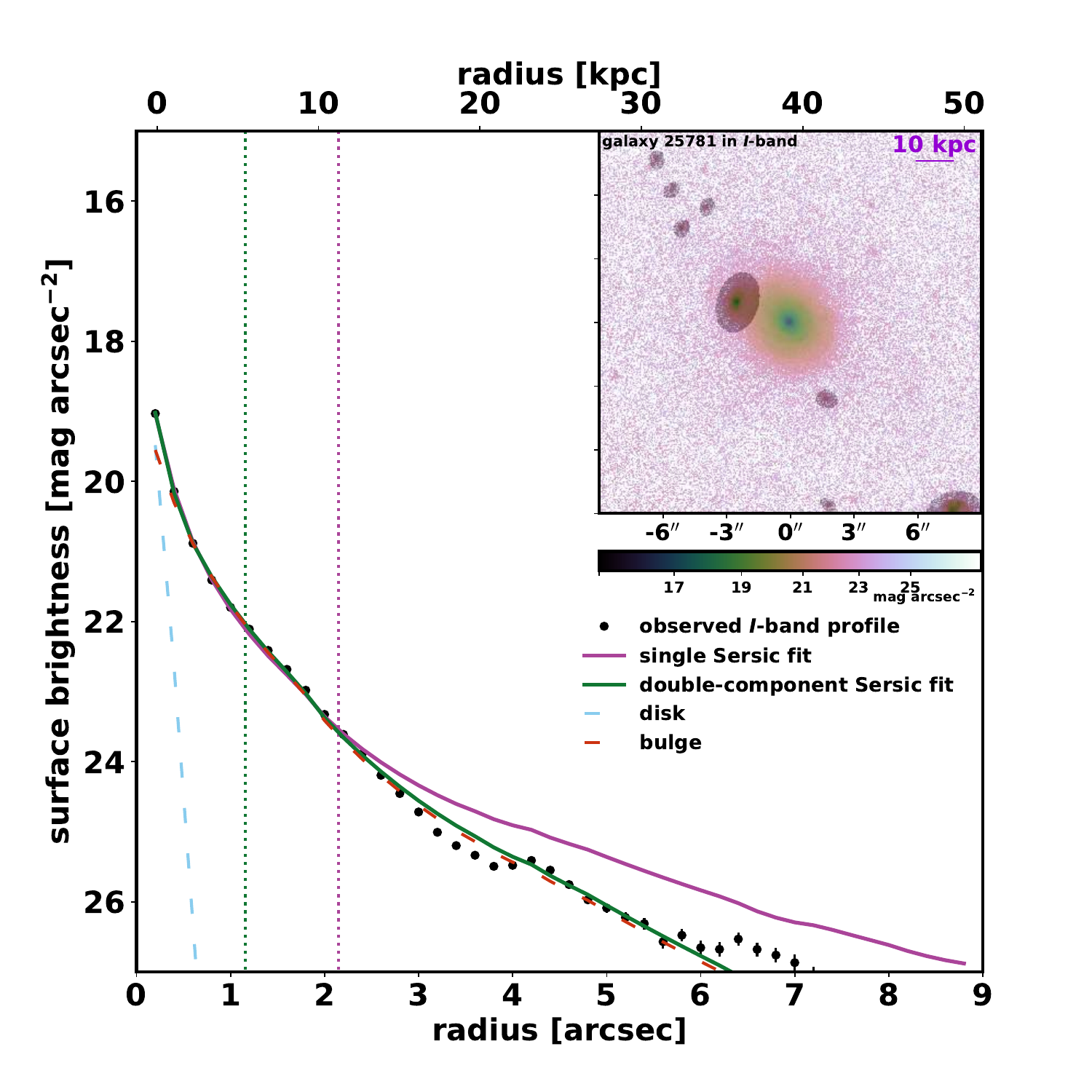}
 \caption{Surface brightness profiles of early-type galaxy \#25781 in the $H$-band (left panel) and $I$-band (right panel).}
 \vspace{-10pt}
 \label{fig:25781_SB}
\end{figure*}

\begin{figure*}[h]
\centering
    \includegraphics[page=5,width=0.4\textwidth]{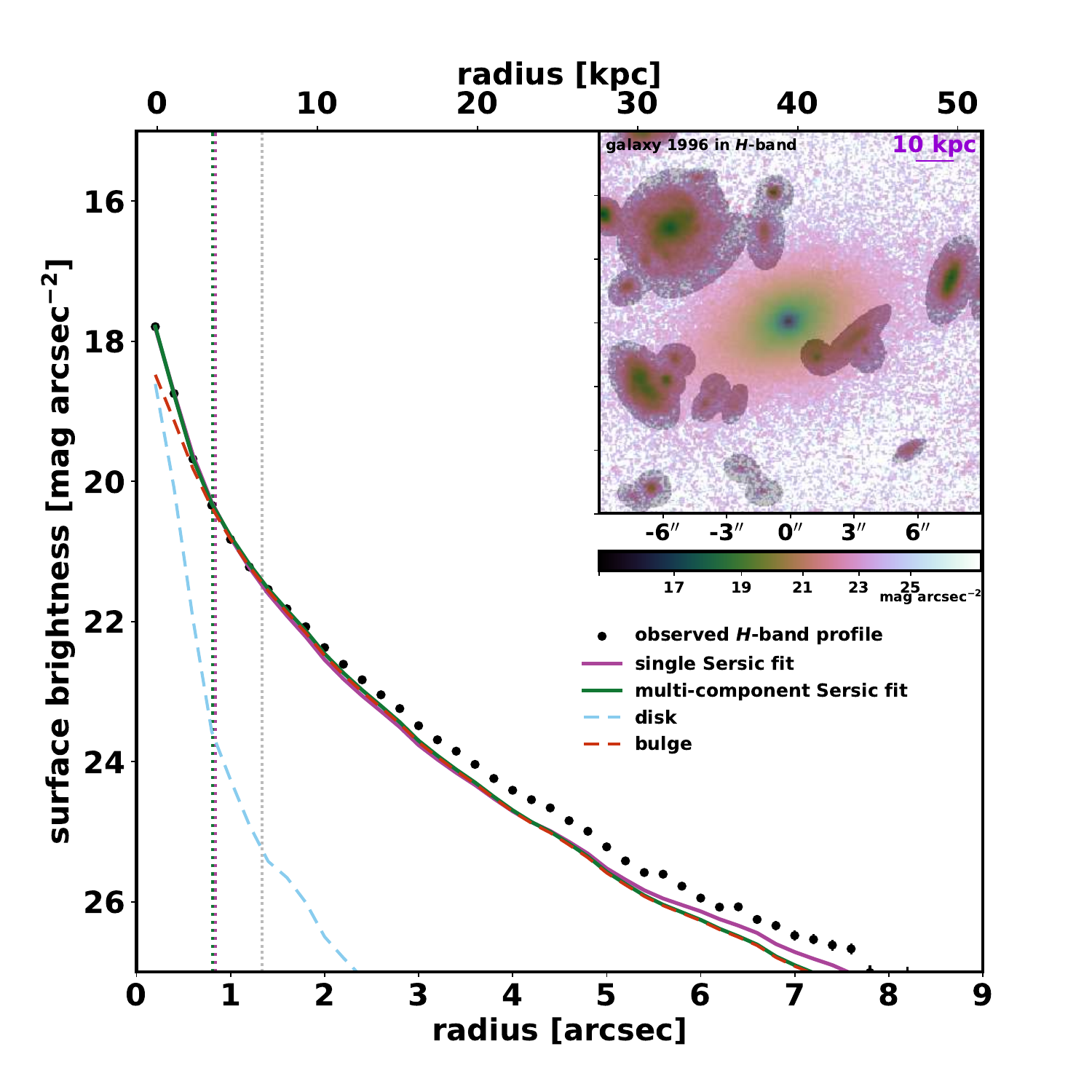}
	\includegraphics[page=5,width=0.4\textwidth]{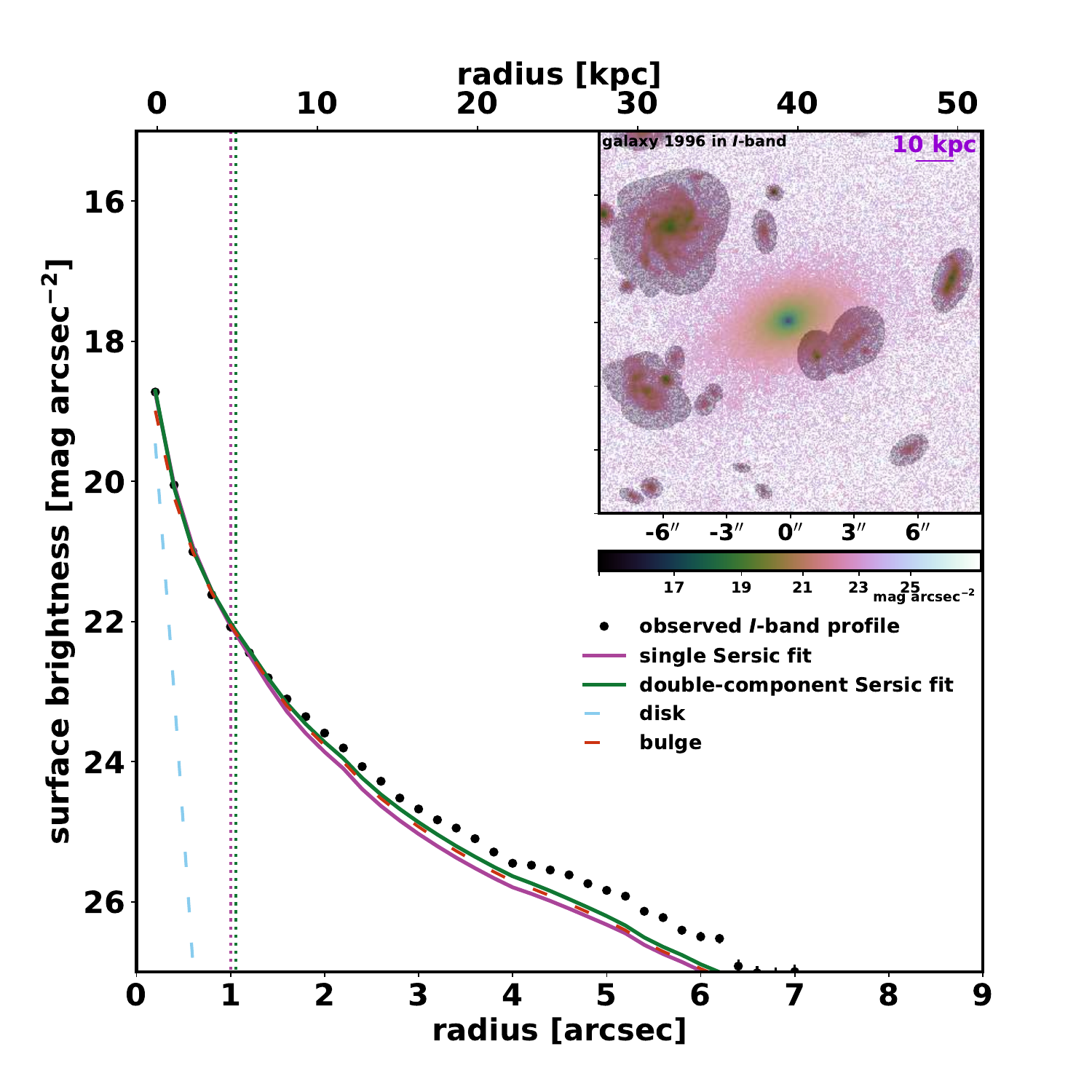}
 \caption{Surface brightness profiles of early-type galaxy \#1996 in the $H$-band (left panel) and $I$-band (right panel).}
 \vspace{-10pt}
 \label{fig:1996_SB}
\end{figure*}

\begin{figure*}[h]
\centering
    \includegraphics[page=7,width=0.4\textwidth]{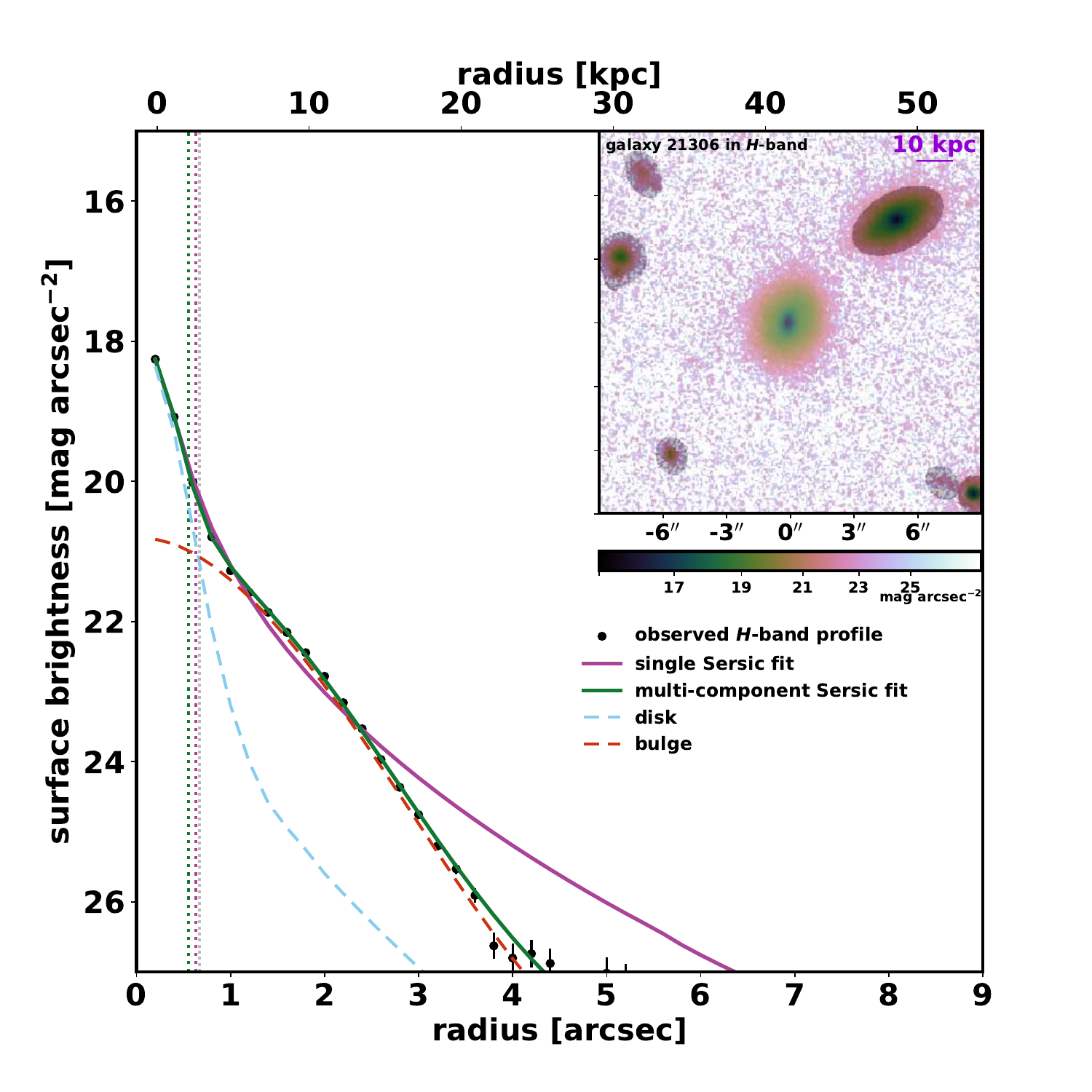}
	\includegraphics[page=7,width=0.4\textwidth]{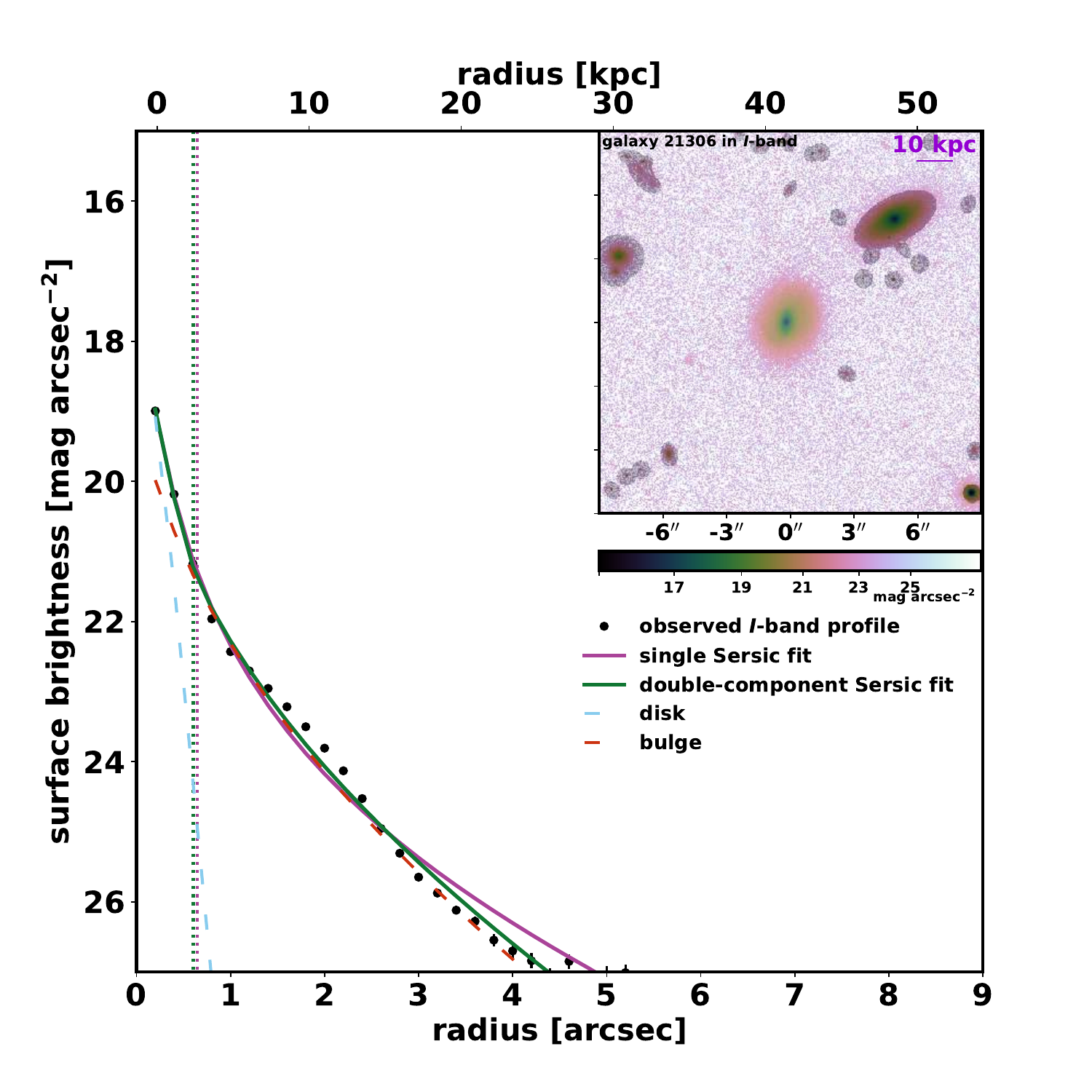}
 \caption{Surface brightness profiles of early-type galaxy \#21306 in the $H$-band (left panel) and $I$-band (right panel).}
 \vspace{-10pt}
 \label{fig:21306_SB}
\end{figure*}

\begin{figure*}[h]
\centering
    \includegraphics[page=16,width=0.4\textwidth]{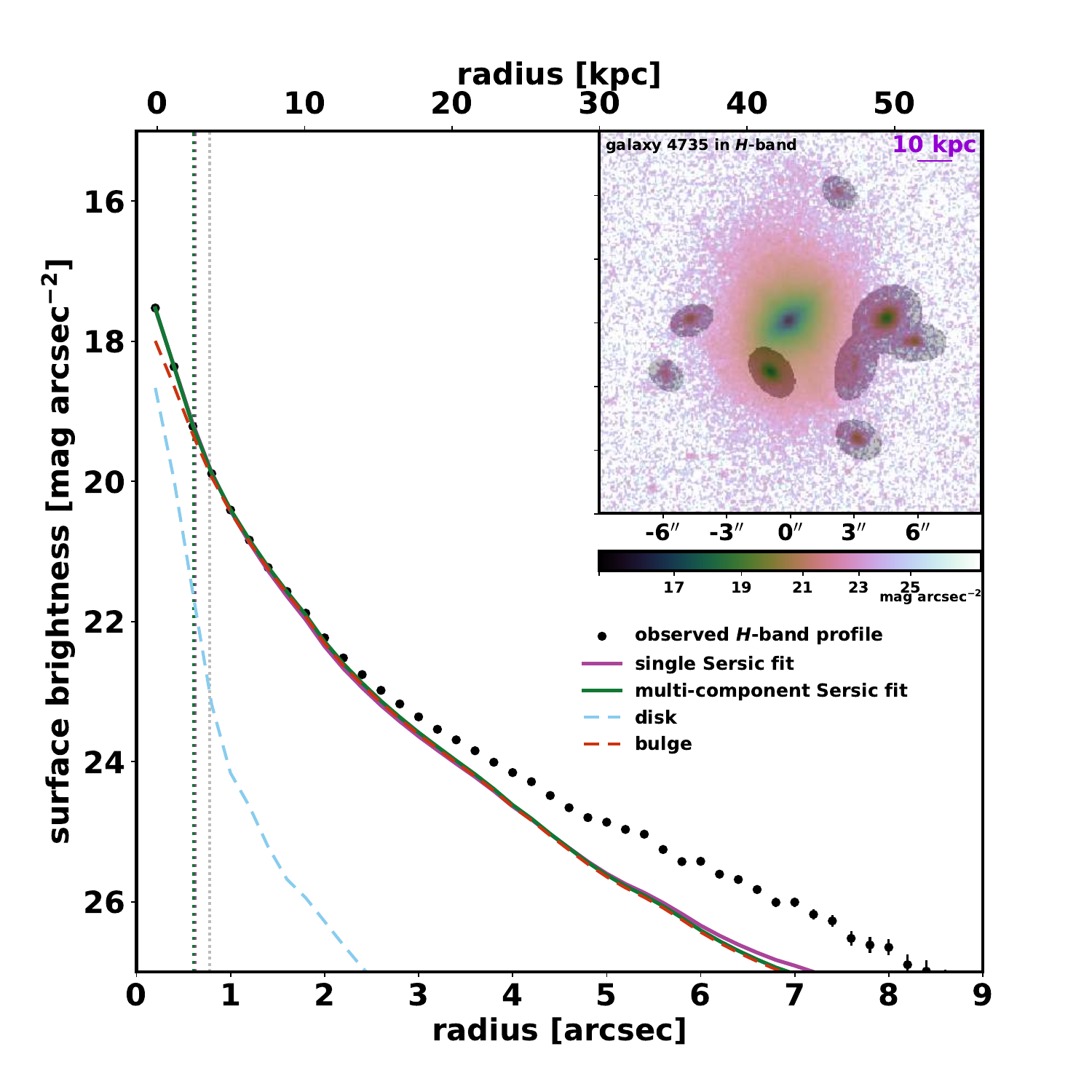}
	\includegraphics[page=15,width=0.4\textwidth]{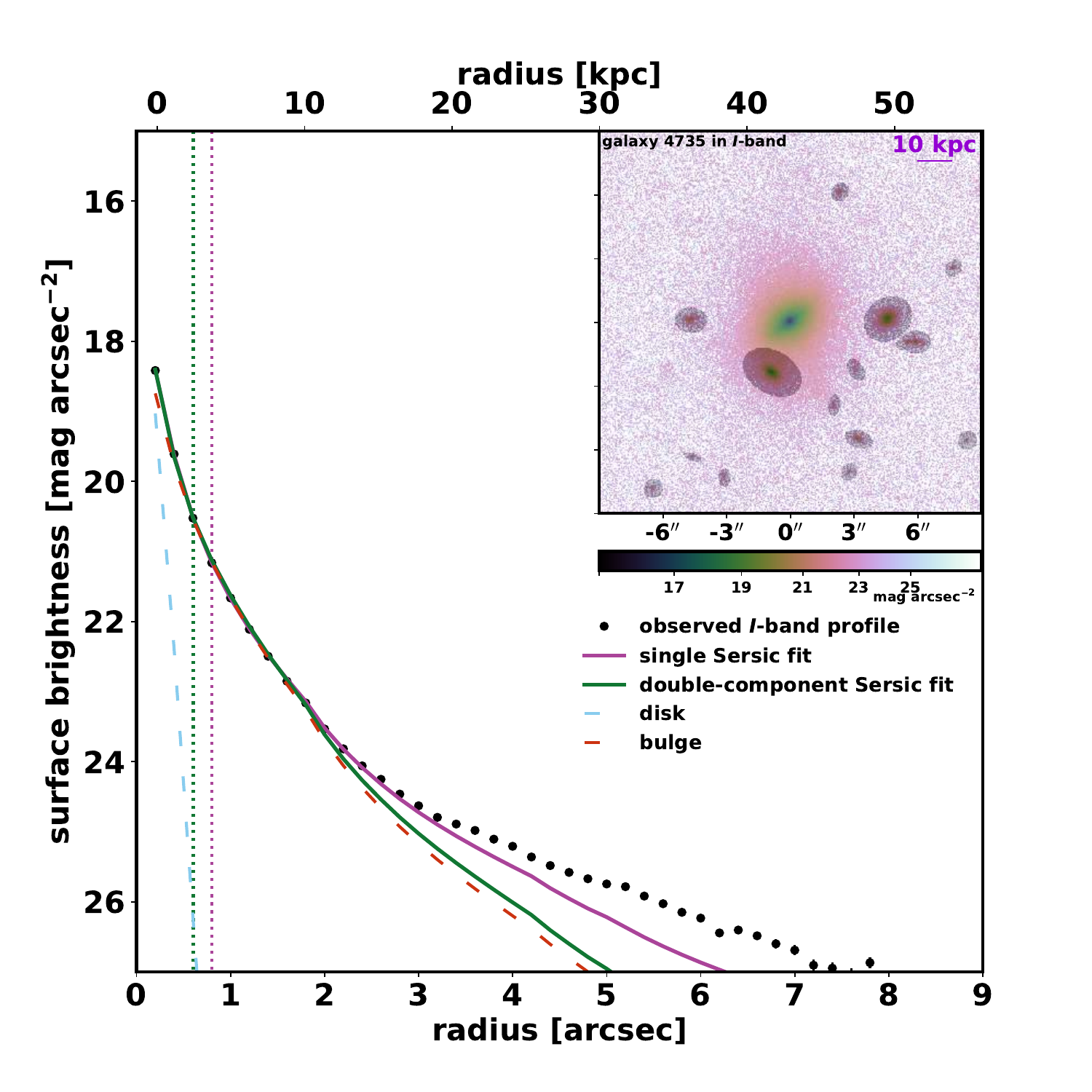}
 \caption{Surface brightness profiles of early-type galaxy \#4735 in the $H$-band (left panel) and $I$-band (right panel).}
 \vspace{-10pt}
 \label{fig:4735_SB}
\end{figure*}

\begin{figure*}[h]
\centering
    \includegraphics[page=10,width=0.4\textwidth]{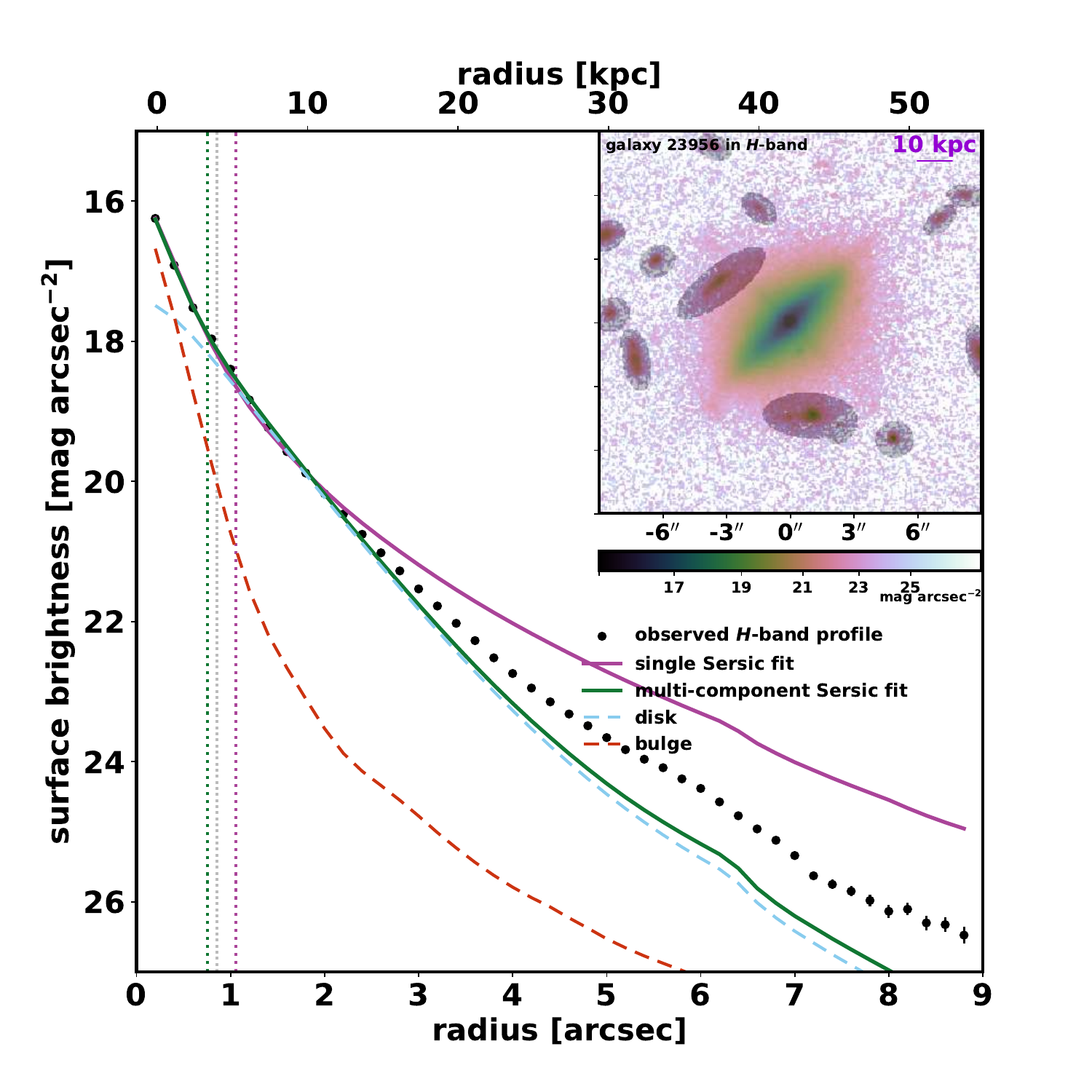}
	\includegraphics[page=9,width=0.4\textwidth]{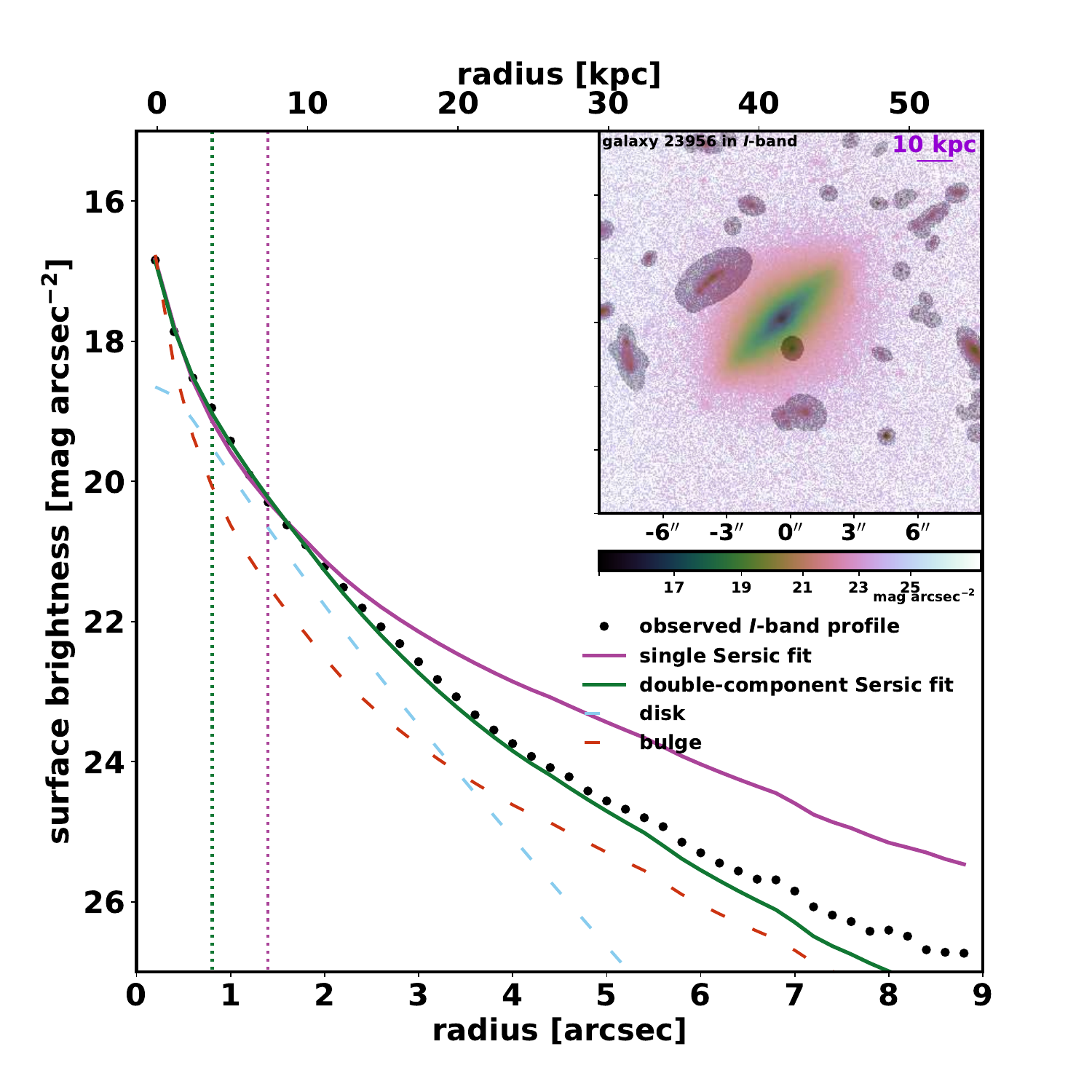}
 \caption{Surface brightness profiles of late-type galaxy \#23956 in the $H$-band (left panel) and $I$-band (right panel).}
 \vspace{-10pt}
 \label{fig:23956_SB}
\end{figure*}

\begin{figure*}[h]
\centering
   \includegraphics[page=17,width=0.4\textwidth]{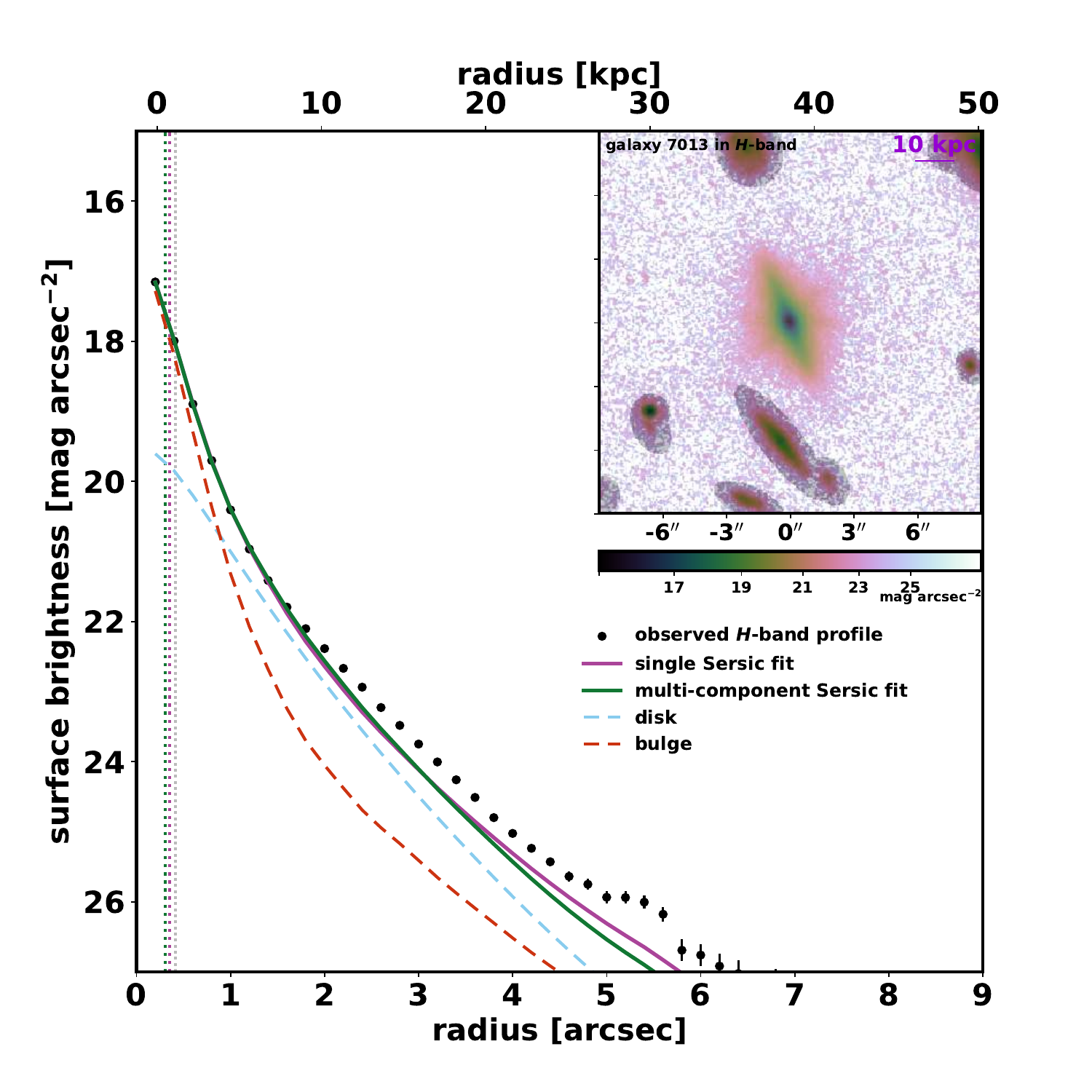}
	\includegraphics[page=16,width=0.4\textwidth]{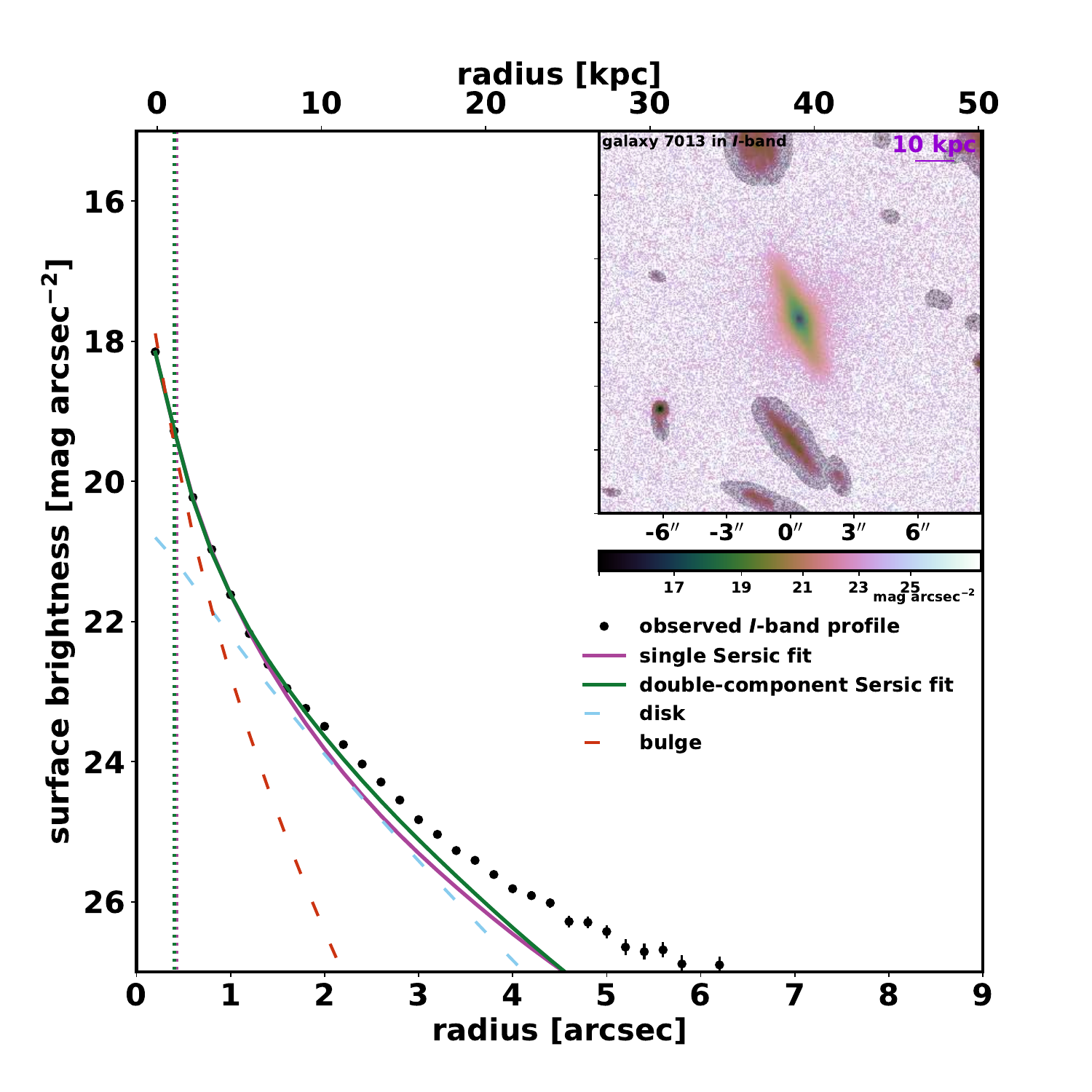}
 \caption{Surface brightness profiles of late-type galaxy \#7013 in the $H$-band (left panel) and $I$-band (right panel).}
 \vspace{-10pt}
 \label{fig:7013_SB}
\end{figure*}

\clearpage

\section{Color profiles}
\label{appen:colors}

\begin{figure*}[h]
 \centering
    \includegraphics[page=12,width=0.28\textwidth]{figures/colors_profiles.pdf}
    \includegraphics[page=11,width=0.28\textwidth]{figures/colors_profiles.pdf}
    \includegraphics[page=14,width=0.28\textwidth]{figures/colors_profiles.pdf}
    \includegraphics[page=13,width=0.28\textwidth]{figures/colors_profiles.pdf}
    \includegraphics[page=6,width=0.28\textwidth]{figures/colors_profiles.pdf}
    \includegraphics[page=1,width=0.28\textwidth]{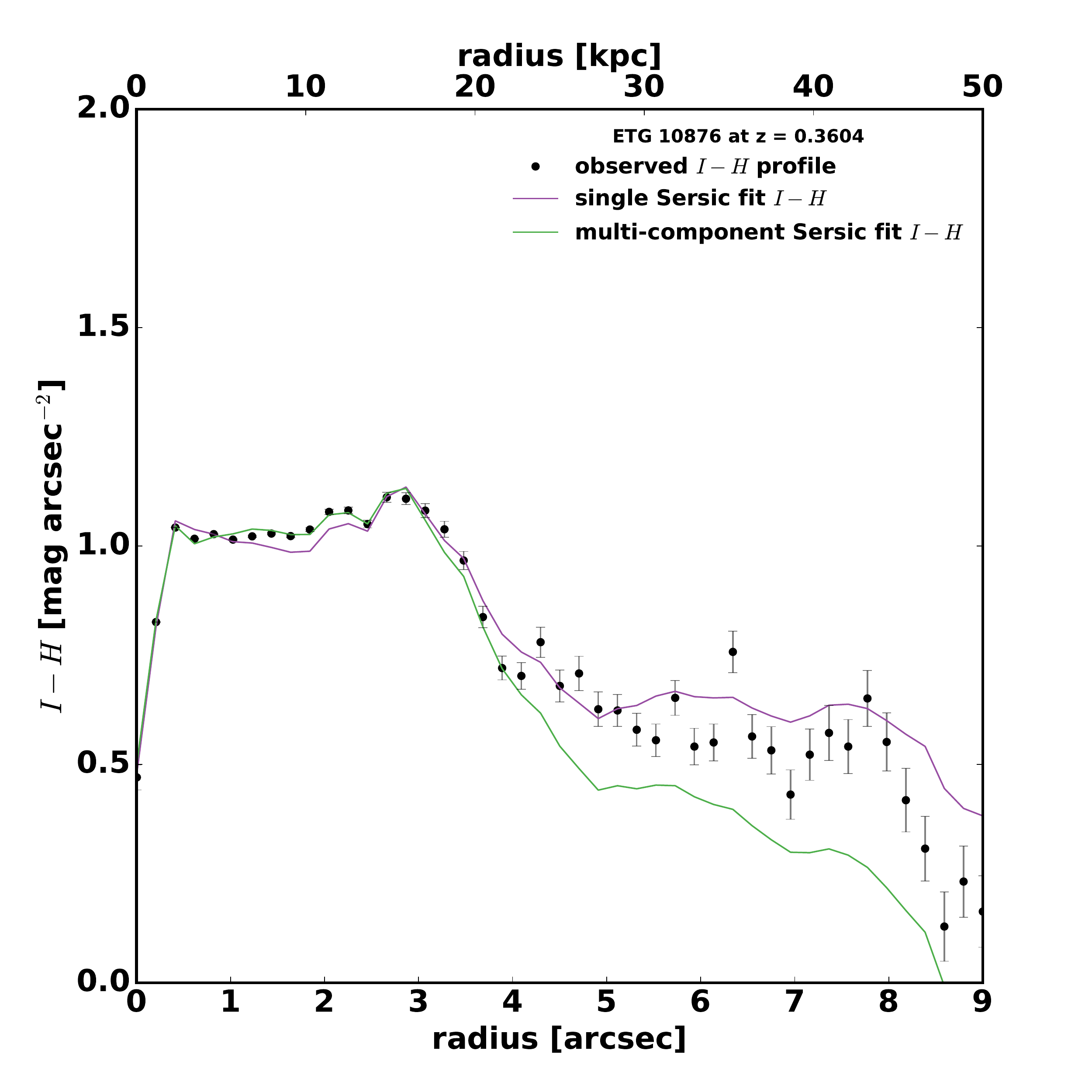}
    \includegraphics[page=3,width=0.28\textwidth]{figures/colors_profiles.pdf}
    \includegraphics[page=10,width=0.28\textwidth]{figures/colors_profiles.pdf}
    \includegraphics[page=5,width=0.28\textwidth]{figures/colors_profiles.pdf}
    \includegraphics[page=7,width=0.28\textwidth]{figures/colors_profiles.pdf}
    \includegraphics[page=15,width=0.28\textwidth]{figures/colors_profiles.pdf}
    \includegraphics[page=9,width=0.28\textwidth]{figures/colors_profiles.pdf}
    \includegraphics[page=16,width=0.28\textwidth]{figures/colors_profiles.pdf}
 \caption{Color profiles for our sample. Data points with larger error bars (> 0.3 mag) are chosen not be shown.}
 \label{fig:colors_appendix}
\end{figure*}

\section{Model residuals}
\label{appen:residuals}

\begin{figure*}[h]
 \centering
    \includegraphics[width=0.99\textwidth]{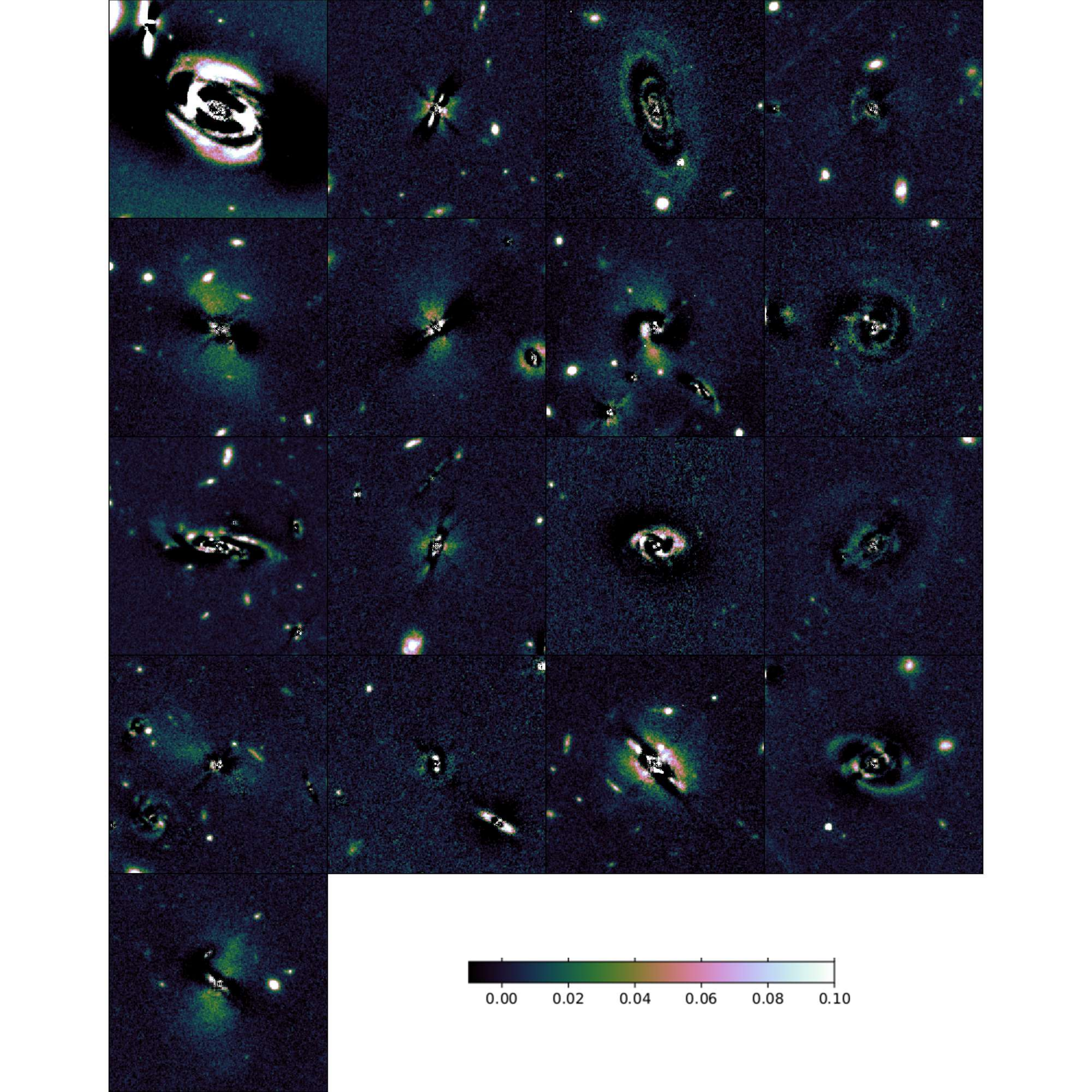}
 \caption{$H$-band residuals for the multi-component fits, after removal of the bulge+disk model from the original image. The images are in units of counts/sec.}
 \label{fig:residuals_appendix}
\end{figure*}

\begin{figure*}[h]
 \centering
    \includegraphics[width=0.99\textwidth]{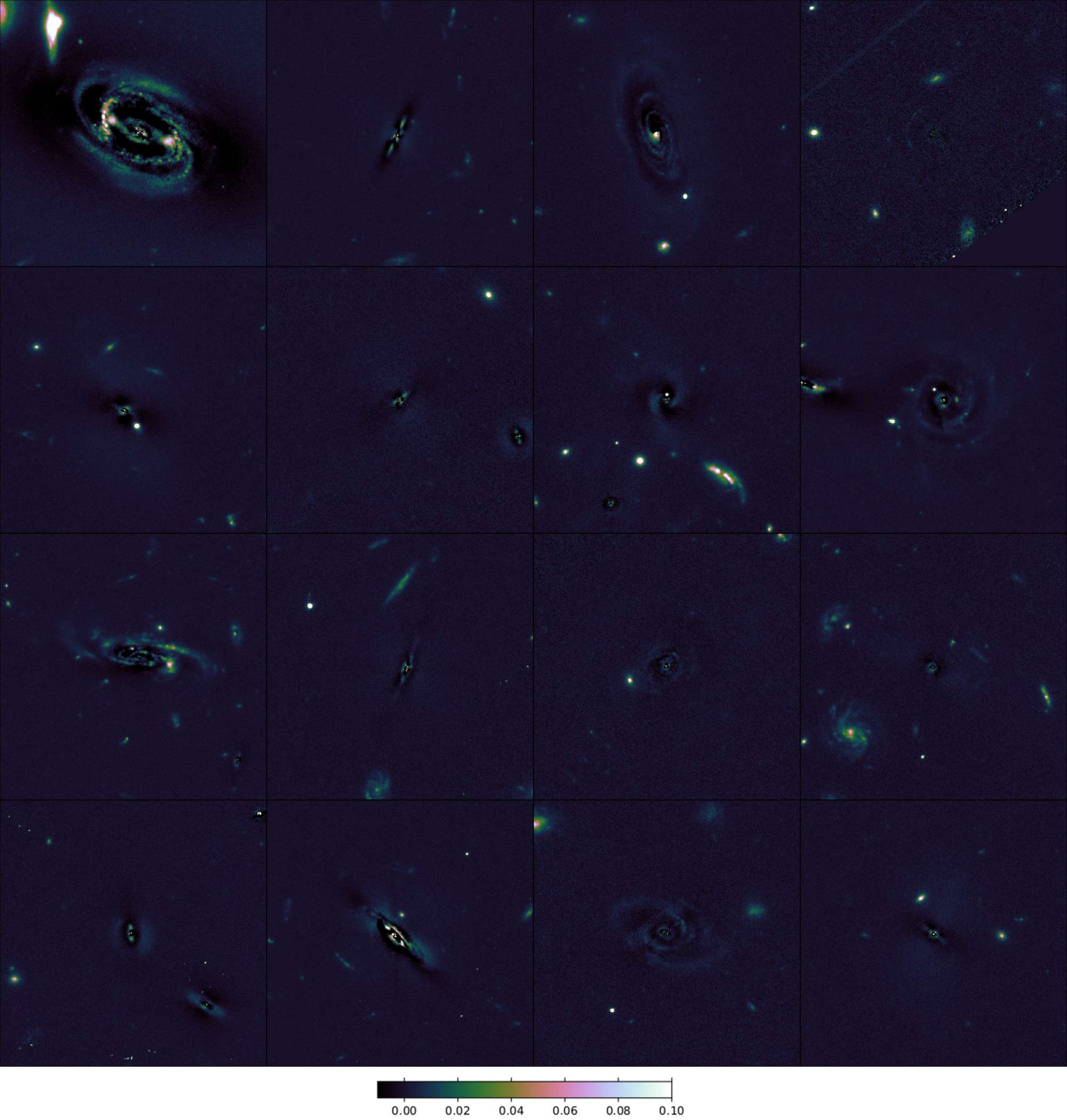}
 \caption{$I$-band residuals for the multi-component fits, in units of counts/sec.}
 \label{fig:residuals_appendix}
\end{figure*}

\end{appendix}

\end{document}